\documentclass[12pt,english,fleqn]{extarticle}

\def\textcoloron{0} 
\def\commentson{0}  

\usepackage{eurosym}
\usepackage{xcolor}
\definecolor{lmagenta}{RGB}{255,225,255}
\definecolor{lcyan}{RGB}{235,235,255}
\definecolor{green}{RGB}{10,150,10}
\usepackage[most]{tcolorbox}
\usepackage{amsmath,amsfonts,amssymb}
\usepackage{setspace}
\usepackage{hyperref}
\hypersetup{
  colorlinks=true,
  citecolor=black,  
  linkcolor=blue,
  urlcolor=blue
}
\usepackage[english]{babel}
\usepackage{graphicx}
\usepackage[errorshow]{longtable}
\usepackage{amstext}
\usepackage{pdfpages}
\usepackage{bbm}
\usepackage{rotating}
\usepackage{url}
\usepackage{array}
\usepackage{geometry}
\usepackage[T1]{fontenc}
\usepackage[utf8]{inputenc}
\usepackage{mathpazo}
\usepackage{afterpage}
\usepackage{caption}
\usepackage{amsmath}
\usepackage{bm}
\usepackage{subfig}
\usepackage{graphicx}
\usepackage{verbatim}
\usepackage{relsize}
\usepackage{booktabs,caption}
\usepackage[flushleft]{threeparttable}
\usepackage{booktabs,dcolumn}
\usepackage{amsmath}
\usepackage{float}
\usepackage{placeins}
\usepackage{lscape} 
\usepackage[round]{natbib}
\def\ts{\hspace{0.05em}}
\def\mts{\hspace{-0.05em}}

\def\ori{{\hbox{\scriptsize{\em O}}}}
\def\des{{\hbox{\scriptsize{\em D}}}}
\usepackage{bbm}
\def\smid{\,|\,}

\ifnum\textcoloron=1
 \def\magenta#1{\textcolor{magenta}{#1}}

\else
 \def\magenta#1{{#1}}

\fi

\usepackage{pdfcomment}
\pdfcommentsetup{date={}}
\ifnum\commentson=1
  \def\carlo#1{\pdfmargincomment[author=Carlo, color=blue,hoffset=-7.7in,date=" "]{#1}}
  \def\dennis#1{\pdfmargincomment[author=Dennis, color=magenta, hoffset=-7.7in,date=" "]{#1}}
  \def\ale#1{\pdfmargincomment[author=Alejandra, color=green, hoffset=-7.7in,date=" "]{#1}}
\else
  \def\carlo#1{ }
  \def\dennis#1{ }
  \def\ale#1{ }
\fi

\let\origfootnote\footnote
\renewcommand{\footnote}[1]{%
   \begingroup
   \origfootnote{\baselineskip=1.05\normalbaselineskip #1}%
   \endgroup}

\renewcommand{\baselinestretch}{1.3} 
\small
\normalsize 

\title{\fontsize{20}{20}\selectfont {{Transitivity in International Trade: \\ Evidence from Colombia-U.S. Firm Relationships}}\thanks{\protect\linespread{1.05}\protect\selectfont Novy gratefully acknowledges research support from the Economic and Social Research Council (ESRC grant ES/Z504701/1). We thank participants for helpful comments at the 2025 Midwest Trade Conference at Penn State, the Paris Trade Seminar/Sciences Po, Econometrics Society European Winter Meeting 2025. Martinez: School of Economics, University of Nottingham, Nottingham, UK and CITP; e-mail: alejandra.martinez@nottingham.ac.uk. Novy: Department of Economics, University of Warwick, Coventry, UK and CAGE, CEP/LSE, CEPR, CESifo, e-mail: d.novy@warwick.ac.uk. Perroni: Department of Economics, University of Warwick, Coventry, UK and CAGE, CESifo, e-mail: c.perroni@warwick.ac.uk.
\carlo{Acknowledgements TBC.} 
}\medskip\smallskip}

\author{\setcounter{footnote}{3}{\fontsize{15}{15}\selectfont Alejandra Martinez}
\and \setcounter{footnote}{4}{\fontsize{15}{15}\selectfont Dennis Novy}
\and \setcounter{footnote}{7}{\fontsize{15}{15}\selectfont Carlo Perroni}
}

\date{\vspace*{0.35in}{{\fontsize{13.5}{12.5}\selectfont{
October 2025} \\ \vspace{0.06in} 
}}} 

\begin{document}

{\let\newpage\relax\maketitle}

\renewcommand{\baselinestretch}{1.1}
\small \normalsize

\vspace*{0.05in}
\begin{abstract}
 \noindent 
 A large literature has documented transitivity as a key feature of social networks: individuals are more likely connected with each other if they share common connections with other individuals. We take this idea to trading relationships between firms: firms are more likely to trade with each other if they share common trading partners. Transitivity leads to a clustered pattern of relationship formation and break-up. It is therefore important for understanding how firms meet and how shocks propagate through firm networks.  We describe a method for detecting and quantifying transitivity in firm-to-firm transactions, based on systematic deviations from conditional independence across firm-to-firm relationships. We apply the method to Colombia-U.S. exporter-importer data and 
 show in counterfactuals that transitivity is a significant and economically meaningful factor in how firm networks adjust to cost shocks.  
  
\vspace*{0.15in}

\vspace*{0.3in}

\end{abstract}

\setcounter{page}{0} 

\thispagestyle{empty}\pagebreak

\smallskip
\smallskip

\medskip
\medskip

\renewcommand{\baselinestretch}{1.22}
\small \normalsize
\thispagestyle{empty}

\section{Introduction\ \hspace*{1in}\ 
\label{sec:intro}}

There is growing recognition that firm-to-firm relationships form complex networks rather than merely collections of independent buyer-seller pairs. In the context of international trade, firm-level networks play a central role in shaping cross-border flows of goods and services and in propagating shocks through global value chains \citep{antras2013organizing}. These relationships matter for performance: better supplier matches, for example, can reduce marginal costs and raise output \citep*{bernardmoxnes2019}.

\ale{Brownbag notes: This paper Chaney "The Network Structure of International Trade" I think might be important to add. This was brought up in the presentation.}
Despite this emphasis on the network-like features of the relationships between firms, a large literature has relied on a \textit{dyadic} paradigm of export link formation that omits explicit network effects \citep*{antrascostinot2011, BernardMox2018, eaton2022two, herkenhoff_et_al_2024}, treating a firm's decision to trade with one partner as being unrelated to its relationships with others---apart from general-equilibrium effects operating through the price system. This paradigm underpins both structural models and statistical approaches to firm matching, including the \emph{balls-and-bins} model of \citet{armenterkoren2014}. Even so, such frameworks have proven empirically successful in reproducing key qualitative features of production networks—such as their sparsity, skewed degree distribution, and variation across heterogeneous firms \citep*{BernardDhyne2022, CarvalhoNireiSaitoTahbazSalehi2021}---while retaining tractability. 

In contrast, research on social networks has shown that agent-to-agent links are typically not independent. For example, \citet*{Banerjee2013} emphasize the role of shared social ties in supporting trust in loan arrangements; and \citet*{dustmannetal2016} document how referrals from prior employers mitigate information frictions in job matching. A common feature of these examples from the social networks literature is that agents are more likely to engage with one another when they share a connection with a third party, leading to the formation of \textit{triads}. This \textit{transitivity} mechanism is a hallmark of social networks \citep*{newman2003, jacksonrogers2007, jackson2008social, currarini2009economic}, corresponding to the familiar principle that ``the friend of my friend is also my friend.''

In this paper, we take this paradigm to firm-to-firm networks in international trade, using microdata on Colombian flower exports to the United States to test for the presence of transitivity and assess its quantitative importance. Investigating the role of transitivity in this context is not only of conceptual interest. How firms connect has direct implications for economic performance. It determines how bilateral trade costs are shaped by network structure and influences how trade responds to shocks. Transitivity provides a concrete mechanism for the emergence of agglomeration effects. In standard trade models, clustering is often introduced through reduced-form assumptions about external economies, whereas here it arises endogenously: if a seller’s connection to one buyer makes it more likely that another nearby seller also connects to that buyer---unlike in the firm matching models prevalent in the literature---then relational spillovers are embedded directly in link formation. This mechanism can help explain why trade networks exhibit localized clustering and why shocks to one set of links propagate unevenly through the system. In this sense, transitivity offers a microfoundation for the types of agglomeration forces long emphasized in  economic geography. Transitive relationships also imply that link formation is path-dependent, potentially amplifying the clustering of gains or losses from policy interventions. They also influence network resilience and the pace of adjustment to external shocks.

\ale{Brownbag notes: Contribution in terms of other workhorse models of trade. Partial vs general equilibrium effects.}Yet, given that dyadic matching models can successfully explain key features of the data, any departure from the independence assumption requires empirical justification: it must be guided by patterns in the data which models featuring independent link formation are unable to explain. But detecting these patterns requires looking in the right place, with diagnostics that are specifically targeted to uncovering evidence of transitivity. One such diagnostic relates to the observed number of triads. If links are formed independently across dyads, then the expected number of triads in the network is pinned down by the number of observed dyads. An excess number of measured triads relative to this prediction indicates that a dyadic representation alone is insufficient to account for the observed network structure---\textit{no} specification based solely on independent link formation is capable of explaining this discrepancy, irrespective of its particular formulation.

Detecting symptoms of transitivity in real-world networks, however, is significantly less straightforward than the above simple triad count might suggest. Nodes differ mark\-ed\-ly in size, and larger or more central firms naturally appear in more triangles, even in the absence of transitivity. The empirical challenge, then, is to construct a statistical test for transitivity that can account for the effects of observed heterogeneity in node characteristics. This requires comparing the actual number of closed triads in the network to a benchmark where each link is formed independently, but in a way that reproduces the empirical heterogeneity in dyadic linking probabilities. To this end, we build on the ideas in \citet{graham2017}, \citet{dzemski2019}, and \citet{hughes2022},  to develop a nonparametric simulation approach that generates a benchmark distribution of triads for heterogeneous nodes under conditional independence. Our test then examines whether the triadic closures observed in the data significantly exceed this benchmark. If so, we interpret the excess as evidence of transitivity, meaning that the link formation process violates conditional independence and is shaped, at least in part, by shared linkages with third parties.

We apply this test to microdata on flower exports from Colombia to the United States. The dataset, provided by DIAN (Colombia’s National Directorate of Taxes and Customs), includes all transactions between exporters and U.S. importers from 2007 to 2019. We use this data to construct a bipartite network of Colombian exporters and U.S. importers and examine the extent of triadic closure over time and across regions using geographical distance between exporters as a proxy for latent links between them. Exporters in our sample are concentrated in two major, geographically distinct regions: the Savannah region near Bogot\'{a} and the Antioquia region near Medell\'{i}n. These regions differ in exporter characteristics, product variety, infrastructure, logistics networks, and access to international air freight, which may lead to differences in transitivity effects across the two regions. To account for these differences, we conduct our tests separately for each region.

Our test results show strong evidence of transitivity in the Savannah region. The observed number of triads in 2019 exceeds the 95th percentile of the benchmark distribution generated by synthetic samples by a margin of 5 to 10 standard deviations. This remains true under both continuous and threshold-based measures of proximity. In contrast, we find no statistically significant evidence of transitivity in Antioquia. This divergence suggests that transitivity is not a universal feature of trade networks but is conditional on institutional and logistical context.

\ale{Brownbag notes: We discuss calling "triadic closures" or something "transitive triangles" to the transitivity effect in this part to be more precise on what we do.} To arrive at a quantification of transitivity effects, we turn to a linear probability panel regression with fixed effects based on the entire panel dimension of the data from 2007 to 2019. Specifically, we derive estimates of the effect of past shared trading partnerships on the formation of a connection between a Colombian exporter and a U.S. importer in a current period. Since the formation of shared links is endogenous, we instrument shared trading partnerships with a shift-share-like instrument based on the exposure of Colombian exporters' trade to exchange rate changes for countries other than the U.S., focusing on the next five main flower export destinations and excluding flower-producing countries. We show that in the Savannah region, an increase in the number of common partners, when exporters are no more than 7 km apart (median of the distribution), has a statistically significant and sizable effect on the likelihood of linking to the same buyers. In the Antioquia region---where exporters are comparatively more geographically dispersed and production is more specialized---we cannot reject the no-transitivity null hypothesis.

Our panel estimates also address the concern that transitivity in cross-sectional tests may instead reflect \textit{latent homophily} arising from unobserved characteristics \citep{graham2017}. If a group of traders are connected because of latent homophily (e.g., a group of individuals sharing the same language), we would observe a higher density of connections within this group and, mechanically, more triangles than in the rest of the network. However, a shock to a specific link between two traders would have no knock-on effects on other links. 
The finding that an exogenous shock to a trade link affects related links directly points to a transitivity mechanism.

\ale{Brownbag notes: Policy implications} Finally, we gauge the role of transitivity in shaping how buyer-supplier networks respond to changes in economic conditions---such as trade cost shocks or trade policy shifts. For that purpose, we embed our panel estimates into a stochastic link formation model augmented for transitivity and conduct illustrative counterfactual simulations of a reduction in trade costs. The results highlight both the quantitative and qualitative implications of transitivity responses: the effects are sizable but unevenly distributed. For more connected traders, transitivity responses nearly doubles the resulting increase in link formation, while having weaker effects on less connected traders. These results underscore that transitivity can have significant implications for network responses even when the associated patterns are not readily apparent in the data and can only be uncovered through targeted diagnostics. 

Our analysis contributes to multiple strands of literature. First, we add to the literature on international trade and firm networks by identifying a specific empirical mechanism—transitivity—that complements models of matching and search (\citealp{BernardMox2018}; \citealp*{dhynekikawa2021}; \citealp{BernardDhyne2022}). 

Second, we relate our findings to literature on trade and information frictions (\citealp{allen2014}; \citealp{bernardmoxnes2019}; \citealp*{Morales2019}; \citealp{spray2021search}; \citealp*{Bailey2021};) and learning in trade \citep*{bernardjensen2004, koenig2010, fernandes2014}.\footnote{\citet{Bailey2021} find evidence that social connectedness boosts trade by reducing information asymmetry. \citet{bernardjensen2004} consider the role of exporter-to-exporter spillovers and find negligible effects from regional export activity. In our analysis, we use firm location coordinates, allowing us to measure geographical spillovers more precisely. \citet{fernandes2014} develop a statistical decision model in which a firm updates its prior belief about demand in a foreign market based on  the number of neighbors selling there. \citet{koenig2010} document how the influence of other exporting firms on exporters declines with distance.}




Third, we connect to the literature on endogenous transaction costs and, specifically, to recent work on endogenous trade costs and infrastructure \citep*{brancaccio2020, ganapatietal2024}.\footnote{\citet*{allen2024traveling} provide a further application of this approach.} This literature shows how the network-level implications of cost changes can depend on relational patterns not captured by conventional dyadic models. Our paper adopts a similar perspective of spillovers at the relationship level, but we explicitly focus on externalities arising from transitivity.

Finally, we bridge insights from sociology and management, where the role of triadic relationships in organizational networks has long been emphasized \citep*{choiwu2009, havila2004}. While these fields have focused on qualitative case studies, our approach uses large-scale administrative data to quantify and test the strength of triadic structures. To our knowledge, this is the first paper to formally test for and quantify transitivity in international firm-level trade networks using panel data.


The remainder of the paper is organized as follows. Section 2 discusses a statistical characterization of transitivity in firm-to-firm transactions that is applicable to any model of trade where firm-to-firm link formation involves a stochastic component (e.g., arising from random matching or directed search) and describes a statistical test that can be use to detect transitivity. Section 3 describes firm-level Colombia-U.S. exporter-importer data, which we use to carry out the test and derive quantitative estimates of transitivity effects. Section 4 presents results of illustrative counterfactuals experiments with a calibrated model. Section 5 concludes.

\section{Transitivity in firm-to-firm link formation}
\label{sec:baseline}

Our discussion will focus on a network of exporting and importing firms (the \textit{nodes} or \textit{vertices} of the network) that form links with each other (also referred to as \textit{edges}; \citealp{newman2003}). The links in this network are \textit{directed} links---importers purchasing goods from exporters. The outcome over a potential directed link from node $i \in I$ to node $j \in J$ is denoted by $y_{ij}\in\{0,1\}$, with 1 indicating a link (one or more sale transactions) and 0 absence of a link (no sale transaction). 

\subsection{Dyadic independence}
\label{sec:independence}

A general representation of a probabilistic model of directed dyadic link formation is
\begin{equation}
\Pr\big(y_{ij} = 1 \smid 
x_i, x_j, h_{ij} \big), 
\label{dyadic}
\end{equation}
where $x_i$ and $x_j$ are node-specific covariates and $h_{ij}$ are pair-specific covariates. One example of this is a logistic model:
\begin{equation}
\Pr\big(y_{ij} = 1 \smid 
(x^\ori_i, x^\des_i), (x^\ori_j, x^\des_j),\ts h_{ij} \big)
= \left(1 + \exp\big(-(\alpha + x^\ori_i + x^\des_j + \delta\ts h_{ij})\big)\right)^{-1},
\label{logistic_bl}
\end{equation}
where $x^\ori_i$ is often referred to in the network literature as $i$'s \emph{attractiveness} ($i$'s propensity to accrue new links), $x^\des_j$ is referred to as $j$'s \emph{gregariousness} ($j$'s expected degree or number of links) and $h_{ij}$ measures \emph{homophily} between $i$ and $j$ (similarities or costs between $i$ and $j$ that make them attractive or unattractive to each other, e.g., geographical distance between $i$ and $j$ or trade costs more generally). Another is the balls-and-bins model of \citet{armenterkoren2014}---which abstracts from homophily---with $x_i$ and $x_j$ being unidimensional covariates representing size:
\begin{equation}
 \Pr\big(y_{ij} = 1 \smid x_i, x_j\big) = 1 - \left(1 - \frac{x_i}{\sum_{k \in I} x_k}\right)^{\beta\ts x_j},
 \label{ballsbins_bl}
\end{equation} 
where $\beta >0$. This corresponds to the probability of obtaining at least one success from $n = \beta\ts x_j$ independent trials---the number of balls thrown by $j$---with a probability of success $x_i/\sum_{k \in I} x_k$ in each trial---the size of $i$' bin---expressed as the complementary probability of all trials failing.\footnote{This specification abstracts from homophily but can be generalized to account for it by incorporating $h_{ij}$ in the expression that determines the number of balls that can be thrown, e.g., $n = \beta\ts x_j {h_{ij}}^{\hspace{-0.03in}\kappa}$.}
Search-theoretic models such as the outsourcing model of \cite{antras2017margins}---where buyers incur a cost to sample potential suppliers delivering uncertain match surplus---are other examples. 

What all models that conform to (\ref{dyadic}) have in common is that they treat binary link formation as resulting from draws (equivalently, structural errors) that are independent across firm pairs. This implies that the status (active or inactive) of links between $i$ and $j$ with a third firm $k$ (an exporter or an importer) have no direct effect on whether $i$ and $j$ actively link with one another. In particular, no matter what the shape of $\Pr\big(y_{ij} = 1 \smid x_i, x_j, d_{ij}\big)$ is, it must be the case that
\begin{equation}
\hspace*{-0.15in}   \Pr\big(y_{ij} = 1 \smid x_i, x_j, h_{ij}, y_{ki}\ts y_{kj}=1  \big) = \Pr\big(y_{ij} = 1 \smid x_i, x_j, h_{ij}, y_{ki}\ts y_{kj}=0 \big)\quad \forall\;(i,j,k),  
\label{ind}
\end{equation}
i.e., having an active link with a common third party, $k$, has no direct relevance for the formation of an active link between $i$ and $j$ (\textit{independence of irrelevant alternatives}; \citealp{Graham2015SocialNetworks}).\footnote{Some components of $x_i$, $x_j$ or $h_{ij}$ (e.g., prices) may be determined endogenously in a way that depends on third links according to some equilibrium restriction $\Omega(y,x,d)=0$ (as in \citealp*{BernardDhyne2022}). However, when looking at equilibrium relationships, provided that all components of $x_i$, $x_j$ and $h_{ij}$ are accounted for (controlling for relevant observed variables or including categorical controls), statistical independence applies. That is, in a standard gravity setup (\citealp{Anderson2003}), multilateral resistance is endogenous but errors in gravity regressions can be taken as being independent across origin-destination pairs when origin and destination controls are included.} 

This independence property also has implications for independence with respect to observables other than $y_{ki}\ts y_{kj}$. For example, if we can observe $y_{kj}$ and not $y_{ki}$, but $d_{ki}$ is positively related to $y_{ki}$, then the above must also imply 
\begin{equation}
\hspace*{-0.15in} \Pr\big(y_{ij} = 1 \smid x_i, x_j, h_{ij}, h_{ki}\ts y_{kj}  \big) = \Pr\big(y_{ij} = 1 \smid x_i, x_j, h_{ij}\big)\quad \forall\;(i,j,k),  
\label{ind2}
\end{equation}
i.e., if $k$ and $j$ are linked, the degree of homophily between $k$ and $i$ can make no difference to whether $i$ and $j$ form a link. 
 
\subsection{Transitivity}
\label{sec:dependence}

Transitivity can be described as a ``triadic'' departure from independence where
\begin{equation}
\hspace*{-0.15in} \Pr\big(y_{ij} = 1 \smid x_i, x_j, h_{ij}, y_{ki}\ts y_{kj} =1 \big) \neq \Pr\big(y_{ij} =1 \smid x_i, x_j, h_{ij}, y_{ki}\ts y_{kj} = 0 \big)\quad \exists\;(i,j,k),     
\end{equation}
i.e., $i$ and $j$ having links with a common third node, $k$, can affect the likelihood of $i$ and $j$ forming a link \citep{jacksonrogers2007}.\footnote{An equivalent way to model transitivity is to posit that the effective degree of homophily between $i$ and $j$ is a latent endogenous variable given by a function ${\tilde h}(\cdot)$ of both $h_{ij}$ and $y_{ki}\ts y_{kj}$, and express the probability of $i$ and $j$ forming a link as $\Pr\big(y_{ij} = 1 \smid x_i, x_j, {\tilde h}(h_{ij}, y_{ki}\ts y_{kj})\big)$. This formulation explicitly represents transitivity as endogenous homophily or trade costs.\vspace*{3pt}} Several mechanisms can be invoked to explain why this might arise: reduction of information frictions (acquiring information through a shared partner; \citealp{calvo-jackson2004}); reduction in search or transaction costs (shared costs for linking to parties that are connected to each other; \citealp{Feld1981}); trust reinforcement (shared connections acting as enforcement devices; \citealp{Banerjee2013}); behavioral contagion (imitating the linking choices of connected partners; \citealp{centola2010}).

In such cases, if $k$ and $j$ are linked, homophily between $k$ and $i$—if relevant to the $y_{ki}$ outcome—\textit{can} make a difference to whether $i$ and $j$ form a link:\footnote{This is the form of violation we will focus on in our empirical analysis.} 
\begin{equation}
\hspace*{-0.15in} \Pr\big(y_{ij} = 1 \smid x_i, x_j, h_{ij}, h_{ki}\ts y_{kj}\big)
\neq \Pr\big(y_{ij} = 1 \smid x_i, x_j, h_{ij}\big) \quad \exists\;(i,j,k).
\end{equation} 

A general representation of models of link formation that allows for systematic transitivity in link formation is
\begin{equation}
\Pr\big(y_{ij} = 1 \smid 
x_i, x_j, h_{ij}, S_{ij} \big), 
\label{triadic}
\end{equation}
where $S_{ij}$ is a count of all third nodes, $k \not\in\{i,j\}$, with which $i$ and $j$ share a common link:
\begin{equation}
S_{ij} = \frac{1}{K} \sum_{k \not\in \{i,j\}}{ y_{ki}\ts y_{kj}},\quad K \equiv \#\big\{k \not\in \{i,j\}\big\}.
\end{equation}
For example, the balls-and-bins model (\ref{ballsbins_bl})  can be augmented by incorporating this transitivity term by writing
\begin{align}
&\hspace*{-0.15in}\Pr\big(y_{ij} = 1 \smid x_i, x_j\big) = 1 - \left(1 - \frac{{x}_i}{\sum_{k \in I} x_k}\right)^{\beta\ts {x}_j + \gamma\ts S_{ij}}.
\label{bb_augmented}
\end{align} 

Unlike (\ref{dyadic}), a structure such as (\ref{triadic}) explicitly incorporates network effects by making link formation directly interdependent across different $(i,j)$ pairs. This can have substantive implications not only for the resulting network structure—e.g., the formation of clusters and cliques \citep*{dekker2019transitivity}—but also for network dynamics—e.g., how the structure responds to exogenous shocks \citep*{navarro2018shock}.

\subsection{Testing for transitivity}
\label{sec:strategy}

Transitivity in link formation is a plausible feature of networks. But even if transitivity is theoretically plausible, invoking it in the context of firm-to-firm networks requires empirical justification---specifically, empirical patterns (statistics) that cannot be explained by a framework adhering to (\ref{dyadic}).

Transitivity patterns cannot be uncovered by generic data moments. Detecting them requires statistics specifically designed to test for transitivity. In a network with homogeneous nodes and homophily (i.e., a uniform random graph; \citealp{erdos1959}), a triad count is sufficient. Specifically, a network with $N$ nodes has $\frac{N(N-1)}{2} \equiv \overline{L}$ possible undirected links. If $L$ links are observed, then under the assumption of independent and uniform link formation, the estimated link probability is $\hat{p} = \frac{L}{\overline{L}}$. The expected number of triads is then $\widehat{T} = \frac{N(N-1)(N-2) \hat{p}^3}{6} = \frac{4L^3(N-2)}{3N^2(N-1)^2}$. This calculation extends to any (known or unknown) link formation process that applies uniformly across the network, i.e., such that  $\Pr\big(y_{ij} = 1\,|\, 
x_i, x_j, h_{ij}\big) = \Pr\big(y_{i'j'} = 1\,|\,x_{i'}, x_{j'}, h_{i'j'}\big)$ whenever $x_{i} = x_{i'},\, x_{j} = x_{j'},\, h_{ij} = h_{i'j'}\ \forall i,j,i',j'$.  Observing a number of triads, $T$, that significantly exceeds $\widehat{T}$ then indicates the presence of transitivity.

However, when nodes and homophily relationships are heterogeneous, a simple triad count can be misleading: patterns that resemble transitivity may arise purely due to variation in node characteristics or matching probabilities. In this case, the specific form of the link formation process also matters. The challenge, then, is to construct a test that both (i) accounts for heterogeneity, and (ii) remains agnostic about the underlying link formation process. Such a test should pick up transitivity patterns that could not be produced by \emph{any} model satisfying (\ref{dyadic})---regardless of heterogeneity and the underlying process of link formation.

To address (ii), we can use a flexible semi-parametric specification capable of approximating \emph{any} model consistent with the general structure in (\ref{dyadic}): \begin{equation}
y_{ij} = \mathbb{1}\{ \Omega_{ij} + \varepsilon_{ij} \geq 0 \},
\quad \varepsilon_{ij} \sim \text{Logistic}(0,s),
\label{flog}
\end{equation}
where $\varepsilon_{ij}$ has mean 0 and variance
$\pi^2 s^2/3$ and
\begin{equation}
\Omega_{ij} \equiv \sum_q
\left(\alpha^\ori_{iq}\, {\mathbb{1}}\big\{{\mathcal X}(x_j) = q\big\} + \alpha^\des_{jq}\, {\mathbb{1}}\big\{{\mathcal X}(x_i) = q\big\}
\right)
+ \sum_q \delta_q\, {\mathbb{1}}\big\{{\mathcal H}(h_{ij}) = q\big\},
\label{nonpar}
\end{equation}
with ${\mathcal X}(\cdot)$ partitioning the relevant ranges of $x_i$ and $x_j$ into $Q^{\mathcal X}$ bins and ${\mathcal H}(\cdot)$ partitioning the relevant range of $h_{ij}$ into $Q^{\mathcal H}$ bins, and with $\alpha_{iq}^\ori$ or $\alpha_{jq}^\des$ representing origin- and destination-specific coefficients that accommodate unobservable characteristics. The interaction of origin indicators with bin indicators for destination observables and of destination indicators with bin indicators for origin observables offers the highest possible level of flexibility---more than would be achieved by separately including additive origin, destination, and bin indicators.\footnote{Additive origin and destination indicators are nevertheless implied by this formulation and would be collinear with the other controls if included separately---as would interactions between origin and destination bin indicators.} As nonlinear models with fixed effects such as (\ref{flog}) produce biased estimates, we opt to use a linear probability model (\citealp{Cox1970}) that does not suffer from the incidental parameters problem:
\begin{equation}
y_{ij} = \Omega_{ij} + \varepsilon_{ij}.
\label{lpm}
\end{equation}
The same semi-parametric specification can be used to provide a flexible approximation for $S_{ij}$ as a function of the observables, $x_i$, $x_j$, $h_{ij}$.
\ale{Brownbag notes: Family of tests like this ones referred to as “randomization testing” (Cox 2006, chap 3). Properties of this tests are not discussed in papers we use. }
We can then construct a test based on the following statistic:
\begin{equation}
T = \sum_i \sum_{j\neq i}\ y^\textit{nr}_{ij}\ts S^\textit{nr}_{ij},
\end{equation}
where $y^\textit{nr}_{ij}$ and $S^\textit{nr}_{ij}$ are, respectively, the min-normalized residuals of regressions of $y_{ij}$ and $S_{ij}$ on observables using specification (\ref{lpm}):
\begin{align}
&y^\textit{nr}_{ij} = y^r_{ij} - \min_{i'j'}\{y^r_{i'j'}\},\ \ \ y^r_{ij} \equiv y_{ij} - {\hat p}_{ij}, \nonumber\\
&S^\textit{nr}_{ij} = S^r_{ij} - \min_{i'j'}\{S^r_{i'j'}\},\ \ \ S^r_{ij} \equiv S_{ij} - {\hat S}_{ij},
\end{align}
where ${\hat p}_{ij}$ and ${\hat S}_{ij}$ are the predicted values from the regressions. If we constructed the statistic directly from the observed $y_{ij}$ and $S_{ij}$ values, rather than from the regression residuals, it would become 
\begin{equation}
T' = \sum_i\sum_{j \neq i}\ y_{ij}\ts S_{ij} =\sum_i\,\ \sum_{j \neq i} \sum_{k \not\in \{i,j\}} {y_{ij}\ts y_{ki}\ts y_{kj}},
\label{triadcount}
\end{equation}
which corresponds to a raw triad count. When nodes are heterogeneous, this would be a misleading statistic. By constructing our statistic on the residuals from (\ref{nonpar})---i.e., the \textit{within-transformed} observations---we purge any systematic variation in $y_{ij}$ and $S_{ij}$ that can be accounted for by heterogeneity in observable characteristics across pairs.\footnote{This is similar to the statistic proposed by \citet{hughes2022}. The difference is that our test also incorporates residuals for $S_{ij}$. The rationale for this departure is that, when nodes are heterogeneous, we should not expect $S_{ij}$ to be independent of the characteristics of $i$ and $j$, even when the formation of links is independent of the characteristics of third nodes. Including the residuals for $S_{ij}$, instead of $S_{ij}$, accounts for any systematic variation across pairs that can be explained by heterogeneity in their characteristics. This is  structurally analogous to deriving a measure of partial correlation between two variables by  regressing each variable on a set of other variables and computing the correlation between the resulting residuals, in order to remove the influence of shared correlations with other variables \citep{cox1996multivariate}. Minimum normalization keeps all the terms in the summation positive, preserving monotonicity (it prevents pairs of negative residuals from being treated as equivalent to pairs of positive residuals of the same magnitude).} \dennis{Would a referee want to see more detail/discussion as to the validity of our procedure, along the lines of Hughes 2022? *** CARLO *** Very possible. I'll need to think a bit more about this.}

After computing $T$ in the data, we then derive a distribution of the same statistic for a comparable set of synthetic samples that are explicitly constructed so as to exclude transitivity, and we then use this synthetic distribution to carry out a one-tailed statistical test on the empirical $T$. Specifically, we obtain a no-transitivity distribution through the following steps:
\begin{enumerate}
\item[(i)] we repeatedly derive synthetic samples of combination of values for the independent covariates from the data via bootstrapping;
\item[(ii)] we run specification (\ref{lpm}) on the data and derive, for each synthetic sample of independent variables, a corresponding prediction for all $y$'s and $S$'s; by construction, this excludes transitivity effects on the outcomes;
\item[(iii)] for each synthetic combination of variables and corresponding predictions, we compute the $T$ statistic.
\end{enumerate} 
This gives us a distribution of $T$ statistics for synthetic samples that are comparable to the data point but involve no transitivity by construction and can therefore be used to carry out a one-sided tail test on the $T$ obtained from the data---the null hypothesis being that the data comes from a distribution of observations generated from a process that does not involve transitivity.\ale{Brownbag notes: I think this point where we run non-residual we also dont need to normalise.}\footnote{An alternative statistic can be constructed by directly replacing $y$ values with the corresponding normalized residuals in (\ref{triadcount}): $\check{T} = \sum_i\ts \ \sum_{j \neq i} \sum_{k \not\in \{i,j\}} {y^\textit{nr}_{ij}\ts y^\textit{nr}_{ki}\ts y^\textit{nr}_{kj}}$. \label{alternative_test}} 
The potential for overfitting bias in a saturated specification such as (\ref{lpm}) is not a concern in the context of our test: by giving the dyadic model the best possible chance to fit the data under the assumption of independence, we are only making it more difficult to reject the null.\footnote{If the aim were instead parameter inference, as in double/debiased machine learning (DDML) estimation \citep*{chernozhukov2018double}, demeaning with high-dimensional controls would need to incorporate regularization to ensure asymptotic consistency. We validate our test procedure through Monte Carlo simulations (results reported in the Online Appendix).}

\subsection{Transitivity vs. latent homophily \label{sec:latent}}

In cross-sectional data, it may be difficult to distinguish between triadic closures that result from transitivity and triadic closures arising from latent homophily \citep{graham2017}. A departure from independence in relation to observables, as described by (\ref{ind}), can also occur if there exists an augmented equilibrium relationship $\Pr(y_{ij} = 1 \smid x_i, x_j, d_{ij}, h_{ij})$ that satisfies conditional independence but includes an additional unobservable dyad-level variable $h_{ij}$. This variable represents a latent dimension of homophily that is indirectly correlated with both $y_{ki}$ and $y_{kj}$ through omitted terms $\omega_{ki}$ and $\omega_{kj}$. In this case, conditional independence still holds when we explicitly condition on $\omega_{ij}$, i.e.,
\begin{equation}
\Pr\big(y_{ij} = 1 \smid x_i, x_j, h_{ij}, \omega_{ij}, y_{ki}\ts y_{kj} \big)
= \Pr\big(y_{ij} = 1 \smid x_i, x_j, h_{ij}, \omega_{ij} \big)\quad \forall;(i,j,k),
\label{indh}
\end{equation}
but (\ref{ind}) is violated when $\omega_{ij}$ remains unobserved. 

Although both transitivity and latent homophily can raise the frequency of observed triadic closures, they imply different predictions for how the network responds to exogenous shocks. Specifically, if the true data-generating process satisfies (\ref{indh}), then a shock to $y_{ki}$ or $y_{kj}$---for instance, via a change in $h_{ki}$ or $h_{kj}$ that leaves $h_{ij}$ unaffected---has no effect on the probability of $i$ and $j$ forming a link. In contrast, if transitivity is present---whether through a violation of (\ref{ind}) when no latent dimension is present, or a violation of (\ref{indh}) when all dimensions of homophily are observable---then such a shock does affect the probability of $i$ and $j$ linking.\footnote{For example, in friendship networks, individuals who share a common interest (homophily) tend to form denser triads. But if such triadic patterns stem from homophily, the idiosyncratic breakdown of one friendship link would not affect the others.}

As this difference relates to dynamic responses, it cannot be easily detected in cross-sectional data, but it is comparatively easier to detect it in longitudinal data. We return to this issue in our empirical analysis where we exploit time variation to estimate the magnitude of transitivity effects using sources of exogenous variation in export links. Because the empirical strategy we use is closely tied to the specifics of our application and dataset, we defer a detailed discussion to the next section.

\section{Transitivity in Colombian-U.S. exporter-importer data}
\label{sec:empirics}

We look for empirical evidence of transitivity in firm-to-firm transactions data connecting Colombian flower exporters with U.S. importers. We begin by describing the data and presenting key descriptive statistics. We then implement the test outlined in Section~\ref{sec:strategy} on cross-sectional data. We then estimate the magnitude of transitivity effects from longitudinal variation. Finally, to illustrate the economic meaning of these transitivity effects, we present counterfactual experiments that incorporate our estimates.

\subsection{Data} \label{sec:data}

We focus on transactions between Colombian flower exporters and their U.S. buyers. The flower export sector is well suited to our analysis due to the perishability of flowers. This limits storage possibilities and necessitates frequent shipments, underlining the strong relational nature of transactions. The flower export sector is also well suited because of a tightly organized but shallow supply chain on both sides of the market---growers account for the majority of value added, and buyers are either commercial end users or sell directly to end users and retailers.

\begin{figure}[t]
\centering
     \includegraphics[
     height=3.5in, trim={20pt 53pt 0 50pt}, clip]{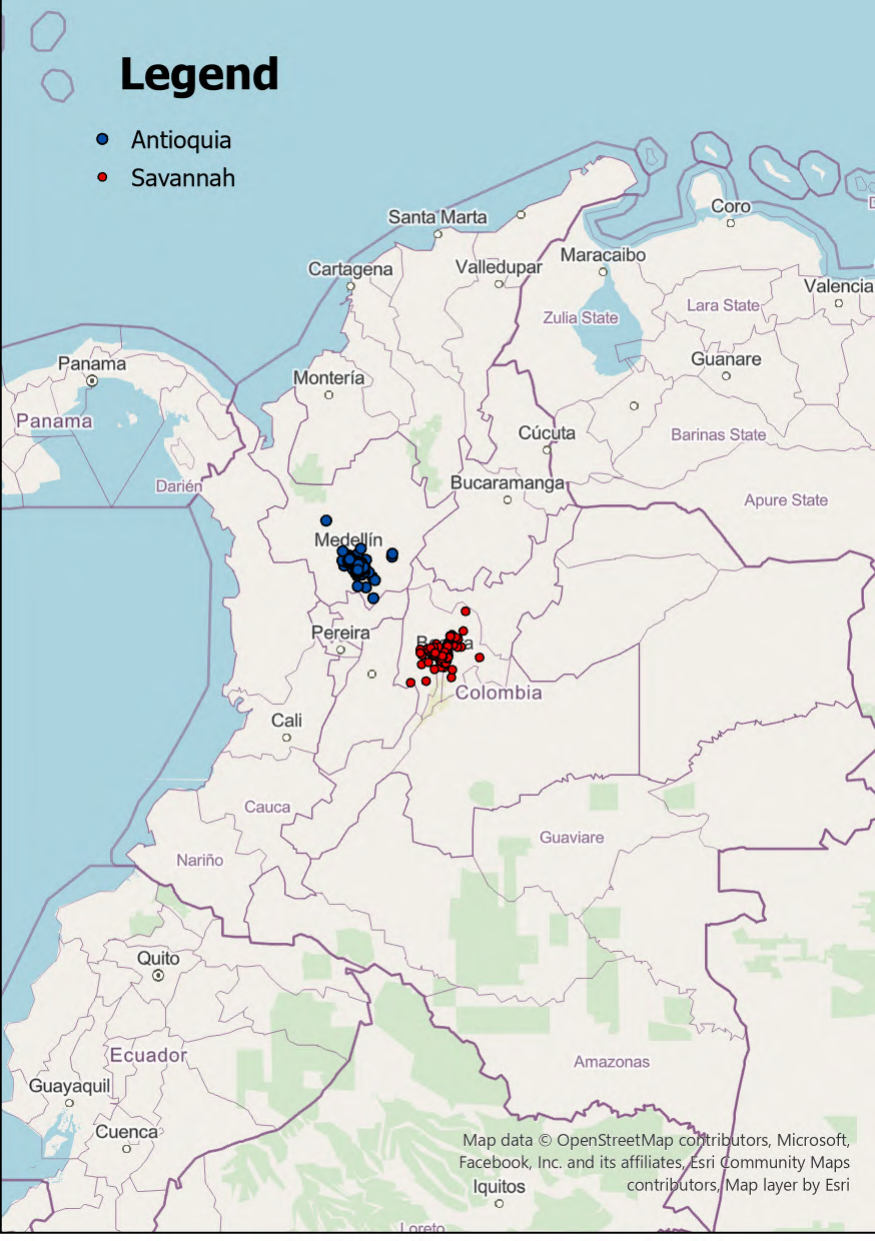}
\vspace*{0.02in}
\caption{Exporters' headquarters locations}
\label{fig:map}
\vspace*{0.1in}
         \scalebox{0.90}{
\begin{minipage}{0.8\textwidth}
\advance\leftskip 0cm
	{{\footnotesize{The figure shows the locations of headquarters for the sample of Colombian flower exporters. The red dots show exporters clustered in the Savannah region, while the blue dots show exporters clustered in the Antioquia region. Source: OpenStreetMap. }\par } }
\end{minipage}
}
\vspace*{0.1in}
\end{figure}

Colombia is the world's second largest flower exporter after the Netherlands. Annual exports exceed USD\,1.7 billion and primarily go to the United States, accounting for roughly 75\% of sales. The industry took off in the 1960s, capitalizing on Colombia’s unique high-altitude tropical climate that allows year-round cultivation—par\-ticularly of roses, carnations, and chrysanthemums---ensuring a steady supply for time-sensitive periods such as the U.S. Valentine's Day and Mother's Day seasons. The main production regions are the Savannah of Bogot\'a (Cundinamarca) and Antioquia (near Medell\'in). They offer ideal growing conditions and proximity to export infrastructure. The sector is composed of a mix of family-owned firms, larger agro-industrial companies, and foreign investors, with production characterized by labor-intensive greenhouse farming. 
\dennis{``foreign investors'': do those have their own farms? This sounds like multinationals and could mislead people. Maybe drop this phrase? *** CARLO ***: We need to look into foreign ownership (Alejandra is doing this). ***ALE*** I try to merge with my Colombian data that report foreign owned firms and none matched. I will try Orbis. *** CARLO *** We have discussed this and found evidence of links through joint ownership. Once we've put the data together, we'll check whether these patterns are correlated with distance and perhaps run a static transitivity test for 2019 with an alternative definition of the T statistic based on these links rather than proximity and see what we get.}

We use proprietary data from the National Directorate of Taxes and Customs (DIAN) in Colombia. The dataset contains detailed records of individual export transactions by Colombian firms to and from foreign firms over the period from 2007 to 2019. For each transaction, the dataset reports trade values in U.S. dollars. We aggregate those to annual frequency. A distinctive feature of the dataset is that it includes unique identifiers for Colombian exporters as well as the names of foreign importers, allowing us to construct firm-to-firm linkages. For our analysis, we focus on all Colombian exporters classified under HS code 06, which covers ``Live trees and other plants; bulbs, roots and the like; cut flowers and ornamental foliage.''

\begin{figure}[p]
\centering
 \vspace*{0.2in}
     \includegraphics[height=6.5in]{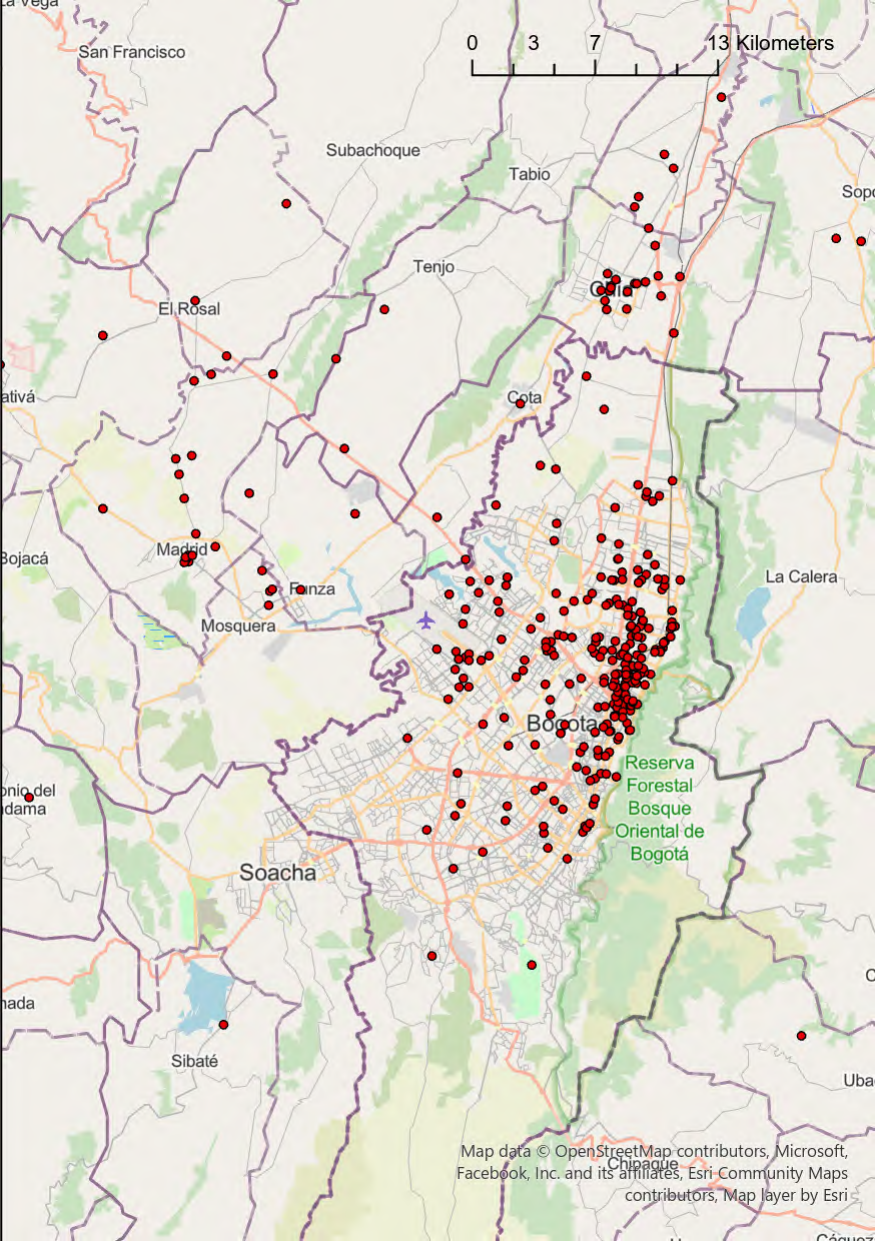}
\vspace*{0.2in}
\caption{Exporters' headquarters locations in the Savannah region}
\label{fig:map_savannah}
\vspace*{0.1in}
         \scalebox{0.90}{
\begin{minipage}{0.8\textwidth}
\advance\leftskip 0cm
	{{\footnotesize{The figure shows the locations of flower exporting firms in the Savannah region. Source: OpenStreetMap.} \par } }
\end{minipage}
}
\end{figure}

\begin{figure}[p]
\centering
 \vspace*{0.2in}
     \includegraphics[height=6.5in]{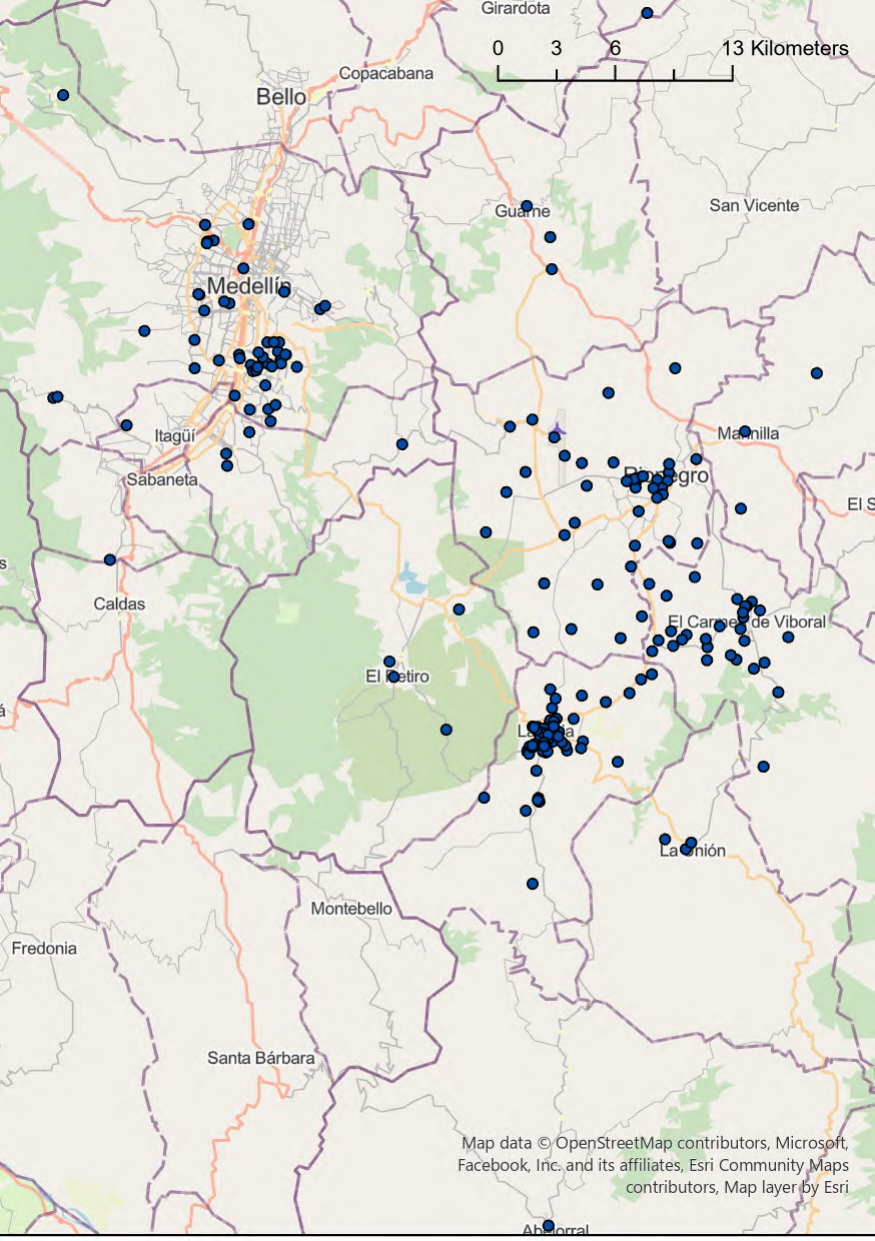}
\vspace*{0.2in}
\caption{Exporters' headquarters locations in the Antioquia region}
\label{fig:map_antioquia}
\vspace*{0.1in}
     \scalebox{0.90}{
\begin{minipage}{0.8\textwidth}
\advance\leftskip 0cm
	{{\footnotesize{The figure shows the locations of flower exporting firms in the Antioquia region. Source: OpenStreetMap.}\par } }
\end{minipage}
}
\end{figure}

Our sample consists of a balanced panel of Colombian flower exporters (sellers) from the Savannah and Antioquia regions and their U.S. importers (buyers). The panel includes 661 sellers and 794 buyers that are active for at least four years during the 2007–2019 period.\footnote{We restrict the buyer-seller network to include only U.S. importers and exclude transactions below USD 100, retaining about 79\% of the total industry export value over the entire period. We also exclude exporters that are active only intermittently, by which we mean exporters active in fewer than four years during the sample period. The 661 sellers break down into 435 in Savannah and 226 in Antioquia. The 794 buyers can buy from firms in both regions. We therefore have 661*794=524,834 buyer-seller relationships per year in our sample. Table~\ref{tab:sumstats} summarizes the active relationships; the remaining relationships are coded as zeros (the panel is rectangularized). Without excluding any sellers and buyers, we would have 1,042 sellers and 1,814 buyers. Table~\ref{tab:coverage} provides statistics on the coverage of the balanced panel relative to the full set of buyers and sellers.} Figure~\ref{fig:map} shows the geographic distribution of exporters across the Savannah region (in red) and the Antioquia region (in blue). Figures~\ref{fig:map_savannah} and~\ref{fig:map_antioquia} zoom into the two regions and provide a more detailed overview of their spatial distribution.

Table~\ref{tab:sumstats} provides descriptive statistics for our balanced network sample. Panel A reports region-level averages, computed by first calculating annual values for each variable and then averaging them across years. For each region, we report the number of active buyer–seller relationships (defined as those with positive transaction values); the number of active buyers and sellers; the number of new relationships (defined as those not active in the previous year); the number of discontinued relationships (those that cease to be active after having been active in the previous year); network density; total transaction values; and average sales per seller. Sellers in the Savannah region record substantially higher total sales, whereas sellers in Antioquia maintain a greater number of active relationships and are linked to more buyers. 

\begin{figure}[p]
\centering
     \hspace*{0.4in}\includegraphics[height=3in]{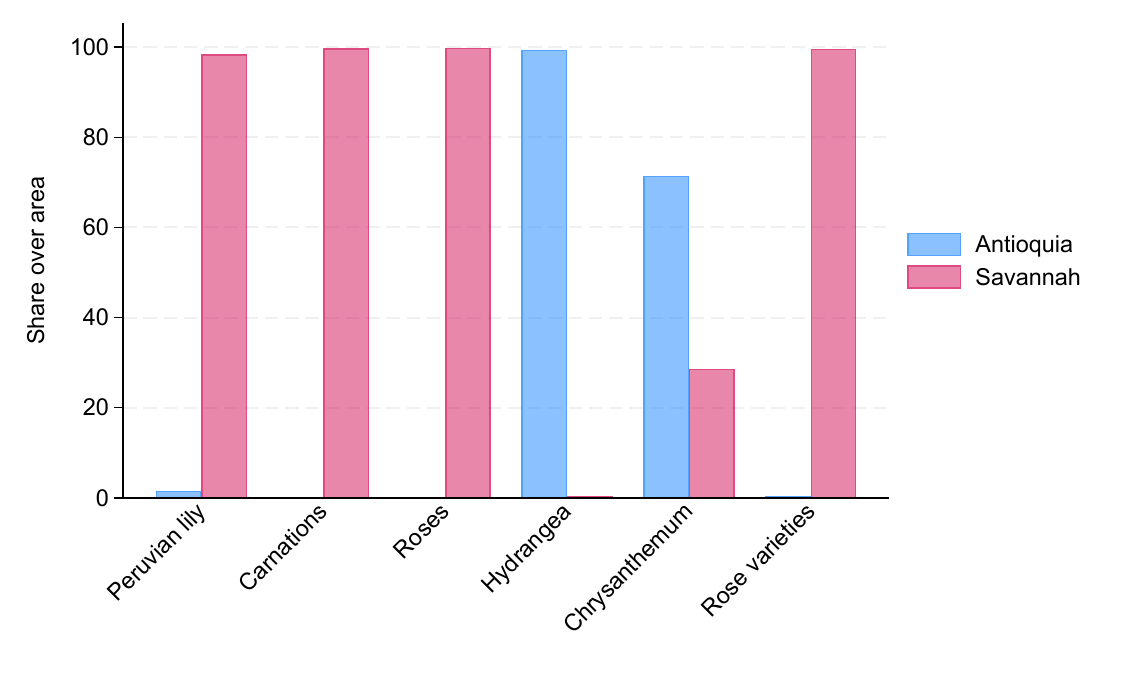}   
\vspace*{-0.2in}
\caption{Commercial flower species by region}
\label{fig:production}
\vspace*{0.1in}     
     \scalebox{0.90}{
\begin{minipage}{1\textwidth}
	\leftskip 0.01cm {{\footnotesize{The figure plots the share of land used for production (in hectares) for each type of flower in the Savannah and Antioquia regions. The remaining percentages represent production from other regions. Production data is sourced from the 2016 report of the Colombian Association of Flower Exporters (Asocolflores). The flowers shown account for 85\% of the total land used for flower production.}\par } }
\end{minipage}
}
\end{figure}

Panel B reports buyer-level averages. These are computed by first calculating, for each buyer, the average across all sellers they are connected to within a given region, and then averaging across all buyers. For each region, we report: the number of active sellers per buyer; the number of new and discontinued relationships (as defined above) per buyer; total transactions per buyer (measured as the buyer’s total annual purchases from all sellers in the region); and average sales per relationship (the average annual purchase amount per seller). Buyers of Savannah-region exports purchase more overall and engage in larger transactions per relationship. The two growing regions are also distinctly specialized in different types of flowers. As illustrated in Figure \ref{fig:production}, Savannah predominantly produces Peruvian lilies, carnations, and roses, while Antioquia tends to focus on hydrangeas and chrysanthemums.

\begin{table}[p]
\centering
\vspace*{-10pt}
\begin{center}
  \caption{Summary statistics of the network}
 \vspace*{0.05in}
\scalebox{0.9}{
  \centering
    \begin{tabular}{lcc}
    \toprule
    \textbf{Statistic} & \multicolumn{1}{l}{\textbf{Savannah}} & \multicolumn{1}{l}{\textbf{Antioquia}} \\
    \midrule
    \ \\[-10pt]
    \textit{A.\ Region-level averages (per year)} &       &  \\[3pt]
    Active buyer-seller relationships & 2,091 & 2,159 \\
    Number of active buyers & 379   & 392 \\
    Number of active sellers & 256   & 140 \\
    New relationships (not active in previous year) & 672   & 763 \\
    Discontinued relationships (not active in current year) & 658   & 648 \\
    Network density (active/possible relationships) & 0.006 & 0.012 \\
    Total transactions (USD millions) & \$796 & \$169 \\
    Sales per seller (USD millions) & \$3.0 & \$2.3 \\
          &       &  \\[-4pt]
    \textit{B.\ Buyer-level averages (per year)} &       &  \\[3pt]
    Number of active sellers per buyer & 4.24  & 4.45 \\
    Number of new relationships per buyer & 1.54  & 1.69 \\
    Number of discontinued relationships per buyer & 1.34  & 1.32 \\
    Total transactions per buyer (USD millions) & 1.36 & 0.31 \\
    Sales per relationship (USD millions) & 0.37 & 0.07 \\[5pt]
    \bottomrule
    \end{tabular}%
    }
  \label{tab:sumstats}
 \vspace*{0.1in}

    \scalebox{0.90}{
\begin{minipage}{0.93\textwidth}
\advance\leftskip 0cm
	{\footnotesize {
\textit{Notes}: The table reports summary statistics for the balanced network sample. Panel A reports region-level averages across years. Statistics include the number of active buyer-seller relationships; the number of active buyers and sellers; the number of new relationships (relationships that were inactive in the previous year and became active in the current year, excluding the first period); the number of discontinued relationships (relationships that were active in the previous year but became inactive in the current year, also excluding the first period); the network density defined as the ratio of active buyer-seller relationships over all possible buyer-seller relationships in the region; total transactions referring to the total of all transactions across buyer-seller relationships in the region; and sales per seller, measured as the average total sales of a seller with all their buyers. Panel B reports buyer-level averages. For each region, values are first averaged over years at the buyer level, then averaged across buyers. Statistics include the number of active sellers per buyer; the number of new and discontinued relationships per buyer; total transactions per buyer; and sales per relationship, defined as the buyer's average purchases per active seller. All monetary values are in constant 2018 USD millions.
 
} \par }
\end{minipage}
}
\vspace*{0.1in}
\end{center}
\end{table}

\begin{figure}[p]
\centering
  \vspace*{-.75in}
\textbf{\small Savannah\hspace*{1.9in}Antioquia}\\[-1pt]  
     \includegraphics[height=3.2in,trim= {0 0 0 18pt}, clip]{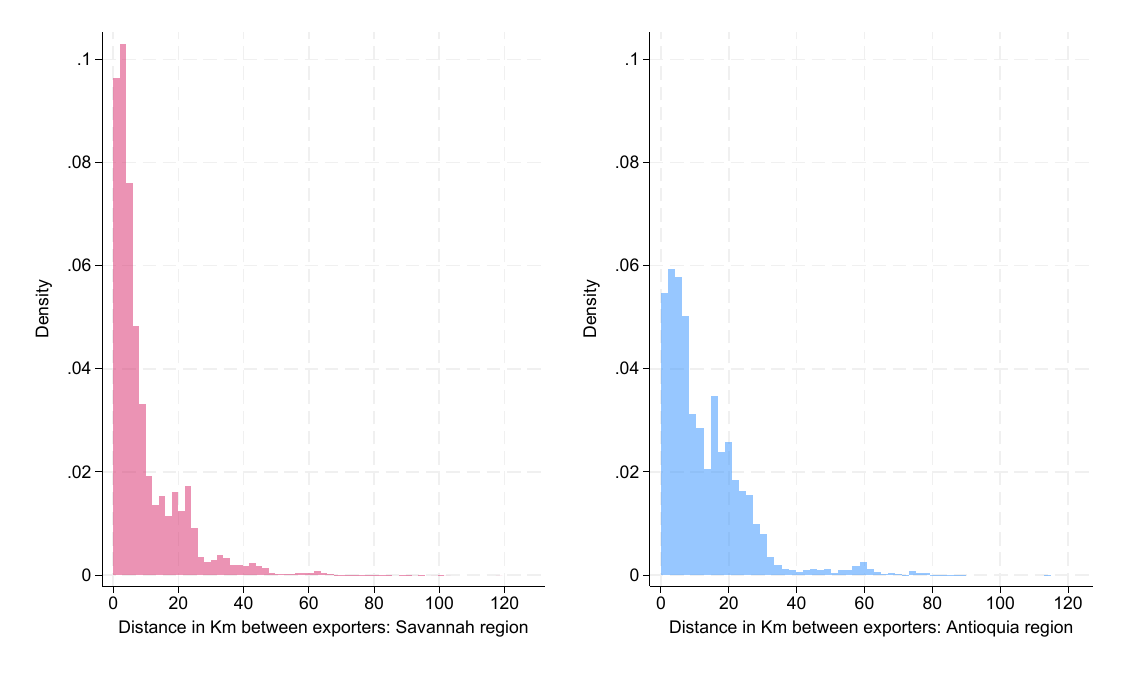}
\vspace*{-0.1in}
\caption{Distribution of distances between exporters}
\label{fig:distr_dist}
\vspace*{0.05in}      
     \scalebox{0.90}{
\begin{minipage}{1\textwidth}
\advance\leftskip 0.5cm
	{{\footnotesize{The figure shows the distribution of distances (in kilometers) between Colombian flower exporters within each region.}\par } }
\end{minipage}
}
\end{figure}

Our analysis also relies on pairwise distances between flower exporters. This is based on information on the geographical location of firms' headquarters from the Single Business and Social Registry (RUES), which contains the addresses of all Colombian firms.\footnote{The address reported for a firm in RUES is its legal address, used for all legal and taxation purposes, which we take as the firm's headquarters location. A firm has to update its legal address if it changes the location from which it operates.} We use the unique tax ID, sourced from the customs data, to match the DIAN and RUES records. Using ArcGis, we then geocode the addresses and find the latitude and longitude coordinates for all exporters that report headquarters addresses.\footnote{ArcGis reports the level of precision with which it can geocode an address. It can be the street associated with the address, a point of interest, or otherwise the municipality. For 110 exporters where the match is at the municipality level, we are nevertheless able to pinpoint coordinates from Google Maps. For 15 exporters, we are not able to identify the municipality and therefore omit those from the sample.} Our final dataset only contains exporters that are located in the Savannah or Antioquia regions, and we are able to geocode 98\% of them at least to the point of interest level.\footnote{Exporters in rural areas are typically geocoded at the \textit{vereda} level. There are approximately 30,000 \textit{veredas} and, they can usually be crossed on foot in two hours. For exporters with two addresses in the same year---208 firms---we use the mean latitude and longitude coordinates of the two locations. In case of two addresses per firm, we use different ways of selecting the headquarters location. If one address is in the main city of a region (Bogot\'{a} or Medell\'{i}n), we take this as the headquarters location (this applies to 108 exporters). For the remaining exporters, if both addresses are in the same municipality and less than 5\,km apart, we average coordinates. For exporters with two addresses above the 5\,km threshold, we select the one with the lowest numerical municipality code as the headquarters location. Only 29 firms change their headquarters locations in our sample period. In 2009, one seller moved its headquarters from Savannah to Antioquia.} 

Figure~\ref{fig:distr_dist} shows the distribution of great-circle pairwise distances (in kilometers) between exporters located within each region. Exporters in the Savannah region are on average located closer to each other, with a median and mean distance respectively equal to 6\,km and 9\,km. Exporters in Antioquia are further apart, with a median and mean distance equal to 9\,km and 12\,km. This difference reflects underlying geographic patterns. While Savannah exporters are mostly concentrated around Bogot\'a, Antioquia exporters are more spatially dispersed, with production sites located near multiple cities (Medell\'in, Rionegro, La Ceja, El Carmen de Viboral).

\subsection{Cross-sectional evidence: Transitivity test} \label{sec:crosssectionaltest}

\begin{figure}[t]
\centering
\vspace*{-0.15in}
\hspace*{-0.55in}     \includegraphics[height=2.4in]{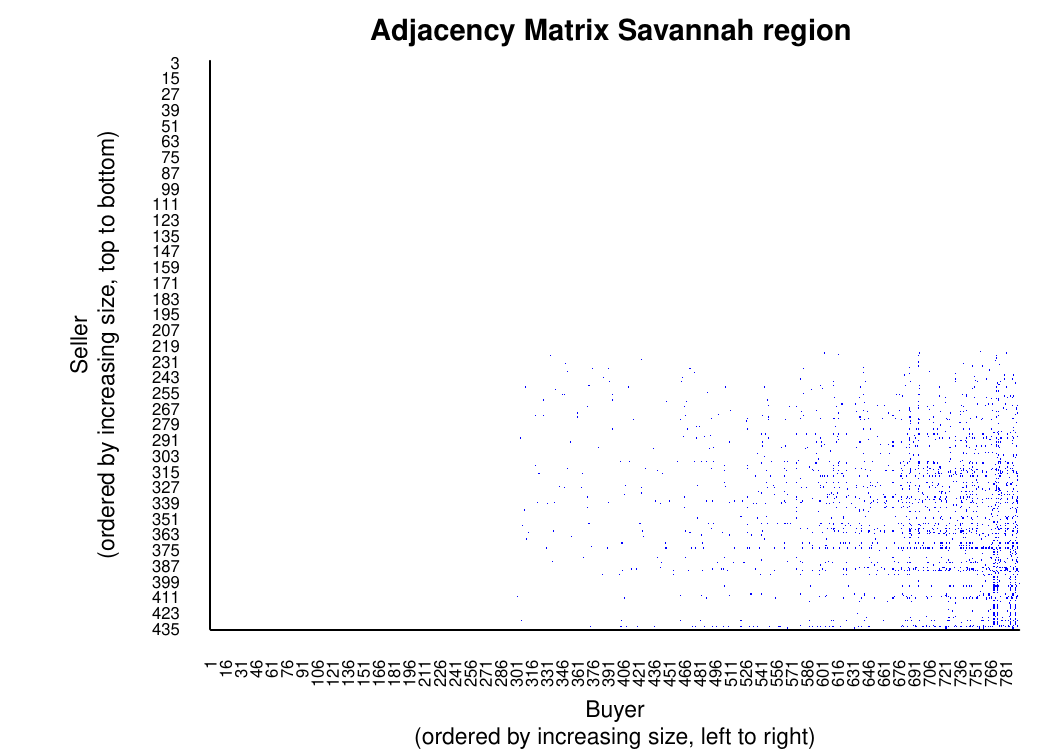}   \includegraphics[height=2.4in]{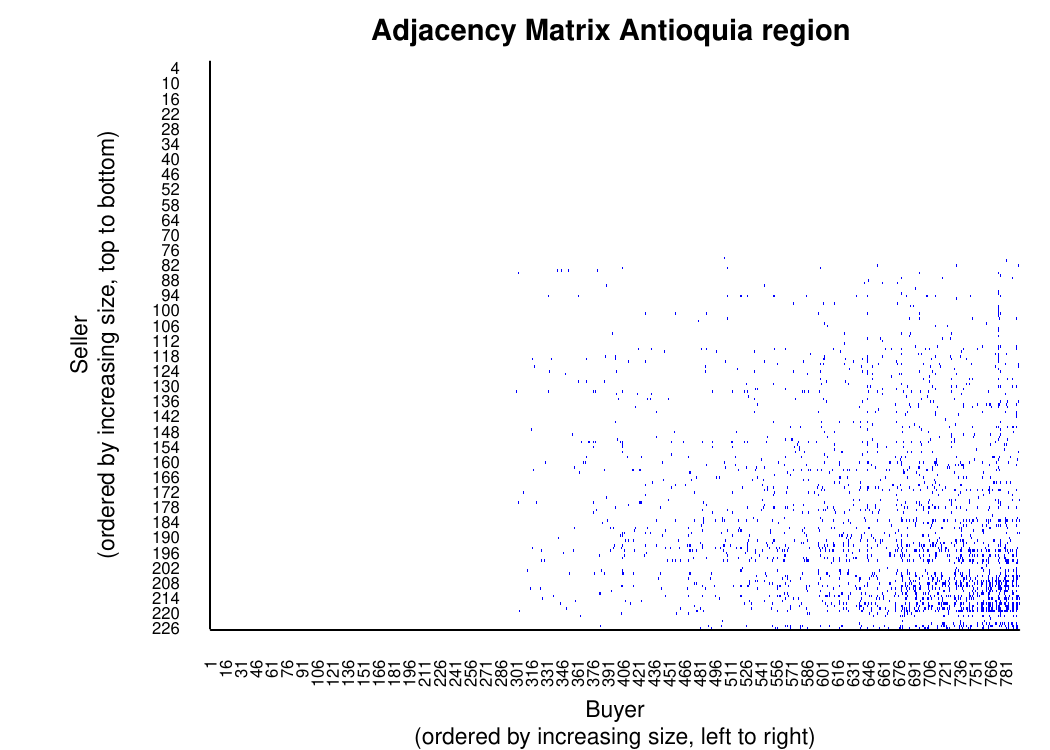}\hspace*{-0.3in}
\vspace*{0.05in}
\caption{Buyer-seller adjacency matrices}
\textbf{for the Savannah and Antioquia regions}\vspace*{0.1in}
\label{fig:adj}
\vspace*{0.03in}
     \scalebox{0.90}{
\begin{minipage}{1\textwidth}
	{\footnotesize \leftskip 0.5cm {The figure shows the adjacency matrices for the Savannah region (left) and the Antioquia region (right) for the 2019 cross-section. Buyers are on the horizontal axis and sellers on the vertical axis. The blue dots indicate relationships being active. Buyers and sellers are sorted from largest to smallest based on their total transactions.}\par }  
\end{minipage}
}
\vspace*{0.125in}
\end{figure}

Figure \ref{fig:adj} shows the adjacency matrix ($y_{ij}$) for exporters and importers by region in the 2019 cross-section. Sellers ($i$) are on the vertical axis and buyers ($j$) on the horizontal axis, ordered by size (total value of transactions) from smallest to largest from left to right and from top to bottom. There is a clear positive correlation between size and number of connections, particularly on the buyer side of the network. 

\begin{figure}[h]
\centering
\textbf{\small Savannah }\vspace*{0.1in}

\hspace*{-0.45in}
\includegraphics[height=1.95in]{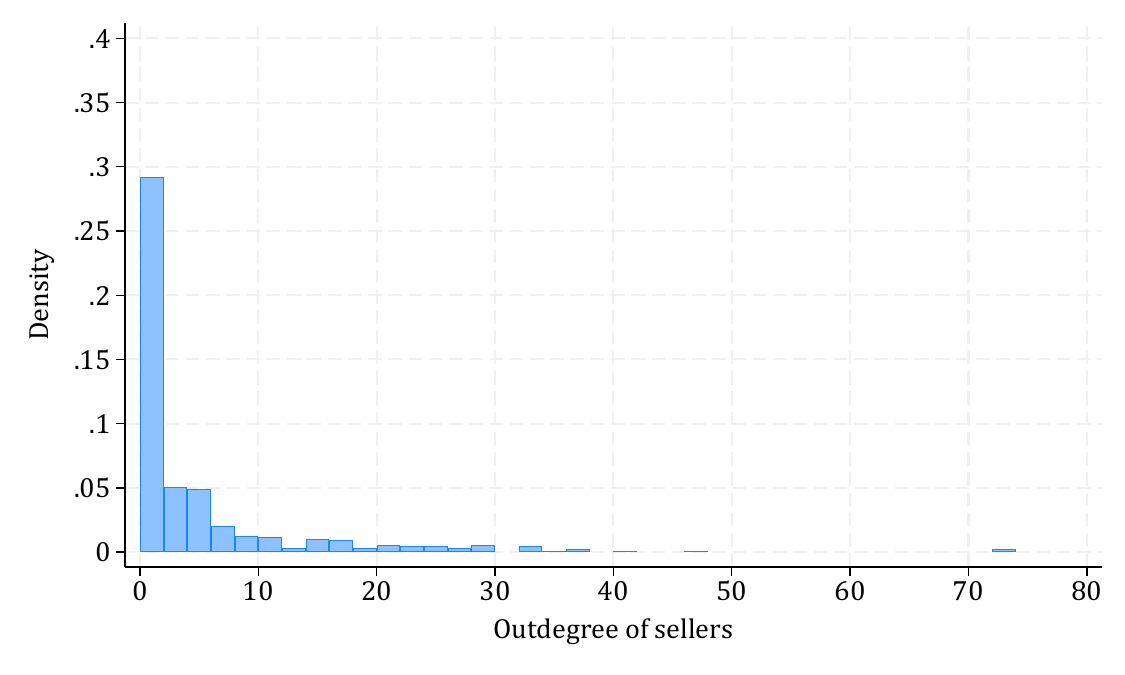}
\includegraphics[height=1.95in]{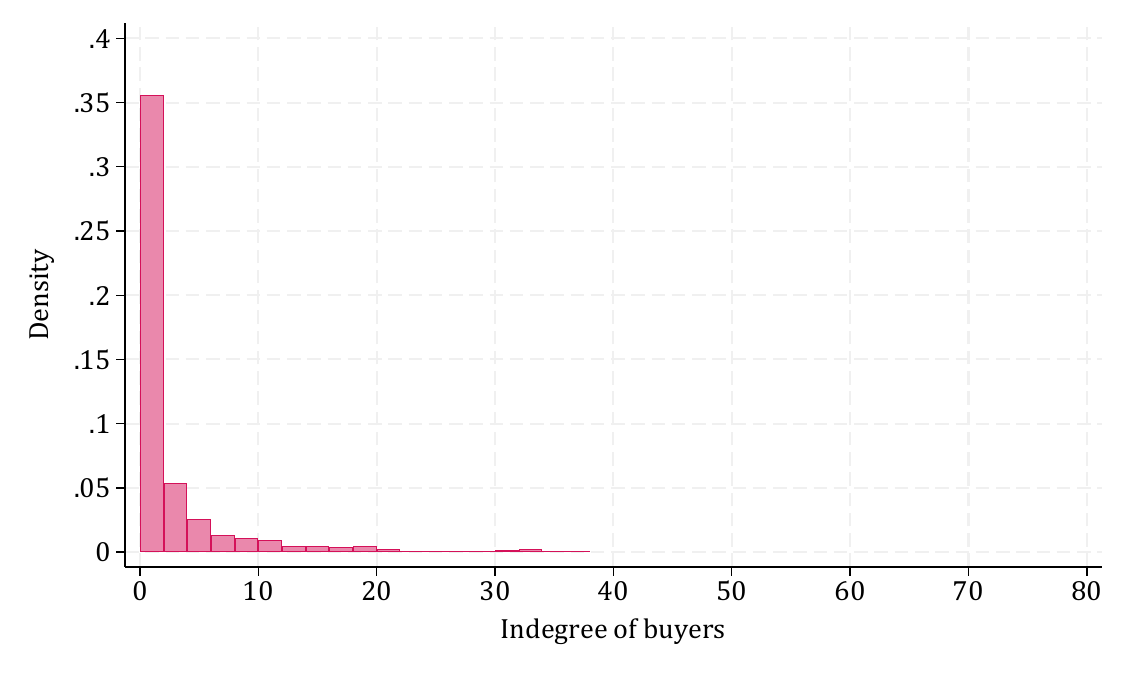}  
\hspace*{-0.4in} 
\vspace*{0.1in}

\textbf{\small Antioquia }\vspace*{0.1in}

\hspace*{-0.45in}
\includegraphics[height=1.95in]{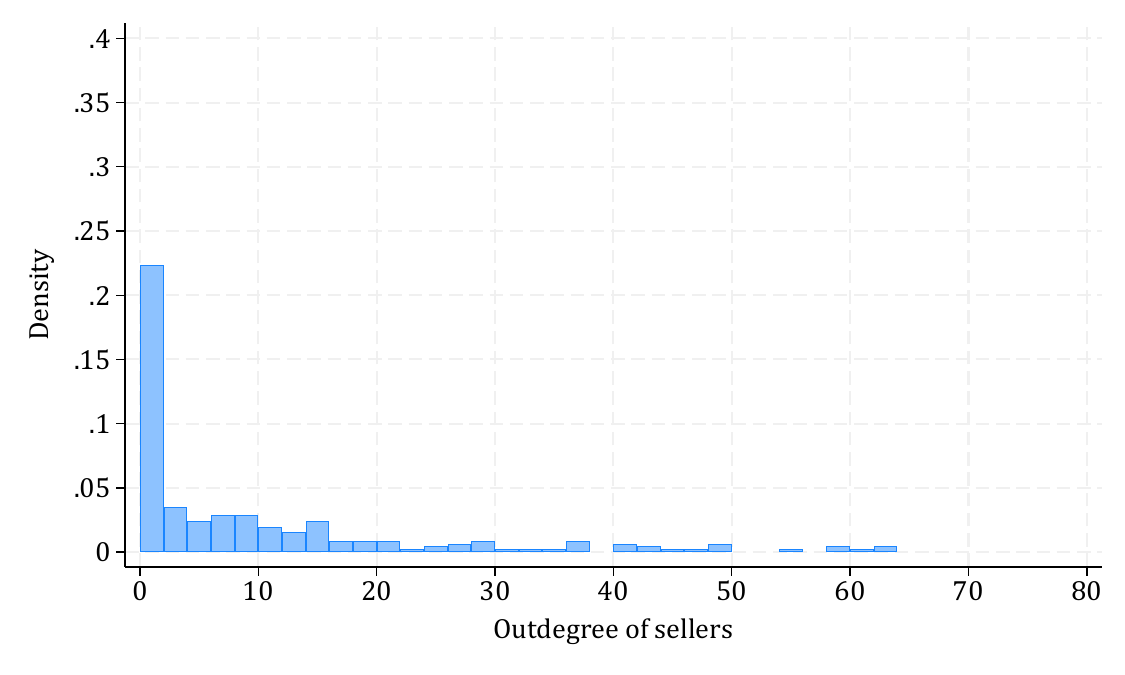}
\includegraphics[height=1.95in]{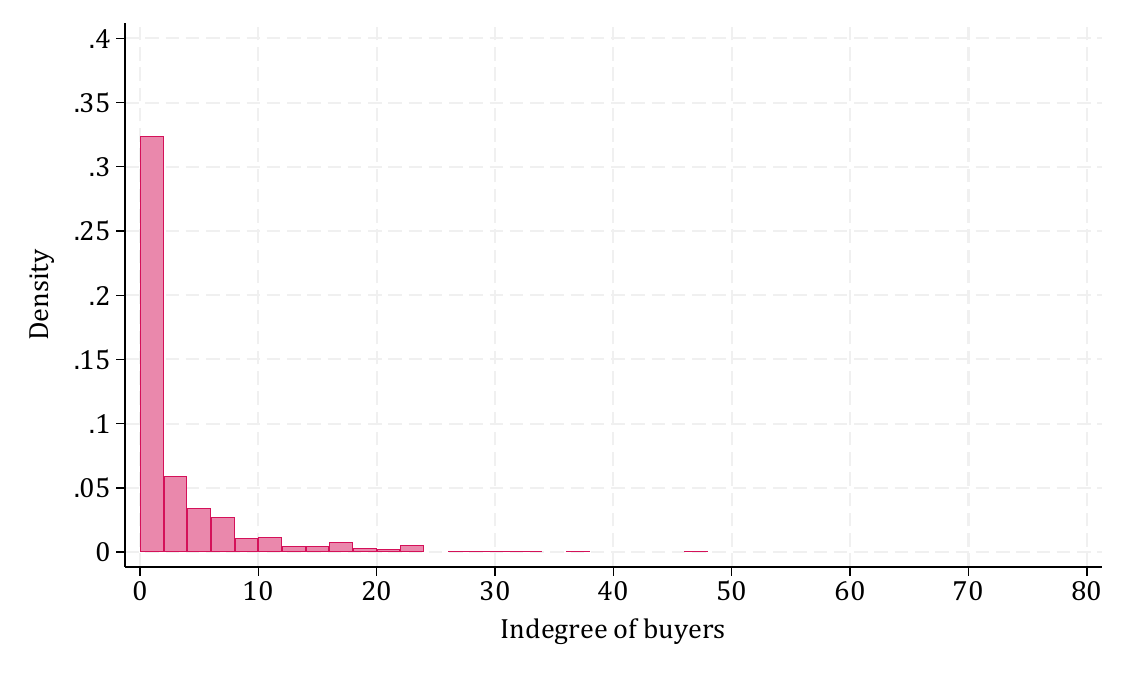}
\hspace*{-0.4in} 
\caption{2019 outdegree and indegree distributions}
\label{fig:degree}
\vspace*{0.1in}
     \scalebox{0.90}{
\begin{minipage}{1\textwidth}
	{\footnotesize \leftskip 0.5cm {The figure shows the outdegree (left) and indegree (right) distributions for the Savannah and Antioquia regions for 2019. 
    }\par} 
\end{minipage}
}
\vspace*{0.1in}
\end{figure}

For the same cross-section, the left panels of Figure \ref{fig:degree} show the outdegree distributions for sellers (number of buyers per seller). For both regions, the outdegree distributions exhibit a similar pattern: the vast majority of sellers have few connections, and a very small number of sellers are highly connected. The right panels of Figure \ref{fig:degree} show the indegree distributions for buyers (number of sellers per buyer). Here, we see different patterns across the two regions: the Savannah distribution is skewed further to the left. That is, buyers connected to sellers in the Savannah region connect to relatively few sellers on average, whereas buyers connected to sellers in the Antioquia region connect with more sellers.

We test for the presence of transitivity in 2019 separately for the Savannah and Antioquia regions.  We assume no connections between importers ($y_{jj'} = 0\ \forall j,j'$). Since the distance between a given U.S. importer and any Colombian exporter is effectively constant, we abstract from this dimension. We do not observe edges between two exporters $k$ and $i$ but we know their pairwise geographical distances $d_{ki}$. We use these distances as a proxy for an active edge $y_{ki}$\ts\ts---based on the standard gravity intuition that $d_{ki}$ and $y_{ki}$ are negatively correlated. The goal of our test is to detect systematic violations of the conditional independence property in (\ref{ind2}). To this end, we construct a summary statistic, denoted $\widetilde{T}$, using a distance-based proxy for common support $S_{ij}$:
\begin{equation}
\widetilde{S}_{ij} = \frac{1}{K}\sum_{k \neq i}\ r_{ki}\ts y_{kj}, \nonumber
\end{equation}
where $r_{ki}$ measures proximity between $k$ and $i$, with higher values interpreted as a proxy for a higher (latent) probability of $k$ and $i$ being linked. This proximity metric is either: (i) a rank-normalized continuous measure $r^n_{ki}$ with $r^n_{ki} = 0$ denoting the top distance percentile and $r^n_{ki} = 1$ the bottom distance percentile; or (ii) a binary indicator $r_{ki}^q=\mathlarger{\mathbb{1}}\{d_{ki}\leq d^q\}$, where $q$ is a distance quantile threshold.

 \begin{table}[t]
  \centering
  \caption{Pairwise correlations between size, degree, and $\widetilde{S}_{ij}$}
  \scalebox{0.9}{
    \begin{tabular}{lccccc}
    \toprule
         & \multicolumn{2}{c}{\sc{buyers}} & &\multicolumn{2}{c}{\sc{sellers}} \\
\sc{pairwise correlation}\hspace*{0.2in}\quad                   &Savannah &Antioquia & &Savannah &Antioquia \\     
    \cmidrule{2-3} \cmidrule{5-6}
    \ \ Size and degree                  &  0.4025* &  0.1530* & & 0.0642 &\hspace*{5pt}0.5717*  \\
    \ \ Size and $\widetilde{S}_{ij}$    &0.4115*   &  0.1489* & &0.0446 &0.0361  \\
    \ \ Degree and $\widetilde{S}_{ij}$   &0.9982*  & 0.9982* & &0.0895 & \hspace*{-9pt}$-0.0423$\\
    \midrule
    \end{tabular}
    }
     \vspace*{0.05in}
     
           \label{tab:correlations}%
\scalebox{0.90}{
\begin{minipage}{0.85\textwidth}
\advance\leftskip 0cm
	{{\footnotesize{The table reports correlation coefficients between (i) firm size (measured as total transaction value), (ii) node degree (number of distinct trading partners), and (iii) the $\widetilde{S}_{ij}$ common support index. Correlations are shown separately for buyers and sellers in the Savannah and Antioquia regions. $^*$ indicates significance at the 5\% level.}\par } }
\end{minipage}
\vspace*{0.5in}}\\[5pt]
\end{table}

\begin{figure}[p]
\centering
\vspace*{-0.3in}
   \includegraphics[height=3.1in]{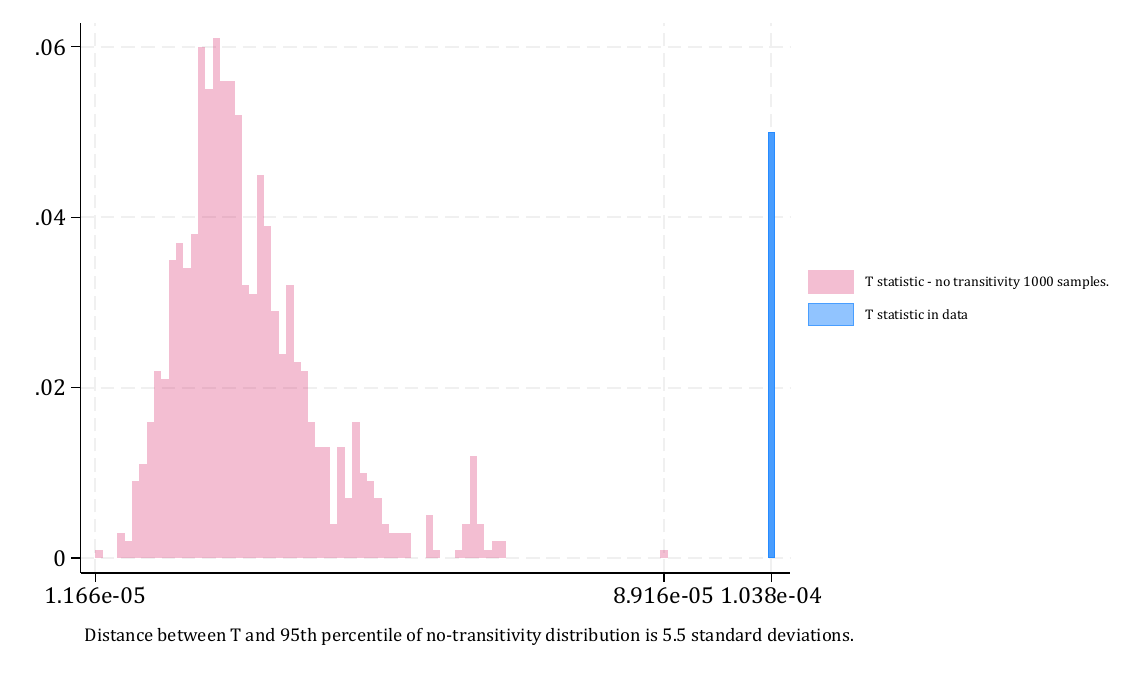}
\vspace*{-0.1in}
\caption{Distribution of $\widetilde{T}$ statistic from our test: Savannah} 
\textbf{(continuous proximity measure)}\vspace*{0.1in}
\label{fig:bog_trans_cont}
\scalebox{0.90}{
\begin{minipage}{0.95\textwidth}
\advance\leftskip 0cm
	{{\footnotesize{The figure shows the distribution of the $\widetilde{T}$ statistic for 1,000 samples without accounting for transitivity in magenta (under the null hypothesis), and the value of the $\widetilde{T}$ statistic obtained from the data in blue. The number of importers is 794, and the number of exporters in the Savannah region is 435. The ${\widetilde S}_{ij}$ index is constructed using the rank-normalized proximity measure $r^n_{ki}$. 
    }
    \par } }
\end{minipage}
}
\end{figure}

\begin{figure}[p]
\centering
\vspace*{-0.3in}
    \includegraphics[height=3.1in]{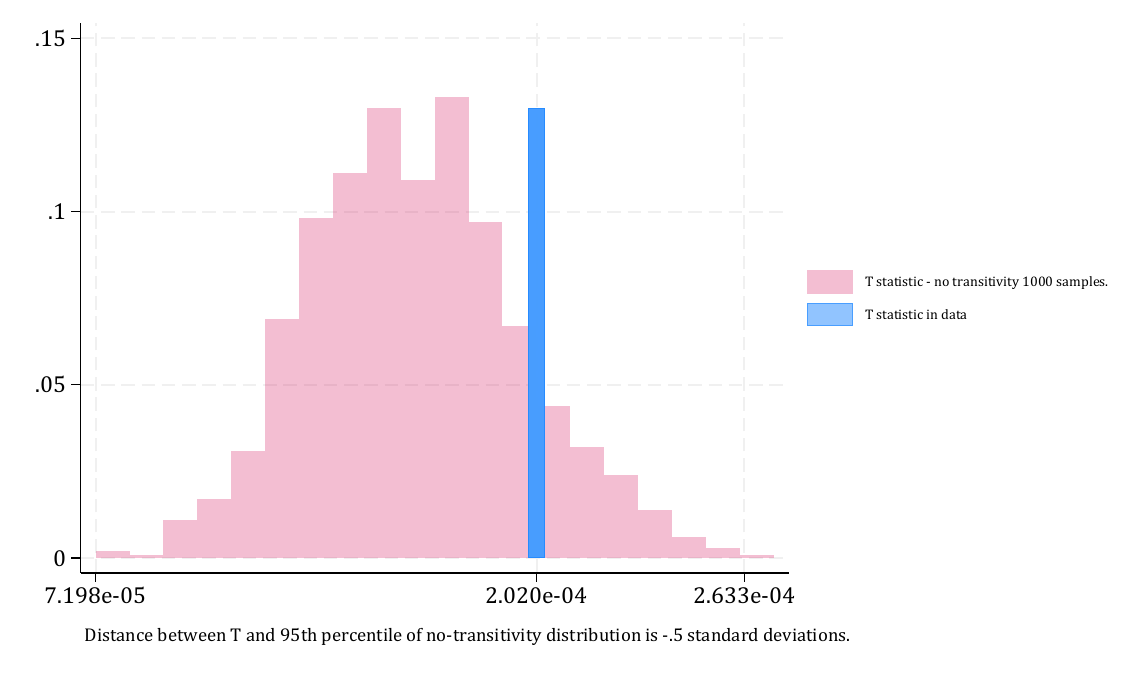}
\caption{Distribution of $\widetilde{T}$ statistic from our test: Antioquia}
\textbf{(continuous proximity measure)}\vspace*{0.1in}
\label{fig:med_trans_cont}
     \scalebox{0.90}{
\begin{minipage}{0.95\textwidth}
\advance\leftskip 0cm
	{{\footnotesize{The figure shows the distribution of the $\widetilde{T}$ statistic for 1,000 samples without accounting for transitivity in magenta, and the value of the $\widetilde{T}$ statistic obtained from the factual data in blue. The number of importers is 794, and the number of exporters in the Antioquia region is 226. The ${\widetilde S}_{ij}$ index is constructed using the rank-normalized proximity measure $r^n_{ki}$.  
    }
    \par } }
\end{minipage}
}
\end{figure}

We perform two tests for each region: one using the continuous measure $r^n_{ki}$, and the other using the discrete measure $r^q_{ki}$. Figures~\ref{fig:bog_trans_cont} and~\ref{fig:bog_trans_50} present the test results for the Savannah region, using the continuous and discrete (50th percentile) distance measures, respectively. In each figure, the distribution of the $\widetilde{T}$ statistic under the synthetic (null) model is shown in magenta, and the corresponding statistic from the data is shown in blue. The results indicate strong evidence of transitivity: in the case of the 50th percentile distance measure, the data statistic lies five standard deviations above the 95\textsuperscript{th} percentile of the synthetic distribution.

Figures~\ref{fig:med_trans_cont} and~\ref{fig:med_trans_50} report analogous results for the Antioquia region. Here, we are unable to reject the null hypothesis of no transitivity in this region. This does not necessarily imply that transitivity effects are absent; rather, they may be weaker or more difficult to detect with our test.\footnote{If we use the alternative test described in Footnote~\ref{alternative_test}, we obtain very similar results (reported in the Online Appendix).}

\subsection{Longitudinal evidence: Panel regression estimates}
\label{sec:causality}
In this section, we use the full panel from 2007 to 2019 to examine longitudinal variation for evidence of transitivity. This approach allows us not only to quantify the magnitude of transitivity effects, but also provides us with a means of discriminating between transitivity and latent homophily. This is possible because we can observe how links evolve over time in response to exogenous variation affecting neighboring connections. 
\dennis{On latent homophily, refer to the discussion at the end of this section here already? The underlying assumption is that latent homophily is time-invariant. State this upfront, at least briefly? *** CARLO *** I think we can assume sufficient "memory retention" from our earlier discussion.}

\subsection*{\normalfont\textit{Empirical strategy}}

\def\asinh{\text{\,asinh\,}}

To estimate transitivity effects, we use the following linear probability panel specification:
\begin{equation}
y_{ijt}\ = \ \theta\ts \asinh {\widetilde S}_{ij,t-l}  +  \alpha_{it} + \alpha_{j}+ \varepsilon_{ijt}, 
\label{eq:iv}
\end{equation} 
where ${\widetilde S}_{ij,t-l} \equiv \frac{1}{K} \sum_{k \neq i} r_{ki}\, y_{kj,t-l}$. This corresponds to the measure used in the test, but normalized by the number of exporters $k\neq i$ and lagged. Because the distribution of this measure is highly right-skewed and contains many zeros, we apply an inverse hyperbolic sine transformation.\footnote{The inverse hyperbolic sine is defined as $\asinh x = \ln\big(x + \sqrt{1 + x^2}\,\big)$. Unlike the logarithm, this transformation accommodates zero values; but unlike the $\ln(1 + x)$ transformation, it behaves approximately linearly near zero.} We consider two variants of this specification: one where ${\widetilde S}_{ij,t-l}$ is constructed using a median-based discrete proximity measure, $r^{q=50}_{ik}$, and one where it is constructed from a rank-normalized continuous measures, $r^n_{ik}$.

We include $(i,t)$ effects and $j$ fixed effects. These control for any seller-specific shocks that vary over time---including geographically localized disruptions that simultaneously affect clustered sellers---as well as period-specific common shocks affecting all sellers---such as fluctuations in overall U.S. demand for flowers---and time-invariant, buyer-specific unobservables such as size and product specialization.\footnote{We do not include $jt$ controls as these are correlated with our instrument (introduced below), which exploits time variation in importer relationships. We also omit $ij$-specific effects: although these are time-invariant, the extreme sparsity of the network (with fewer than 1\% of links active) means that including them would absorb much of the observed time variation, making inference infeasible \citep{jochmans2016fixed}.}

Estimating the linear probability specification in (\ref{eq:iv}) by OLS would yield biased estimates. This is because the exporter–importer connections $y_{kj,t-l}$ that enter the transitivity index $\widetilde{S}_{ij,t-l}$ are endogenous by construction \ale{? I think putting simultenaity in the classical way is not correct because  is not the classical system of simultaneous equations and what we have is more concern that $y_{jk,t-1}$ is serially correlated to $e_{ijt}$ in a network way. (Wooldrige Chp 9 ``Econometric analysis of cross section an panel data''. Read example 6.2 pg 121 on generated instruments -interactions.)} (due to simultaneity). To address this, we adopt an instrumental variables (IV) strategy, using a shift–share instrument that interacts exchange rate changes in export markets outside the U.S. with exporters' pre-existing exposure to those markets. The idea is that, given production capacity constraints, exchange-rate–driven shifts in foreign demand can lead exporters to reallocate sales across destinations, initiating or severing links with U.S. importers. Such reallocations are then reflected in changes in $\widetilde{S}_{ij,t-l}$. 
\dennis{We will need to explain this IV in more detail (both the motivation at the beginning of the paragraph and the proposed IV solution). Hard to understand for readers otherwise. *** CARLO *** I have tweaked the paragraph and added some upfront intuition about how we build the instrument; I have also added a reference to the exclusion restriction in the next paragraph.}

Specifically, we construct the instrument for $\asinh \widetilde{S}_{ij,t-1}$ as 
\begin{equation}
Z^c_{ij,t-1} \equiv \asinh {\overline r}_{-i}\ \asinh {\overline x}_j\ \asinh \bigg(\frac{1}{K}\sum_{k \neq i} z^c_{ki,t-1}\bigg),
\label{instrument}
\end{equation}
where ${\overline r}_{-i}$ denotes an average measure of geographical proximity between exporter $i$ and exporters $k \neq i$ between 2007 and 2019, ${\overline x}_j$ the average annual imports by $j$, and $z^c_{ki,t-1}$ is a shift-share index capturing exposure to exogenous variation in export links.\footnote{${\overline r}_{-i}$ and ${\overline x}_j$ are defined as ${\overline x}_j = \frac{1}{T} \sum_t x_{jt}$ and ${\overline r}_{-i} = \frac{1}{TK} \sum_t \sum_{k \neq i} r_{ikt}$. 
Constructing the instrument as an interaction with average-level measures helps mitigate mechanical correlation with the endogenous regressor. Interacting asinh-transformed values rather than levels ensures that the instrument is not collinear with the included controls.
}  
\ale{I think in (20) if we use $\sigma$ we will loose 2 years because we can observe $\chi_c$ in 2006 but not $\sigma_{kj}$.} We construct the latter as follows. Let $c$ denote a non-U.S. export destination, $X_{kct}$ the value of exports by $k$ to $c$ importers at $t$, $X_{kt}$ total exports by $k$ at $t$, $\chi_{kct} = X_{kct}/X_{kt}$  exporter $k$'s relative exposure to destination $c$ at $t$, $\hbox{\textit{CEX}}_{ct}$ $c$'s currency exchange rate at $t$. Then:\
\begin{align}
\hspace*{-0.13in} z^c_{kit} \equiv\ 
\left\{
\begin{array}{ll}
\big(1 - \chi_{kc,t-1}\big) + \chi_{kc,t-1}\,\frac{\hspace*{-0.025in}\hbox{\textit{CEX}}_{ct}}{\ \hbox{\textit{CEX}}_{c,t-1}}\quad &\text{if}\ \chi_{ic,t-1} = 0 \text{\ and\ } \chi_{ict}  = 0,\\
\ 1 &\text{otherwise}.
\end{array}
\right.
\label{eq:zdef}
\end{align}
In this formulation, only exchange rate shocks in non-U.S. markets to which exporter $k$ is exposed but exporter $i$ is not are used as a source of exogenous variation for the index ${\widetilde S}_{ij}$ (thus satisfying the exclusion restriction). The normalization we apply ensures that if $\textit{CEX}_{ct} = \textit{CEX}_{c,t-1}\ \forall c,t$, the index $z^c_{kit}$ remains constant across $k$, $j$, and $t$, and it is unaffected by time-varying export shares of exporters $i$ or $k$ to country $c$. As a result, variation in $Z^c_{it}$ arises solely from changes in the relative exchange rate $\textit{CEX}_{ct}/\textit{CEX}_{c,t-1}$ (for a destination market $i$ is not exposed to), mediated by the export market exposure of exporters other than $i$.\footnote{The inclusion of time effects controls for co-movements in the exchange rate used in the instrument and in the exchange rates of other destinations that exporter $i$ may be exposed to. This implies that our identification strategy relies solely on deviations in the movement of a specific exchange rate relative to a common trend.}
\dennis{This paragraph needs to be rephrased/explained in simpler language. A simple example might help: ``Imagine exporter $k$ is active in the Japanese market but exporter $i$ is not. The instrument exploits exogenous yen exchange rate variation that affects $k$ but not $i$...'' We also need to spell out the sign we would expect. Imagine $k$ gets a favorable exchange rate shock in the Japanese market (i.e., an appreciation of the yen against the dollar such that Japanese demand increases) [need to mention that flower exports are invoiced in dollars]. Then we would expect $k$ to shift flower exports towards the Japanese market, reducing flower exports to the US and other destinations. That is, we have a ``capacity constraint'' logic in mind whereby flower production cannot easily be adjusted in the short run. Note we will reiterate this interpretation when we discuss Table 3. *** CARLO *** See the previous changes.}

\begin{figure}[t]
\centering
 \vspace*{-0.2in}
     \hspace*{-0.25in}\includegraphics[height=2.9in]{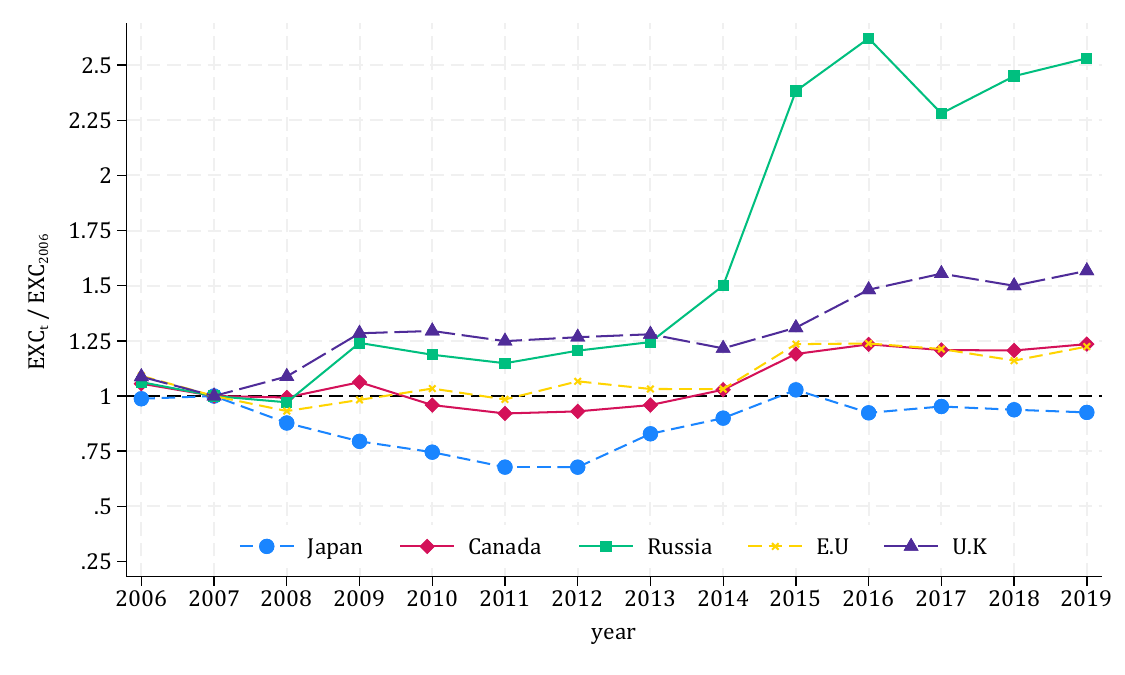}
\vspace*{-0.15in}
\caption{Evolution of exchange rates for main export markets}
\label{fig:exch_rate}
\vspace*{0.1in}
\scalebox{0.90}{
\begin{minipage}{0.95\textwidth}
\advance\leftskip 0cm
	{{\footnotesize{The figure shows the average exchange rate in local currency units per US dollar for the Japanese Yen, Canadian Dollar, Russian Ruble, Euro, and British Pound from 2006 to 2019, normalized relative to their respective 2006 values.
    }\par } }
\end{minipage}
}
\vspace*{0.15in}
\end{figure}

The main non-U.S. export destinations for Colombian flower exporters are Europe, Japan, the United Kingdom, Russia, and Canada. Within Europe, we exclude the Netherlands as a major flower producer.\footnote{The included Eurozone countries that adopted the Euro between 1999-2001 are Austria, Belgium, Finland, France, Germany, Italy, Greece, Ireland, Portugal, Spain and Luxembourg.} We use period-averaged exchange rates, expressed in local currency units (LCU) per USD, obtained from the World Development Indicators database. Figure~\ref{fig:exch_rate} plots the evolution of these exchange rates---relative to their 2006 levels---for the five selected currencies.

\magenta{To prevent overfitting, we estimate (\ref{eq:iv})-(\ref{eq:zdef}) by IV-DDML \citep{chernozhukov2018double} with linear learners and 5-fold cross-validation (three repetitions).}
To ensure robust first-stage inference, we define 100 two-dimensional percentile bins in $(x_i,x_j)$ space and cluster errors at this level.\footnote{Instrumenting an interaction term that includes an exogenous variable (size in our case) with another interaction-based instrument introduces mechanical correlation between the two. While this does not invalidate the instrument, it can artificially inflate the first-stage $F$-statistic. Clustering standard errors in this context yields more conservative and reliable first-stage diagnostics \citep{wooldridge2010econometric}.} 
\dennis{We will need to add a brief explanation of DDML. Why not use standard 2SLS? We cannot assume that referees will feel confident here. *** CARLO *** DDML is fairly mainstream now. I think we can leave the burden with the reviewers.}

\subsection*{\normalfont\textit{Estimation results}}

Table~\ref{tab:ddml_bog} presents the estimation results for equation~(\ref{eq:iv}) in the Savannah region using the continuous proximity-based regressor. In Panel A, column (1) displays the DDML (no IV) estimate, which is positive and statistically significant, indicating that a higher value of the transitivity index is associated with an increased likelihood of a link being formed between exporter $i$ and importer $j$. Columns (2) through (6) report the IV-DDML estimates using the shift-share instrument. These estimates remain statistically significant and are of a similar magnitude to the DDML estimate---about 20\% smaller---indicating robustness of the transitivity effect to potential endogeneity concerns. The direction of the bias is as expected: positive simultaneity between the regressor and the outcome---each positively affecting the other---leads to an upward bias. Panel B shows the corresponding first-stage results. The coefficient estimates have the expected sign: an appreciation of the destination country's currency makes imports more attractive there, leading exporters $k \neq i$ to shift toward that market and discontinue relationships with U.S. importers. The instrument is significant and relevant.

Table~\ref{tab:ddml_bog_dis} presents results using the median proximity-based regressor. The pattern is similar: the IV-DDML estimates are consistently significant across all specifications and remain of the same order of magnitude as the DDML estimates, though roughly 20\% smaller. The instrument remains significant and relevant in all cases. As expected, the DDML estimate is upward biased since the (lagged) regressor is a function of the (contemporaneous) dependent variable, leading to simultaneity bias.

Tables \ref{tab:ddml_med} and \ref{tab:ddml_med_dis} report the corresponding results for exporters in the Antioquia region. The DDML coefficient estimates from Panel A in column (1) are positive and significant, but IV-DDML estimates are insignificant, even if the instrument remains relevant and significant (Panel B). Thus, for the Antioquia region, we are unable to reject the no-transitivity null hypothesis. These results are in line with the results from our cross-sectional test in Section \ref{sec:crosssectionaltest} where we found no evidence of transitivity in the Antioquia network. 

\begin{table}[p]
\vspace*{0.3in}
  \centering
  \caption{ 
  DDML and IV-DDML transitivity effect estimates for the Savannah region\\ (continuous proximity measure)}
  \scalebox{0.835}{
\hspace*{-10pt}    \begin{tabular}{lcccccc}
    \toprule
    
      & DDML (no IV) &   \multicolumn{5}{c}{IV-DDML} \\
           & (1)   & (2)   & (3)   & (4)   & (5) & (6) \\
     Country Exchange Rate in IV & & EU    & JPN   & GBR   & RUS   & CAN \\
    \midrule
     \multicolumn{3}{l}{\textit{\textbf{Panel A.  Probability of Linking}}}              &       &             &       &  \\
  $\asinh \widetilde{S}_{ij,t-1}$ &  1.189$^a$ &  0.949$^a$ &0.950$^a$ & 0.950$^a$ & 0.954$^a$ & 0.949$^a$ \\
  & (0.1957) & (0.1341) & (0.1339) & (0.1339) & (0.1327) & (0.1339) \\
    \midrule
    
   \multicolumn{3}{l}{\textit{\textbf{Panel B. First Stage}}}                &       &       &  \\
    $Z_{ij,t-1}^c$ &    & -0.0002$^a$    & -0.0002$^a$    &   -0.0002$^a$   &  -0.0002$^a$    &  -0.0002$^a$     \\
          &         & (0.0000)  &  (0.0000)   & (0.0000)   &  (0.0000) & (0.0000)  \\
          \midrule
    Observations &   4,145,474
&  4,145,474  &   4,145,474  &   4,145,474 &    4,145,474  &   4,145,474 \\
 F-statistic &            & 35.53  & 35.55 & 35.53 & 35.54  &  35.54 \\
    k-folds & 5    &5 & 5     & 5     & 5     & 5 \\
    Repetitions  & 3 & 3    & 3     & 3     & 3     & 3 \\
    \bottomrule
    \end{tabular}
    }
    
    \vspace*{15pt}
     \scalebox{0.90}{
\begin{minipage}{1.09\textwidth}
\advance\leftskip 0cm
	{{\footnotesize{\textit{Notes}: The table presents DDML and IV-DDML estimates
    \citep{chernozhukov2018double}
    of the effect of $\text{asinh}\, S_{ij,t-1}$ on the probability of link formation between buyers ($j$) and sellers ($i$) in the Savannah region.  Column (1) reports DDML (no IV) estimates. Columns (2) to (5) report IV-DDML estimates. We split the sample into five folds and repeat the estimation 3 times for each cross-validation iteration. The coefficients and standard errors are the averages across each repetition. Each fold excludes observations for all years for sets of randomly drawn $(i,j)$ pairs.
    All regressions include buyer and seller-time fixed effects. Standard errors are clustered by buyer size (percentile) and seller distance (percentile) groups. Standard errors are in parentheses. Significance levels: $^{a}p<0.01$, $^{b}p<0.05$, $^{c}p<0.10$.}\par } }
\end{minipage}
}
  \label{tab:ddml_bog}%
\end{table}%

\begin{table}[p]
  \centering
  \caption{
  DDML and IV-DDML transitivity effect estimates for the Antioquia region\\ (continuous proximity measure) }
  \scalebox{0.835}{
\hspace*{-10pt}    \begin{tabular}{lcccccc}
    \toprule
      &  DDML  (no IV) &   \multicolumn{5}{c}{IV-DDML} \\
           & (1)   & (2)   & (3)   & (4)   & (5) & (6) \\
     Country Exchange Rate in IV & & EU    & JPN   & GBR   & RUS   & CAN \\
    \midrule
 \multicolumn{3}{l}{\textit{\textbf{Panel A.  Probability of Linking}}}              &       &             &       &  \\
  $\asinh \widetilde{S}_{ij,t-1}$ & \multicolumn{1}{c}{0.827$^a$} & \multicolumn{1}{c}{$-0.640$} & \multicolumn{1}{c}{$-0.639$} & \multicolumn{1}{c}{$-0.639$} & \multicolumn{1}{c}{$-0.639$} & \multicolumn{1}{c}{$-0.638$} \\
          & \multicolumn{1}{c}{(0.0653)} & \multicolumn{1}{c}{(0.4878)} & \multicolumn{1}{c}{(0.4877)} & \multicolumn{1}{c}{(0.4876)} & \multicolumn{1}{c}{(0.4876)} & \multicolumn{1}{c}{(0.4877)} \\
 \midrule
   \multicolumn{3}{l}{\textit{\textbf{Panel B. First Stage}}}                &       &       &  \\
    $Z_{ij,t-1}^c$ &    & -0.0003$^a$     &  -0.0003$^a$   &   -0.0003$^a$  & -0.0003$^a$   &  -0.0003$^a$   \\
          &         &  (0.0000) &  (0.0000) & (0.0000)  &  (0.0000) &  (0.0000) \\

    \midrule
    Observations &  2,150,946  &    2,150,946 & 2,150,946  & 2,150,946  & 2,150,946 & 2,150,946  \\
     F-statistic &            &  116.79 & 116.77 & 116.84 & 116.86  & 116.67  \\
    k-folds &  5 &5 & 5     & 5     & 5     & 5 \\
    Repetitions  & 3 & 3    & 3     & 3     & 3     & 3 \\
    \bottomrule
    \end{tabular}
    }
    
    \vspace*{15pt}
     \scalebox{0.90}{
\begin{minipage}{1.09\textwidth}
\advance\leftskip 0cm
	{{\footnotesize{\textit{Notes}: The table presents DDML and IV-DDML estimates
    \citep{chernozhukov2018double}
    of the effect of $\text{asinh}\, S_{ij,t-1}$ on the probability of link formation between buyers ($j$) and sellers ($i$) in the Antioquia region.  Column (1) reports DDML (no IV) estimates. Columns (2) to (5) report IV-DDML estimates. We split the sample into five folds and repeat the estimation 3 times for each cross-validation iteration. The coefficients and standard errors are the averages across each repetition. Each fold excludes observations for all years for sets of randomly drawn $(i,j)$ pairs.  All regressions include buyer and seller-time fixed effects. Standard errors are clustered by buyer size (percentile) and seller distance (percentile) groups. Standard errors are in parentheses. Significance levels: $^{a}p<0.01$, $^{b}p<0.05$, $^{c}p<0.10$.}\par } }
\end{minipage}
}
  \label{tab:ddml_med}%
  \vspace*{0.3in}
\end{table}%

\begin{figure}[p]
\centering
\begin{center}
\vspace*{0.2in}
\textbf{\footnotesize{\textsc{sellers}}}
\hspace{2.8in}
\textbf{\footnotesize{\textsc{buyers}}}\\
\hspace*{-0.35in}
\includegraphics[height=2in]{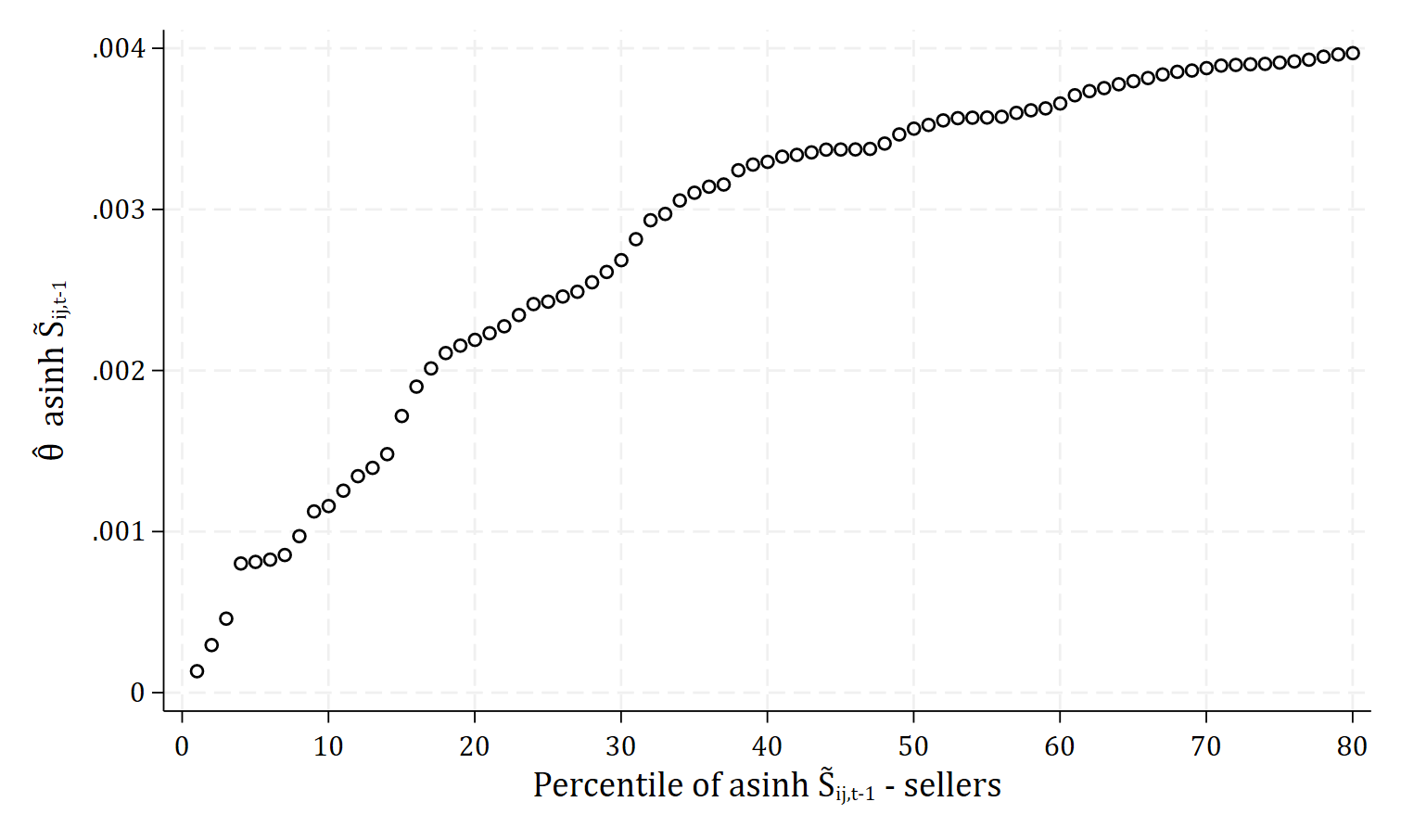}
\includegraphics[height=2in]{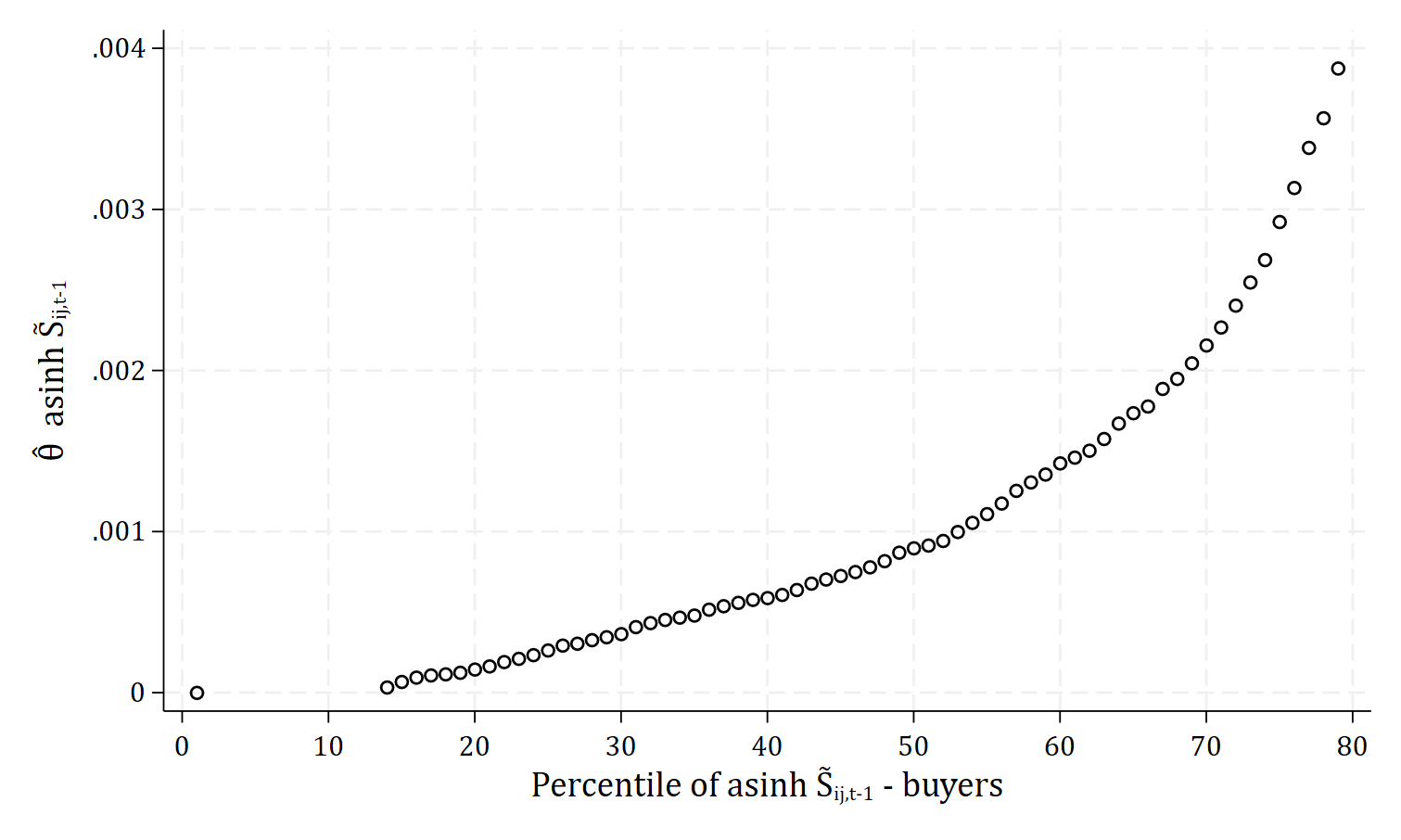}
\end{center}
\vspace*{-0.225in}
\caption{Contribution of $\widetilde{S}_{ij,t-1}$ to predicted probability }
\textbf{by $\widetilde{S}_{ij,t-1}$ percentile (Savannah)}
\label{fig:inter_theta_s}
\vspace*{0.06in}
\scalebox{0.90}{
\begin{minipage}{0.95\textwidth}
\advance\leftskip 0cm
	{{\footnotesize{The figures show the contribution of ${\hat \theta} \asinh \widetilde{S}_{ij,t-1}$ to predicted linking probability by $\widetilde{S}_{ij,t-1}$ percentile for Savannah sellers in the left panel and Savannah buyers in the right panel (EU exchange rate-based instrument). Percentiles are computed at the firm level, first by averaging $\widetilde{S}$ across links, and then over time.}\\   \par } }
\end{minipage}
}
\end{figure} 

\begin{figure}[p]
\vspace*{-0.9in}
\centering
\textbf{\footnotesize{\textsc{sellers}}}
\hspace{2.8in}
\textbf{\footnotesize{\textsc{buyers}}}\\
\hspace*{-0.2in}
   \includegraphics[height=2in]{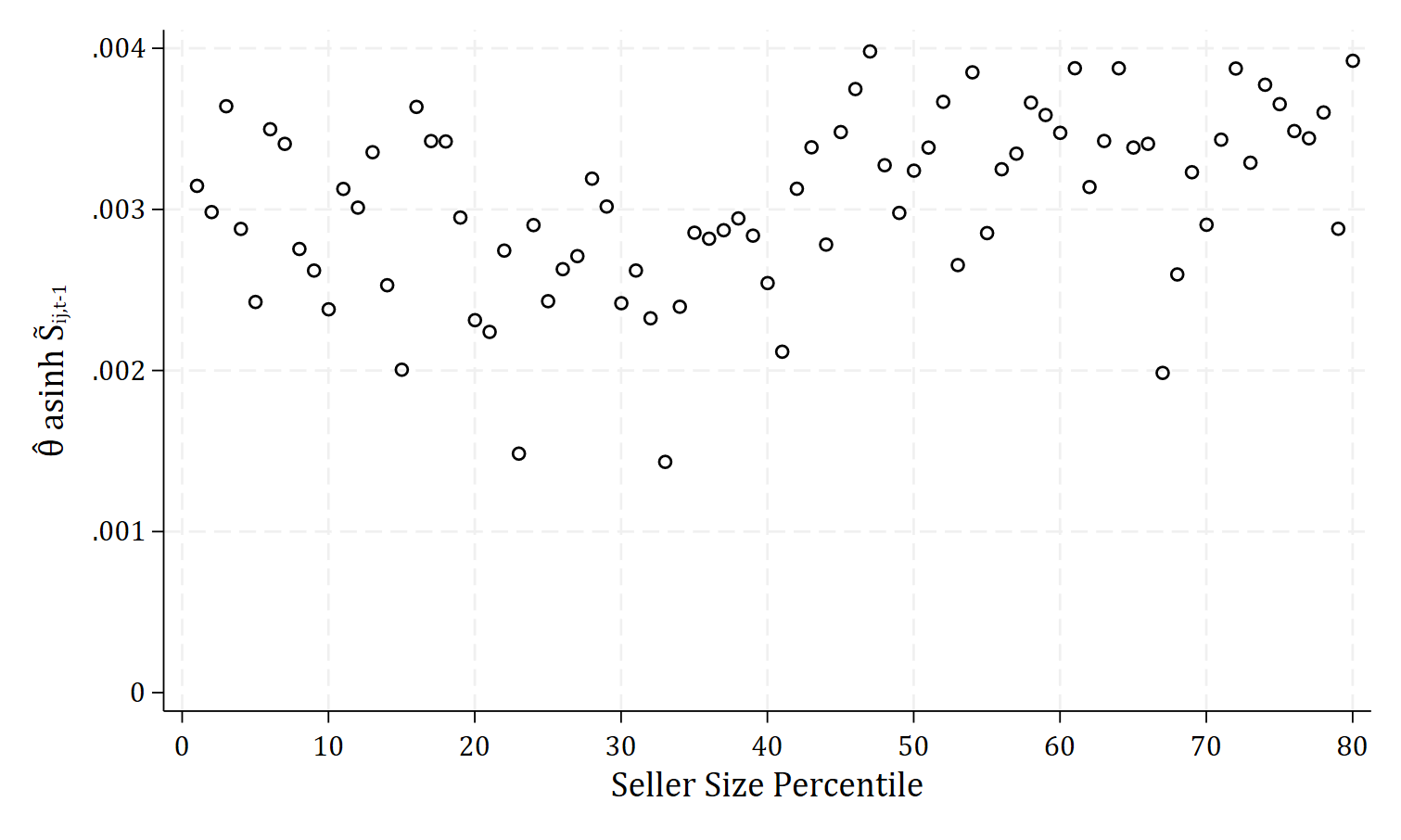}
   \includegraphics[height=2in]{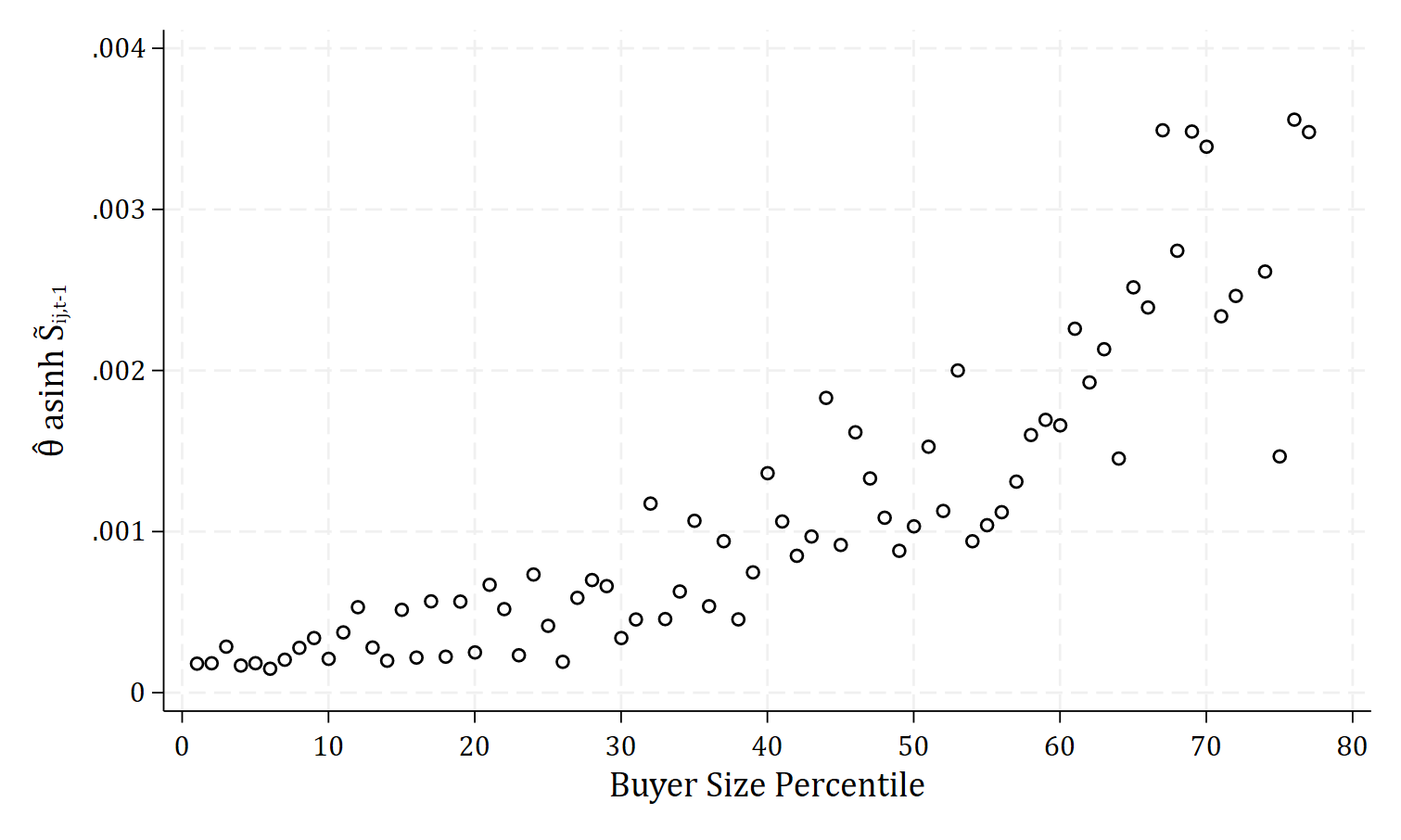}
\vspace*{-0.075in}
\caption{Contribution of $\widetilde{S}_{ij,t-1}$ to predicted probability}
\textbf{by firm size percentile (Savannah)} \\[4pt]
\label{fig:inter_theta_size}
\vspace*{0.01in}
     \scalebox{0.90}{
\begin{minipage}{0.9\textwidth}
\advance\leftskip 0cm
	{{\footnotesize{The figures show the  contribution of ${\hat \theta} \asinh \widetilde{S}_{ij,t-1}$ to predicted linking probability by firm size percentile for Savannah sellers in the left panel and for Savannah buyers in the right panel (EU exchange rate-based instrument). Percentiles are computed at the buyer or seller level, by averaging size over time.}   \par } }
\end{minipage}
}
\vspace*{0.1in}
\end{figure} 

To help interpret these findings---focusing on the Savannah region---we use our estimates to illustrate how ${\hat \theta} \asinh \widetilde{S}_{ij,t-1}$ contributes to linking probabilities at different percentiles of the $\widetilde{S}_{ij,t-1}$ distribution, with results (corresponding to column (2) in the tables) shown for sellers and buyers in the left and right panels of Figure~\ref{fig:inter_theta_s}, respectively. Patterns are nearly identical for all the different IV specifications. The effect is sizable---averaging over 0.003 percentage points, compared to an overall mean linking probability (network density) of 0.006 (Table~\ref{tab:sumstats}). That is, on average the linking probability increases by half. For sellers, the relative effect at the 80th percentile is approximately 1.8 times greater than at the 20th percentile for both the discrete and continuous proximity-based measures. For buyers, the corresponding ratios are 28 for the continuous proximity measure and 35 for the discrete one. Figure~\ref{fig:inter_theta_size} shows how this contribution varies with seller and buyer size, measured by annual sales or purchases. While ${\hat \theta}\asinh \widetilde{S}_{ij, t-1}$ exhibits no systematic variation with seller size, it increases markedly with buyer size, becoming largest for the largest buyers.\footnote{We also ran our specification on the full unbalanced sample (24 million observations), and the results are similar. \magenta{Robustness checks with OLS and 2SLS estimators also deliver similar results.}} 
\dennis{Do we have an intuition for the buyer size effect? *** CARLO *** not yet. *** We also need to explain the 28 and 35 numbers (they are very large).}

\subsection*{\normalfont\textit{Discussion}}

Returning to the distinctive implications of transitivity and latent homophily (Section \ref{sec:latent}), an example can help to clarify how latent homophily would operate in our context. This would involve some unobservable trait shared by exporters $i$ and $k$ and importer $j$---for instance, $j$ is a U.S. resident of Colombian origin who shares cultural or community ties with both $i$ and $k$. Suppose that we observe $j$ purchasing from both $i$, and $k$ and that this is because of the shared heritage.\footnote{The role of cultural ties in international trade was first highlighted by \citet*{RauchTrinidade2002}.} If an exogenous shock disrupts the relationship between $k$ and $j$, the relationship between $i$ and $j$ should remain unaffected unless transitivity is also present alongside homophily. The fact that in the Savannah network, we observe systematic changes in $y_{ij,t}$ following shocks to $y_{kj,t-1}$ directly points to a transitivity mechanism.

\magenta{Our findings provide clear evidence of transitivity among Savannah exporters but not among Antioquia exporters, both in cross-sectional and longitudinal analyses. Examining patterns in the data for clues that might account for this divergence, we find some notable differences. In the 2019 cross-section, the correlation between size and degree is sizable and significant in Antioquia but insignificant in the Savannah. The correlations between size and $\widetilde{S}$ in the Savannah full panel (Figure~\ref{fig:corr_size_S}) are consistent with this picture. Since the fixed effects absorb size, identification of $\theta$ relies on residual variation in $y$ and $\widetilde{S}$ that is \textit{not} size-related. Such variation is only evident among Savannah exporters, which may help explain the contrasting results across regions.}
\dennis{I find the discussion of these correlations hard to understand. Why do they explain the different transitivity results for Savannah vs. Antioquia? *** CARLO *** I have tweaked this and removed the discussion of patterns that are not strictly relevant to the main point. Hopefully this makes it clearer now.}

As for other observable characteristics that may account for this disparity, the most notable difference between Savannah and Antioquia exporters lies in the types of flowers they supply: Antioquia producers are mostly specialized in cut hydrangeas and chrysanthemums, while Savannah producers grow cut roses and a wider variety of other species. These product differences translate into distinct buyer profiles and market structures. Cut roses are a standardized, mass-market product with popular varieties (like \emph{Freedom}, a Valentine's Day favorite) graded by clear criteria such as quality and stem length. They are primarily sold to retailers for resale to final consumers, mainly for personal gifting, with customization largely limited to packaging. By contrast, cut hydrangeas and chrysanthemums are highly differentiated and customizable---hydrangeas, for instance, can take a wide range of colors depending on soil pH or via direct dyeing---and are especially popular for large-scale floral installations at weddings, funerals, and corporate/award events, often purchased by commercial organizers \citep{
ceniflores2021sector,
floraldaily2022hydrangea, floraldaily2023grading}.\footnote{Examples of such importers in the data are \emph{The Valley Springs}, which specializes on hydrangeas exclusively grown in Antioquia; and \emph{Trinity Flowers}, which purchases exclusively in Antioquia and specializes in wedding event arrangements.} 
 
This specialization is also reflected in our transaction data: among buyers that source from either region, approximately half---56\% for Savannah and 50\% for Antioquia---source 90\% or more of their annual purchases from a single region. We also observe exporting relationships with U.S. buyers lasting longer in Antioquia---24 months on average---than in Savannah---19 months on average.\footnote{Median durations show a similar pattern: 8 months for Savannah and 11 months for Antioquia.} This suggests comparatively greater relational investment in the Antioquia network---consistent with higher specialization and customi\-zation---which in turn may leave buyers with less flexibility to shift demand to other, indirectly connected sellers.

\section{Counterfactual experiments}

The estimates from our panel regressions suggest that the effect of transitivity is non-trivial in magnitude. However, they do not tell us how transitivity shapes the network's response to shocks. But this is instrumental in understanding how transitivity may alter the network's response to shifts in economic conditions or policy---such as a trade cost shock---and the resulting implications for economic efficiency. For that, we need counterfactual analysis, something the panel specification used for our estimation is ill-suited for.

To this end, we embed our panel estimates in a non-linear model of link formation augmented for transitivity that delivers robustly bounded probability predictions, such as (\ref{bb_augmented}). We then use the calibrated model to conduct illustrative counterfactual simulations for the 2019 cross-section in which the transitivity channel is either left activated or switched off. By doing this, we can gain a picture of how transitivity influences network responses in both quantitative and qualitative terms.

The parameter $\theta$ in our panel specification is the derivative of the linking probability with respect to the inverse hyperbolic sine transformation of $\widetilde{S}$. To obtain parameter values that are consistent with our panel estimate, we impose that ${\hat \theta}$ equal the corresponding marginal effect implied by~(\ref{bb_augmented}) \magenta{in the second stage of our IV-DDML estimation}, while estimating the remaining parameters by maximum likelihood, as detailed in Appendix~\ref{sec:bb_est}. We first apply our method to the generalized balls-and-bins specification described in (\ref{bb_augmented}). As shown in Table~\ref{tab:fit}, with only two parameters, this struggles to replicate even basic within-period moments such as network density.\footnote{We normalize buyer sizes, $x_j$, by dividing them by mean size.} 

\begin{table}[t]
\centering
\caption{Parameter calibration and fit}
\label{tab:fit}
 \centering
{\small 
 \begin{tabular}{lccc}
 \toprule
 \textbf{Statistic / parameter} &\textbf{Data} &\ \ \textbf{ Balls and bins} &\quad \textbf{Poisson}\quad\quad \\
 \midrule \\[-12pt]
    Density                &  0.0141 &  0.0063   &  0.0146 \\
    Outdegree 25th pct.    &  2      &  0.24     &  5.98  \\
    Outdegree 50th pct.    &  4      &  1.09     &  8.6   \\
    Outdegree 75th         &  11     &  4.59     & 11.9   \\
    Indegree 25th pct.     &  0      &  0      &  0.6      \\
    Indegree 50th pct.     &  1      &  0.1      &  1.72   \\
    Indegree 75th pct.     &  3      &  0.52      &  3.4   \\[5pt]
    \hspace*{25pt}$\alpha$ &         &  --       &  0.83  \\
    \hspace*{25pt}$\eta$   &         &  --       &  0.19  \\
    \hspace*{25pt}$\beta$  &         &  2.72    &  0.35  \\
    \hspace*{25pt}$\gamma$ &         & 22.83     & 20   \\   
   \bottomrule
  \end{tabular}
} 
  \vspace{0.25cm}
  
  \scalebox{0.90}{
\begin{minipage}{0.85\textwidth}
\advance\leftskip 0cm
{{\footnotesize{\textit{Notes}: The table reports 2019 statistics for traders that engaged in at least one transaction between 2017 and 2019. For each model, parameters are estimated and 250 simulated networks are generated updating $\widetilde{S}$.
The table reports average link density across simulated samples, as well as averages for selected percentiles (25th, 50th, 75th) of both the indegree and outdegree distributions.
} }}
\end{minipage}
}\vspace*{0.1in}
\end{table} 

\ale{Brownbag notes: Poisson is when BB urns and balls go to infinite, not continuous-time analog.}We can obtain a more flexible specification by modeling link formation as a generalized Poisson process, the continuous-time analog of a balls-and-bins framework \citep{barbour_holst_janson_1992}. In this setting, if the arrival rate of trials over a unit time interval is $\Lambda\big(x_i, x_j, S_{ij}\big)$, then the probability of at least one success---i.e., that a link forms between $i$ and $j$---within the given time interval is
\begin{equation}
\Pr(y_{ij} = 1) = 1 - e^{-\Lambda(x_i,x_j, S_{ij})}.
\label{poisson}
\end{equation}
We specify $\Lambda(x_i,x_j,S_{ij})$ as
\begin{equation}
\Lambda(x_i,x_j,S_{ij}) = \alpha\, x_i^{\,\eta}\ts x_j^{\,\beta}\ts (1+\gamma\ts S_{ij}),\quad \alpha>0,\ \eta>0,\ \beta> 0.
\label{Lambda}
\end{equation}
As shown in Appendix \ref{sec:poisson}, (\ref{poisson}) can be given a discrete-choice microfoundation. Under this interpretation, a change in $\alpha$ can be interpreted as a change in trade costs. If we apply our calibration method to derive parameter values for $\alpha, \eta, \beta, \gamma$, this specification delivers a much better fit than  (\ref{bb_augmented})---as shown in Table~\ref{tab:fit}.\footnote{We also experimented with variants of the balls-and-bins model that incorporate extra parameters, but we found these to be consistently outperformed by the generalized Poisson formulation in terms of fit.} This is the formulation we focus on for our counterfactual experiments.

In our simulations for the 2019 cross-section, we restrict attention to traders who engaged in at least one transaction during the 2017–2019 period and define $x_i$ and $x_j$ as import and export shares over that period. Since the $y_{ij}$ outcomes arising from (\ref{poisson}) are random variables, the equilibrium mapping from parameters to outcomes is stochastic: each set of parameters gives rise to a set of equilibrium configurations $y_{ij}$. These configurations are random draws from a process that satisfies a fixed-point condition whereby the values of $\widetilde{S}_{ij}$ are consistent with the $y_{ij}$ they generate via equation~(\ref{poisson}). To provide an approximate characterization of an equilibrium, we consider a finite number of draws (250), iterating for each until convergence to a fixed point, deriving a distribution of $y_{ij}$ outcomes for each $(i,j)$ pair, one for each draw.\footnote{For each parameter configuration, an equilibrium is defined as a set of matrix pairs $([y_{ij}], [\widetilde{S}_{ij}])$, where $[y_{ij}]$ is drawn from the probabilistic process specified by (\ref{bb_augmented}) given $[\widetilde{S}_{ij}]$, and $[\widetilde{S}_{ij}]$ is derived from $[y_{ij}]$. To approximate the set of equilibria, we select a number of draws $N^D$ and assign to each draw $n \in {1,\ldots, N^D}$ a unique random seed $s_n$. For each draw, convergence to a fixed point is achieved by iteratively drawing $[y_{ij}]$ and updating $[\widetilde{S}_{ij}]$---and thus $\Pr(y_{ij} = 1 \mid \ldots\,)$---re-initializing the pseudo-random number generator to $s_n$ at each step.} 

Our counterfactual experiments involve increasing $\alpha$ from its baseline value $\alpha^0 = \hat{\alpha}$ to a  level $\alpha^1 = \xi\ts \alpha_0,\ \xi > 1$, thereby generating a new equilibrium, also characterized by a distribution of outcomes over $(i,j)$ pairs, and then comparing this counterfactual distribution to the baseline distribution.\footnote{Comparing individual draws from each equilibrium would obscure any systematic differences between them due to the randomness inherent in each realization.} 

We change $\alpha^0$ to $\alpha^1 = \alpha^0/0.9$---corresponding to a 10\% fall in total trade costs---and consider two alternative counterfactual scenarios. Scenario C1 takes transitivity fully into account, while scenario C2 ignores the impact of the trade cost shock on transitivity:
\begin{itemize}
\item[$C1$]  
We iteratively update all $\widetilde{S}_{ij}$'s as described above to derive a new equilibrium.
\item[$C2$]  
We do not update the $\widetilde{S}_{ij}$'s, effectively treating the terms $\gamma\ts \widetilde{S}_{ij}$ as constants and shutting down any knock-on effects from transitivity. 
\end{itemize}
Comparing the outcomes from the first counterfactual with those from the second isolates amplification effects driven by transitivity, as well as their distributional implications. 

\def\??{\textcolor{magenta}{[??]}}

\begin{figure}[p]
\centering
 \vspace*{-0.2in}
     \hspace*{-0.25in}\includegraphics[height=3in]{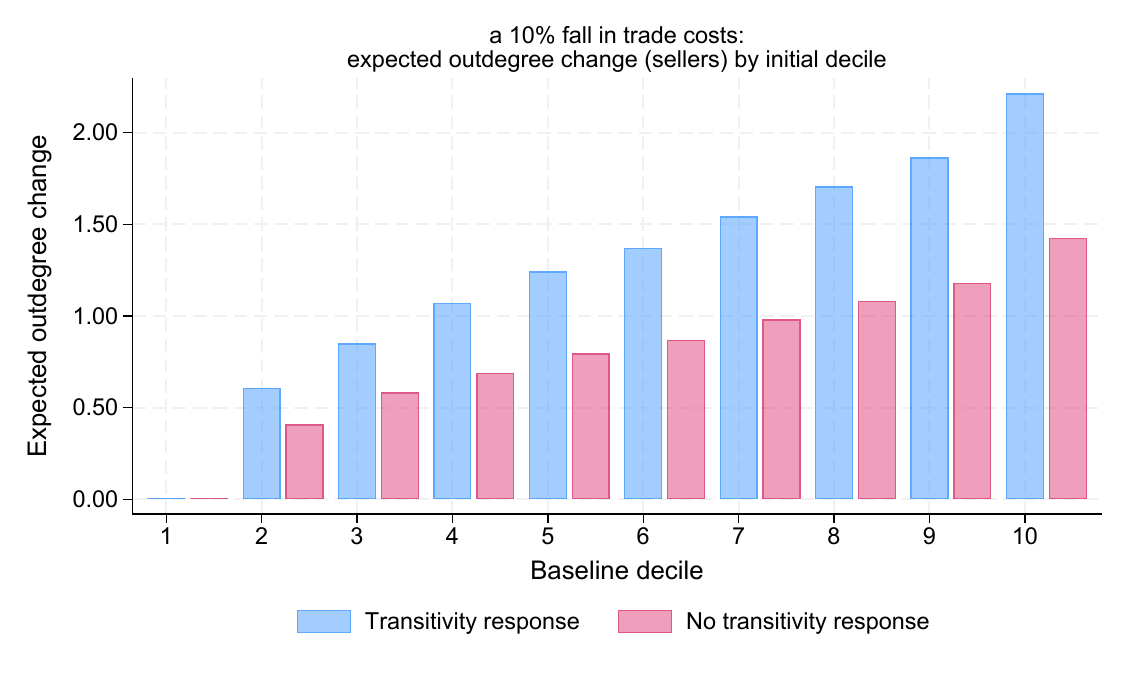}
\vspace*{-0.01in}
\caption{Expected changes in the number of links: Savannah sellers}
\label{fig:outd_change}
\vspace*{0.1in}
\scalebox{0.90}{
\begin{minipage}{0.95\textwidth}
\advance\leftskip 0cm
	{{\footnotesize{The figure shows expected changes in the number of links for Savannah exporters following a 10\% increase in $\alpha$ (corresponding to a 10\% fall in trade costs)---with and without transitivity responses. Changes are reported for each decile of the outdegree distribution in the baseline equilibrium and are based on 250 draws from the baseline and counterfactual equilibria.
    }\par } }
\end{minipage}
}
\vspace*{0.15in}
\end{figure}

\begin{figure}[p]
\centering
 \vspace*{-0.2in}
     \hspace*{-0.25in}\includegraphics[height=3in]{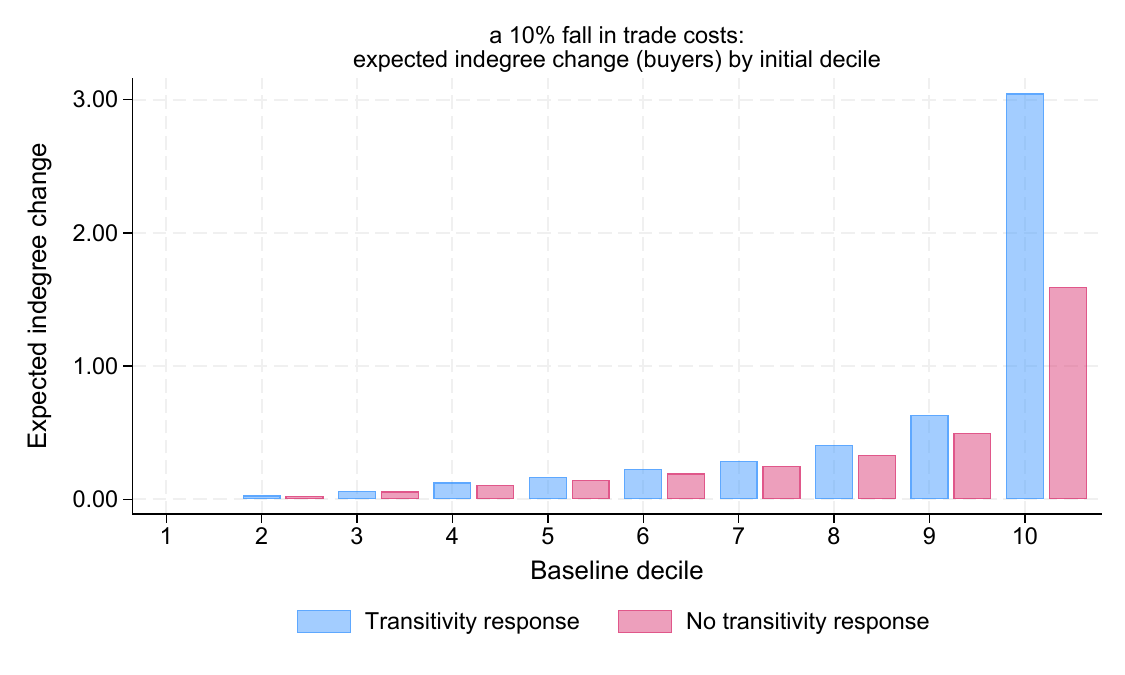}
\vspace*{-0.01in}
\caption{Expected changes in the number of links: Savannah buyers}
\label{fig:ind_change}
\vspace*{0.1in}
\scalebox{0.90}{
\begin{minipage}{0.95\textwidth}
\advance\leftskip 0cm
	{{\footnotesize{The figure shows expected changes in the number of links for importers in the Savannah network following a 10\% increase in $\alpha$ (corresponding to a 10\% fall in trade costs)---with and without transitivity responses. Changes are reported for each decile of the indegree distribution in the baseline equilibrium and are based on 250 draws from the baseline and counterfactual equilibria.
    }\par } }
\end{minipage}
}
\vspace*{0.15in}
\end{figure}

Figures \ref{fig:outd_change} and \ref{fig:ind_change} compare expected changes in outdegree and indegree by baseline decile across the two counterfactuals. The differences are substantial, both quantitatively and qualitatively: the transitivity mechanism more than doubles the impact of trade cost shocks for some traders. But these effects are unevenly distributed, with the most connected traders in the baseline experiencing the largest absolute changes.\footnote{As shown in Appendix \ref{sec:poisson}, if (\ref{poisson}) is given a choice-theoretic interpretation, a higher probability of link formation implies higher expected surplus. It follows that the uneven impact of transitivity responses on link formation translates into similarly uneven effects on expected surplus.} In relative terms, the effects for exporters are evenly distributed, with all deciles forming a similarly higher proportion of connections. For importers, effects are concentrated in the top decile in both absolute and relative terms. This pattern highlights how the transitivity response exerts a compounding effect on clustering. That is, it reinforces the tendency of a node to form additional links if it is independently more connected, further intensifying clustering around already central nodes \citep{jacksonrogers2007}.

\section{Conclusion}
\label{sec:conclusion}

Network effects have been extensively studied in the literature on social networks to understand how connections form between individuals. We apply the same concepts and methods to firm-to-firm trade.

We focus on transitivity---the simplest form of departure from dyadic link formation---and develop a test to detect its presence in the data and apply it to firm-to-firm transaction records between Colombian flower exporters and U.S. importers, using geographic distance between exporters as a proxy for domestic linkages. We find evidence of transitivity in the Savannah, one of Colombia's two main flower-producing regions. Exploiting longitudinal variation and relying on an exchange rate-based shift-share instrument, we estimate that having more shared connections substantially raises the likelihood of link formation---by as much as one-third \ale{is it one third or half?}. Illustrative counterfactual simulations suggest that transitivity can play a central role in shaping how the network responds to external shocks. These findings call into question the common premise in the trade literature that buyer-seller links form independently from one another.

Our analysis is concerned with transitivity effects at the firm level, but these effects have broader implications for aggregate outcomes. In particular, by amplifying the influence of other determinants of link formation such as geographic proximity, transitivity offers a plausible explanation for agglomeration externalities---a staple ingredient of economic geography models that has longed lacked well-defined microfoundations. 

\medskip
\bigskip

\let\oldthebibliography\thebibliography
\let\endoldthebibliography\endthebibliography
\renewenvironment{thebibliography}[1]{
  \begin{oldthebibliography}{#1}
    \setlength{\itemsep}{0.2em}
    \setlength{\parskip}{0em}
}
{
  \end{oldthebibliography}
}

{\small
\bibliography{References}
}

\appendix
\renewcommand{\thetable}{\thesection\arabic{table}}
\setcounter{table}{0}
\renewcommand{\thefigure}{\thesection\arabic{figure}}
\setcounter{figure}{0}

\medskip
\bigskip
\renewcommand{\baselinestretch}{1.1}
\small
\newpage

\section{Appendix}

\subsection{Discrete choice microfoundations of a generalized Poisson process of link formation \label{sec:poisson}}

Link formation can be microfounded in a discrete choice framework where the joint surplus from activating a link between traders $i$ and $j$---to be split according to some given rule---is 
\begin{equation}
 y_{ij}\left( \Lambda(x_i, x_j, S_{ij}) - \varepsilon \right) \equiv \Pi(y_{ij}; x_i, x_j, S_{ij}, \varepsilon),
\end{equation}
with $y_{ij} \in \{0,1\}$ and $\varepsilon$ denoting a random draw, and where $\Lambda(x_i, x_j, S_{ij})$ can be thought of as measuring the gross surplus from link formation prior to accounting for $\varepsilon$, the idiosyncratic component. The optimal linking choice is then given by $y_{ij}(\varepsilon) = \arg\max_{y_{ij}} \Pi(y_{ij}; x_i, x_j, S_{ij}, \varepsilon)$. In this setting, if $\varepsilon$ follows an exponential distribution with CDF $F(\varepsilon) = 1 - \exp(-\varepsilon)$, the probability of a link forming between $i$ and $j$ is
\begin{equation}
\Pr\mts\mts\big(y_{ij} = 1\big) = 1 - \exp\mts\mts\left(-\Lambda(x_i, x_j, S_{ij})\right).
\end{equation}

Expected surplus for a pair of traders is
\begin{align}
\int_0^{\Lambda(W_{ij})}\hspace{-0.06in} \big(\Lambda(W_{ij})-\varepsilon\big)\, \hbox{d}F(\varepsilon) 
\ts =\ts \Lambda(W_{ij})  -1 + \exp\mts\mts\left(-\Lambda(W_{ij})\right) \ts \equiv\ts   \text{E}\Pi\big(\Lambda(W_{ij})\big),
\label{ES}
\end{align}
where $W_{ij} = \big(x_i, x_j, S_{ij}\big)$. This is increasing in $\Lambda(W_{ij})$, as is $\Pr\big(y_{ij}=1\big)$---that is, a higher probability of link formation corresponds to higher expected surplus. This relationship holds regardless of whether $S_{ij}$ adjusts endogenously, although the value of $\Lambda(W_{ij})$ will differ across the two cases.

With proportional (iceberg) trade costs, $\tau \geq 1$, surplus becomes
\begin{equation}
y_{ij} \left( \frac{\Lambda(x_i, x_j, S_{ij})}{\tau} - \varepsilon \right).
\end{equation}
If we define $\Lambda(x_i, x_j, S_{ij}) = \alpha\, x_i^{\eta} x_j^{\beta} (1 + \gamma S_{ij})$, then 
a change in $\alpha$ becomes isomorphic to a change in $\tau$, and can thus be interpreted as a change in proportional trade costs—an $x\%$ increase in $\tau$ corresponding to a reduction in $\alpha$ by a factor of $100/(100 + x)$. 

\smallskip
\subsection{Parameter calibration\label{sec:bb_est}}

Let $W_{ij} = \big(x_i, x_j, \widetilde{S}_{ij}\big)$ and $\Pr(y_{ij}=1) = G(W_{ij}; \alpha,\eta,\beta,\gamma)$, where $x_i$ and $x_j$ denote total annual sales and purchases, respectively. \magenta{Let $y_{ij}^{\,\perp}$ denote the cross-validated residualized outcome from our IV-DDML estimation procedure and $\asinh\widetilde{S}^{\,IV\perp}_{ij}$ the projected residualized regressor from the IV stage of the same procedure (omitting time subscripts).}\footnote{While our panel specification models a lagged effect, we treat the calibrated model used in our counterfactuals as fully static.} The slope coefficient estimate ${\hat \theta}$ from the last stage of the IV-DDML regression must satisfy
\begin{align}
{\hat \theta}\, &\equiv \frac{\text{Cov}\left(\asinh \widetilde{S}^{\,IV\perp}_{ij},\,
y^{\,\perp}_{ij}\right)}{\text{Var}\big(\asinh\widetilde{S}^{\,IV\perp}_{ij}\big)}\nonumber \\
&= \frac{\text{Cov}\left(\asinh \widetilde{S}^{\,IV\perp}_{ij},\
y^{\,\perp}_{ij} - y_{ij} + G(W_{ij};\alpha,\eta, \beta,\gamma) \right)}{\text{Var}\big(\asinh\widetilde{S}^{\,IV\perp}_{ij}\big)}
\equiv\, \Theta(\alpha,\eta,\beta,\gamma).
\end{align}

Hybrid calibration/estimation of ($\hat{\alpha}$, $\hat{\eta}$, $\hat{\beta}$, $\hat{\gamma}$), incorporating the panel estimate $\hat{\theta}$, requires simultaneously satisfying the following two conditions:
\begin{align}
&\ ({\hat \alpha},{\hat \eta},{\hat \beta}) = \arg\max_{\alpha,\eta,\beta}\ \sum_{i,j} 
\,\ln \Big(y_{ij}\ts G(W_{ij};\alpha, \eta, \beta,{\hat \gamma}) + (1-y_{ij})\big(1-G(W_{ij};\alpha,\eta, \beta,{\hat \gamma})\big)\Big),
\label{ahat}
\\
&\ {\hat \theta} = \Theta({\hat \alpha},{\hat \eta},{\hat \beta}, {\hat \gamma}). 
\label{ghat}
\end{align}

We solve the above problem through an iterative process. Starting from an initial parameter guess: (i) we hold $\gamma$ constant and derive new values for $\alpha$, $\eta$ and $\beta$ by MLE through (\ref{ahat}); (ii) we hold $\alpha$, $\eta$ and $\beta$ constant and numerically solve for $\gamma$ from (\ref{ghat}); (iii) if the difference between the new $\gamma$ and the old $\gamma$ is within a specified tolerance range, we stop; otherwise we update $\gamma$ to a weighted average between the two values and cycle back to (i).

\newpage
\subsection{Figure Appendix}

\begin{figure}[H]
\centering
\vspace*{0.1in}
     \includegraphics[height=2.8in]{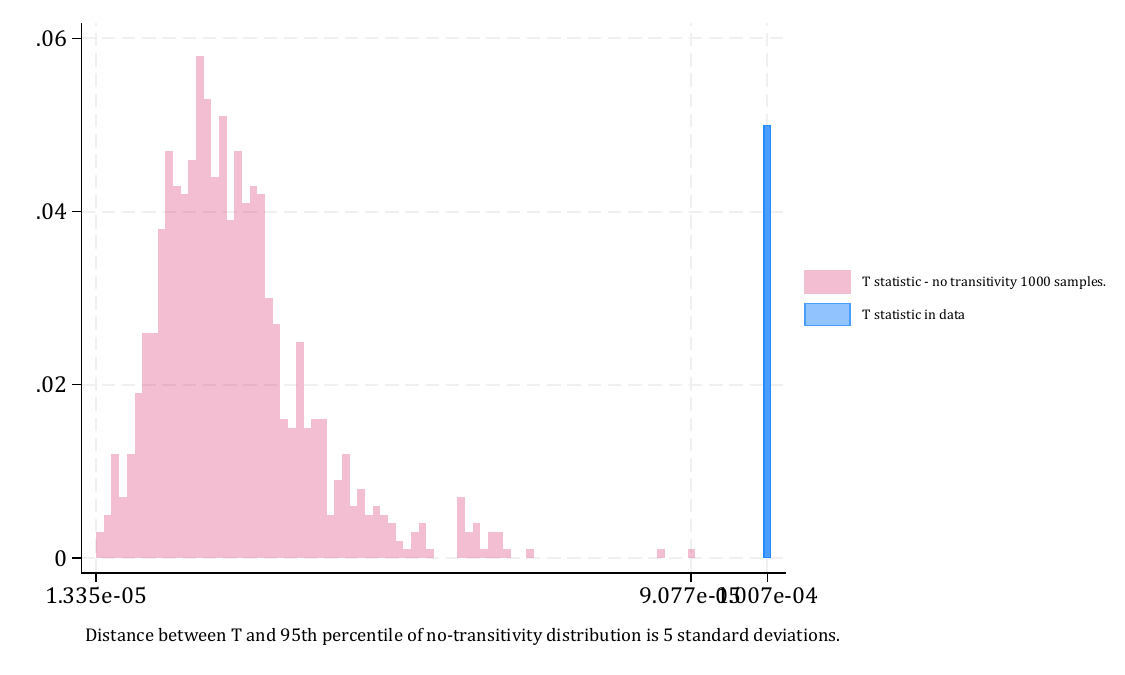}
\vspace*{-0.1in}
\caption{Distribution of $\widetilde{T}$ statistic from our test: Savannah}
\textbf{-- discrete proximity measure}\vspace*{0.1in}
\label{fig:bog_trans_50}
     \scalebox{0.90}{
\begin{minipage}{0.95\textwidth}
\advance\leftskip 0cm
	{{\footnotesize{The figure shows the distribution of the $\widetilde{T}$ statistic for 1,000 samples without accounting for transitivity in magenta, and the value of the $\widetilde{T}$ statistic obtained from the factual data in blue. The number of importers is 794, and the number of exporters in the Savannah region is 435. The ${\widetilde S}_{ij}$ index is constructed using the discrete proximity measure $z^q_{ki}$.}
    \par } }
\end{minipage}
}
\end{figure}

\begin{figure}[H]
\centering
     \includegraphics[height=2.8in]{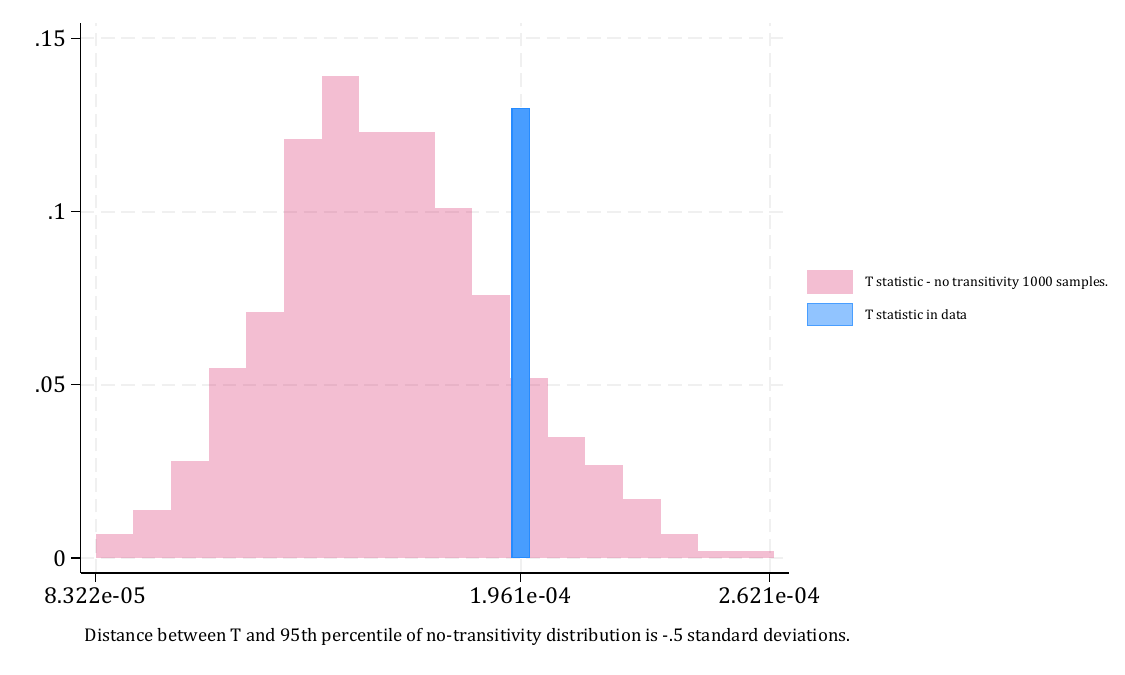}
     \vspace*{-0.1in}
\caption{Distribution of $\widetilde{T}$ statistic from our test: Antioquia}
\textbf{-- discrete proximity measure}\vspace*{0.1in}\label{fig:med_trans_50}
     \scalebox{0.90}{
\begin{minipage}{0.95\textwidth}
\advance\leftskip 0cm
	{{\footnotesize{The figure shows the distribution of the $\widetilde{T}$ statistic for 1,000 samples without accounting for transitivity in magenta, and the value of the $\widetilde{T}$ statistic obtained from the factual data in blue. The number of importers is 794, and the number of exporters in the Antioquia region is 226. The ${\widetilde S}_{ij}$ index is constructed using the discrete proximity measure $z^q_{ki}$.  }\par } }
\end{minipage}
}
\end{figure}

\subsection{Table Appendix}

\begin{table}[H]
  \centering
  \caption{ 
  DDML and IV-DDML transitivity effect estimates for the Savannah region\\ (discrete proximity measure)}
  \scalebox{0.71}{
\hspace*{-10pt}    \begin{tabular}{lcccccc}
    \toprule
    
      & DDML (no IV) &   \multicolumn{5}{c}{IV-DDML} \\
           & (1)   & (2)   & (3)   & (4)   & (5) & (6) \\
     Country Exchange Rate in IV & & EU    & JPN   & GBR   & RUS   & CAN \\
    \midrule
     \multicolumn{3}{l}{\textit{\textbf{Panel A.  Probability of Linking}}}              &       &             &       &  \\
  $\asinh \widetilde{S}_{ij,t-1}$ &  0.779$^a$ &  0.657$^a$ &0.657$^a$ & 0.657$^a$ & 0.660$^a$ & 0.657$^a$ \\
  & (0.1222) & (0.1232) & (0.1232) & (0.1232) &  (0.1222) & (0.1232) \\
    \midrule
    
   \multicolumn{3}{l}{\textit{\textbf{Panel B. First Stage}}}                &       &       &  \\
    $Z_{ij,t-1}^c$ &    & -0.0003$^a$   & -0.0003$^a$  &   -0.0003$^a$   &  -0.0003$^a$    &  -0.0003$^a$      \\
  &       & (0.0000)  &  (0.0000)   &  (0.0000) & (0.0000) & (0.0000)  \\
          \midrule
    Observations &   4,145,474
&  4,145,474  &   4,145,474  &   4,145,474 &    4,145,474  &   4,145,474 \\
 F-statistic &            & 39.71  & 39.71 & 39.71 & 39.71 & 39.71 \\
    k-folds & 5    &5 & 5     & 5     & 5     & 5 \\
    Repetitions  & 3 & 3    & 3     & 3     & 3     & 3 \\
    \bottomrule
    \end{tabular}
    }
    
    \vspace*{12pt}
     \scalebox{0.90}{
\begin{minipage}{1.09\textwidth}
\advance\leftskip 0cm
	{{\footnotesize{\textit{Notes}: The table presents DDML and IV-DDML estimates
    \citep{chernozhukov2018double}
    of the effect of $\text{asinh}\, S_{ij,t-1}$ on the probability of link formation between buyers ($j$) and sellers ($i$) in the Savannah region.  Column (1) reports DDML (no IV) estimates. Columns (2) to (5) report IV-DDML estimates. We split the sample into five folds and repeat the estimation 3 times for each cross-validation iteration. The coefficients and standard errors are the averages across each repetition. Each fold excludes observations for all years for sets of randomly drawn $(i,j)$ pairs.
    All regressions include buyer and seller-time fixed effects. Standard errors are clustered by buyer size (percentile) and seller distance (percentile) groups. Standard errors are in parentheses. Significance levels: $^{a}p<0.01$, $^{b}p<0.05$, $^{c}p<0.10$.}\par } }
\end{minipage}
}
  \label{tab:ddml_bog_dis}%
\end{table}%

\begin{table}[H]
  \centering
  \caption{ 
  DDML and IV-DDML transitivity effect estimates for the Antioquia region\\ (discrete proximity measure)}
  \scalebox{0.71}{
\hspace*{-10pt}    \begin{tabular}{lcccccc}
    \toprule
    
      & DDML (no IV) &   \multicolumn{5}{c}{IV-DDML} \\
           & (1)   & (2)   & (3)   & (4)   & (5) & (6) \\
     Country Exchange Rate in IV & & EU    & JPN   & GBR   & RUS   & CAN \\
    \midrule
     \multicolumn{3}{l}{\textit{\textbf{Panel A.  Probability of Linking}}}              &       &             &       &  \\
  $\asinh \widetilde{S}_{ij,t-1}$ &  0.573$^a$ &  -0.456 & -0.456 & -0.456  & -0.456 & -0.455 \\
  &  (0.0764)    & (0.3307) & (0.3307)  & (0.3307) &  (0.3307) &  (0.3307) \\
    \midrule
    
   \multicolumn{3}{l}{\textit{\textbf{Panel B. First Stage}}}                &    &       &  \\
    $Z_{ij,t-1}^c$ &    &  -0.0004$^a$  &  -0.0004$^a$   &   -0.0004$^a$   &  -0.0004$^a$    &    -0.0004$^a$    \\
  &        &  (0.0001)  &  (0.0001)   &   (0.0001) & (0.0001)  &  (0.0001)  \\
          \midrule
    Observations &  2,150,946
& 2,150,946   &   2,150,946  & 2,150,946   &   2,150,946   &  2,150,946  \\
 F-statistic &            & 127.88 & 127.86 & 127.86 & 127.85 & 127.91 \\
    k-folds & 5    &5 & 5     & 5     & 5     & 5 \\
    Repetitions  & 3 & 3    & 3     & 3     & 3     & 3 \\
    \bottomrule
    \end{tabular}
    }
    
    \vspace*{12pt}
     \scalebox{0.90}{
\begin{minipage}{1.09\textwidth}
\advance\leftskip 0cm
	{{\footnotesize{\textit{Notes}: The table presents DDML and IV-DDML estimates
    \citep{chernozhukov2018double}
    of the effect of $\text{asinh}\, S_{ij,t-1}$ on the probability of link formation between buyers ($j$) and sellers ($i$) in the Antioquia region.  Column (1) reports DDML (no IV) estimates. Columns (2) to (5) report IV-DDML estimates. We split the sample into five folds and repeat the estimation 3 times for each cross-validation iteration. The coefficients and standard errors are the averages across each repetition. Each fold excludes observations for all years for sets of randomly drawn $(i,j)$ pairs. All regressions include buyer and seller-time fixed effects. Standard errors are clustered by buyer size (percentile) and seller distance (percentile) groups. Standard errors are in parentheses. Significance levels: $^{a}p<0.01$, $^{b}p<0.05$, $^{c}p<0.10$.}\par } }
\end{minipage}
}
  \label{tab:ddml_med_dis}%
\end{table}

\afterpage\clearpage

\newpage
\section{Online Appendix}

\subsection{Transitivity test: Monte Carlo simulations \label{sec:mc}}

\begin{figure}[h]
\vspace*{0.35in}
\centering
\hspace*{-0.25in}
\includegraphics[height=2.15in]{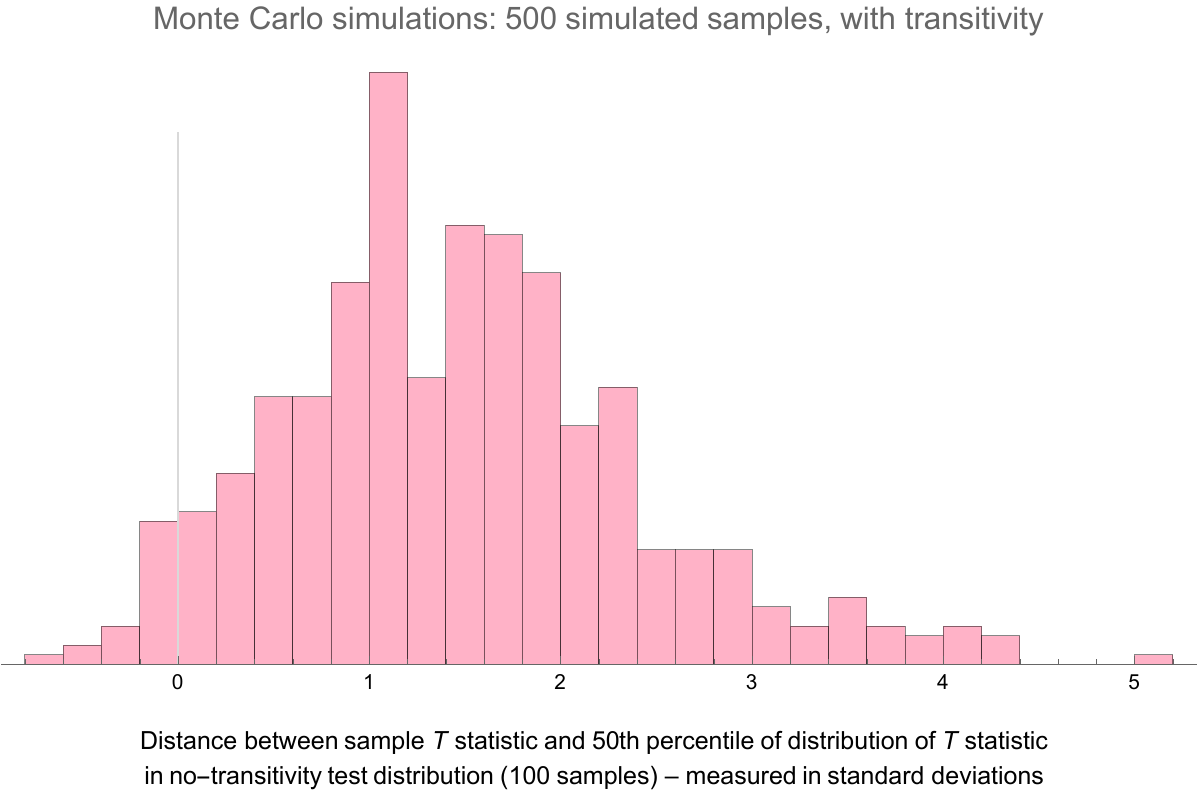}
\includegraphics[height=2.15in]{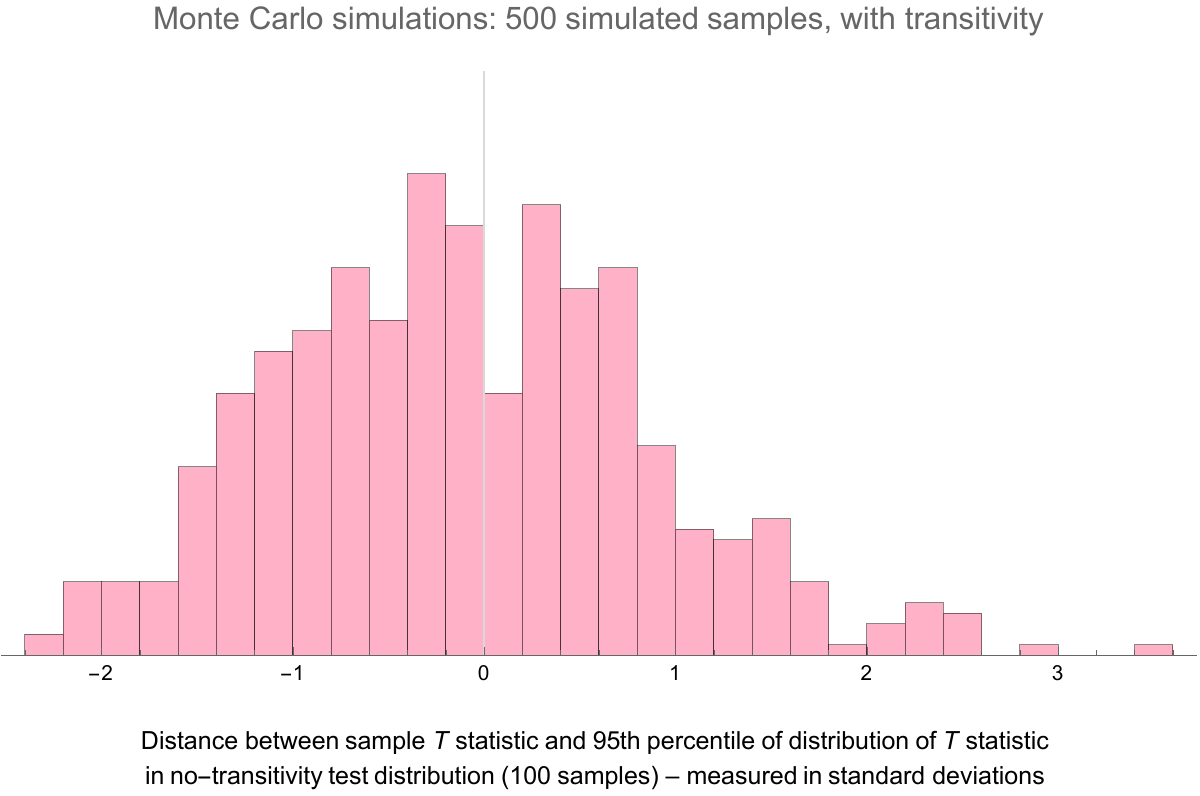}
\hspace*{-0.2in}\\
{\scriptsize\bf Transitivity in the data generating process}\\[20pt]
\hspace*{-0.25in}
\includegraphics[height=2.15in]{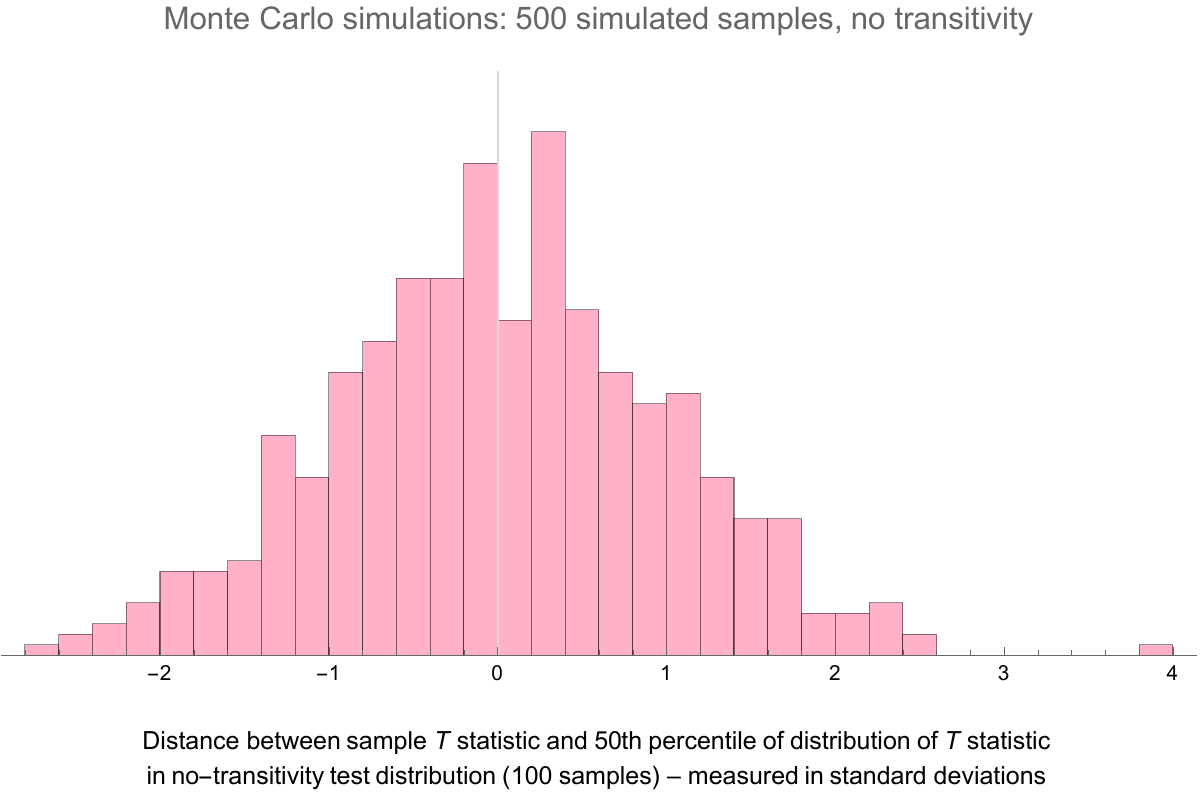}
\includegraphics[height=2.15in]{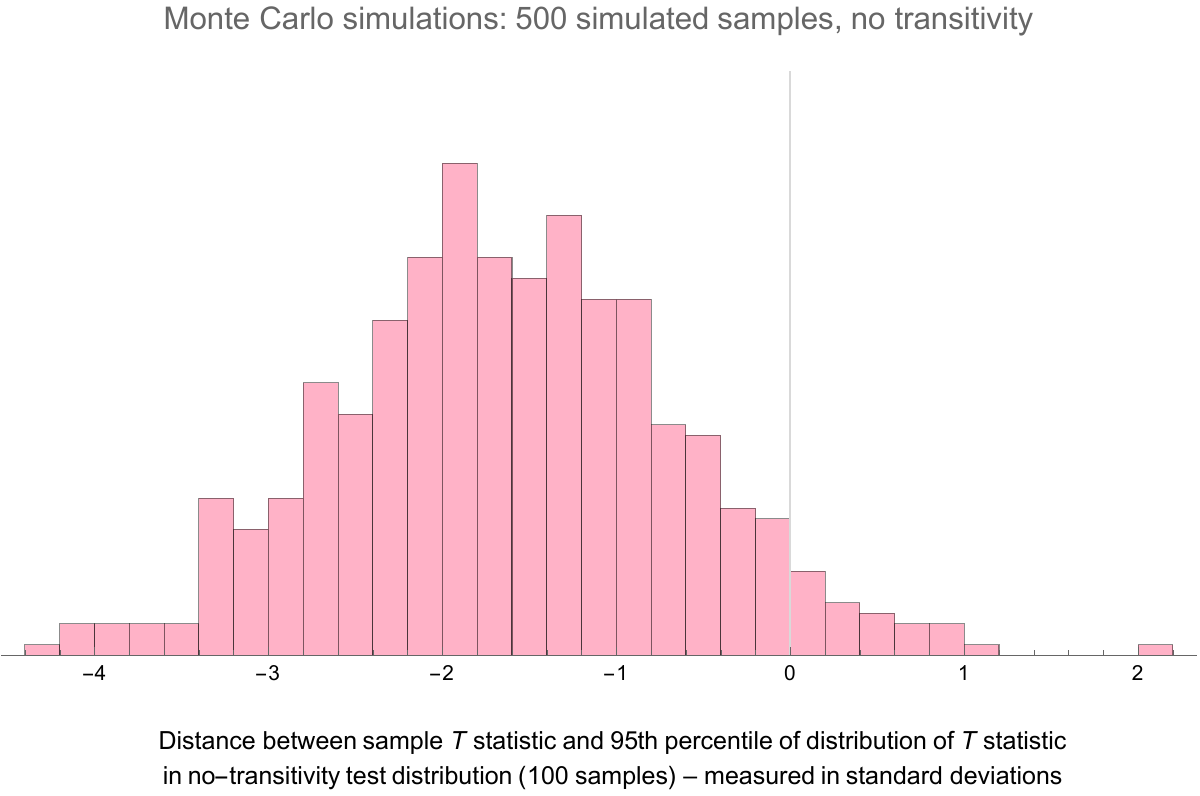}
\hspace*{-0.2in}\\
{\scriptsize\bf No transitivity in the data generating process}
\caption{Monte Carlo simulations}
\label{fig:MC}
\vspace*{-1in}
\end{figure}

We construct synthetic random network samples that either incorporate or exclude transitivity. If we are unable to detect a transitivity effect using our test, there should be no difference between values of the $T$ statistic for data synthetically constructed from a process that features transitivity and a process that does not. We simulate 500 candidate samples and conduct a test for each, generating a distribution of no-transitivity comparable samples for each. Figure \ref{fig:MC} reports the distribution of the distances between the $T$ statistic in each case and the 50th and 95th percentile points of the corresponding comparison distribution for each of the 500 runs. The results indicate a low likelihood of the test delivering false negatives when the test data is from a process featuring transitivity (top panels), and a low likelihood of the test delivering false positives when the test sample is from a process not featuring transitivity (bottom panels). 

{\footnotesize
\begin{landscape}
\begin{table}[p]
  \centering
 \caption{Descriptive statistics for the coverage of the balanced panel}
\vspace*{0.2in}
  \scalebox{0.75}{
    \begin{tabular}{r|ccc|ccc|ccc|ccc|}
          & \multicolumn{3}{c|}{\textbf{Total transaction values}} & \multicolumn{3}{c|}{\textbf{Number of active relationships}} & \multicolumn{3}{c|}{\textbf{Number of active sellers}} & \multicolumn{3}{c|}{\textbf{Number of active buyers}} \\
\cmidrule{2-13}    \multicolumn{1}{c|}{Year} & \multicolumn{1}{p{5.845em}}{Full panel} & \multicolumn{1}{p{4.27em}}{Balanced panel sample} & \multicolumn{1}{p{4.27em}|}{Coverage ratio} & \multicolumn{1}{p{4.885em}}{Full panel} & \multicolumn{1}{p{4.27em}}{Balanced panel sample} & \multicolumn{1}{p{4.27em}|}{Coverage ratio} & \multicolumn{1}{p{4.845em}}{Full panel} & \multicolumn{1}{p{4.27em}}{Balanced panel sample} & \multicolumn{1}{p{4.27em}|}{Coverage ratio} & \multicolumn{1}{p{5.385em}}{Full panel} & \multicolumn{1}{p{4.27em}}{Balanced panel sample} & \multicolumn{1}{p{4.27em}|}{Coverage ratio} \\
    \midrule
    2007  & 1,160 & 1,015 & 0.875 & 3,941 & 2,615 & 0.664 & 536   & 372   & 0.694 & 617   & 359   & 0.582 \\
    2008  & 1,068 & 979   & 0.916 & 4,046 & 2,967 & 0.733 & 523   & 394   & 0.753 & 670   & 397   & 0.593 \\
    2009  & 1,025 & 991   & 0.967 & 4,919 & 4,326 & 0.879 & 494   & 421   & 0.852 & 692   & 484   & 0.699 \\
    2010  & 999   & 992   & 0.993 & 5,325 & 4,987 & 0.937 & 483   & 451   & 0.934 & 681   & 532   & 0.781 \\
    2011  & 948   & 943   & 0.995 & 4,730 & 4,441 & 0.939 & 452   & 415   & 0.918 & 620   & 525   & 0.847 \\
    2012  & 976   & 968   & 0.992 & 4,504 & 4,218 & 0.937 & 418   & 394   & 0.943 & 606   & 522   & 0.861 \\
    2013  & 1,011 & 1,006 & 0.996 & 4,596 & 4,358 & 0.948 & 413   & 392   & 0.949 & 590   & 523   & 0.886 \\
    2014  & 1,012 & 1,010 & 0.998 & 4,717 & 4,560 & 0.967 & 417   & 399   & 0.957 & 602   & 541   & 0.899 \\
    2015  & 955   & 948   & 0.992 & 4,928 & 4,798 & 0.974 & 413   & 399   & 0.966 & 625   & 567   & 0.907 \\
    2016  & 924   & 919   & 0.995 & 5,083 & 4,878 & 0.960 & 415   & 400   & 0.964 & 637   & 561   & 0.881 \\
    2017  & 948   & 941   & 0.993 & 5,145 & 4,595 & 0.893 & 457   & 387   & 0.847 & 668   & 546   & 0.817 \\
    2018  & 949   & 935   & 0.986 & 5,167 & 4,351 & 0.842 & 476   & 379   & 0.796 & 684   & 520   & 0.760 \\
    2019  & 913   & 896   & 0.981 & 5,082 & 4,157 & 0.818 & 454   & 358   & 0.789 & 681   & 487   & 0.715 \\
    \end{tabular}
 }
 \\[5pt]
 \makebox[\textwidth][l]{%
 \scalebox{1.07}{
\begin{minipage}{1.09\textwidth}
\advance\leftskip -2.5cm
	{{\scriptsize{\textit{Notes}: Transaction values are measured in constant USD millions. The balanced panel sample refers to the sellers and buyers for which the panel is rectangularized (exporters that are only active intermittently as well as importers associated with transactions below USD 100 are excluded). The full panel refers to the sample for which no sellers and buyers are excluded. See Section \ref{sec:data} for details.} \par } }
\end{minipage}
}
}
  \label{tab:coverage}%
\end{table}%

\end{landscape}
}

\newpage

\hspace*{-0.3in}
\begin{figure}[h]
\centering
\vspace*{0.4in}
\hspace*{-0.3in}
\includegraphics[height=1.9in]{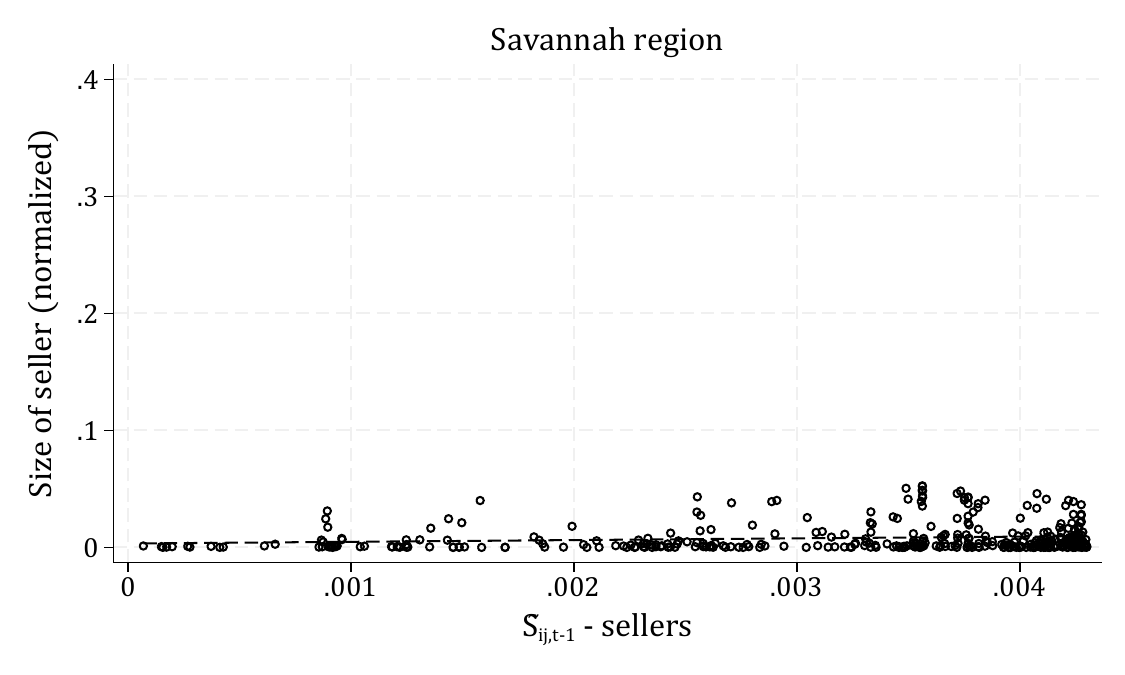}
\includegraphics[height=1.9in]{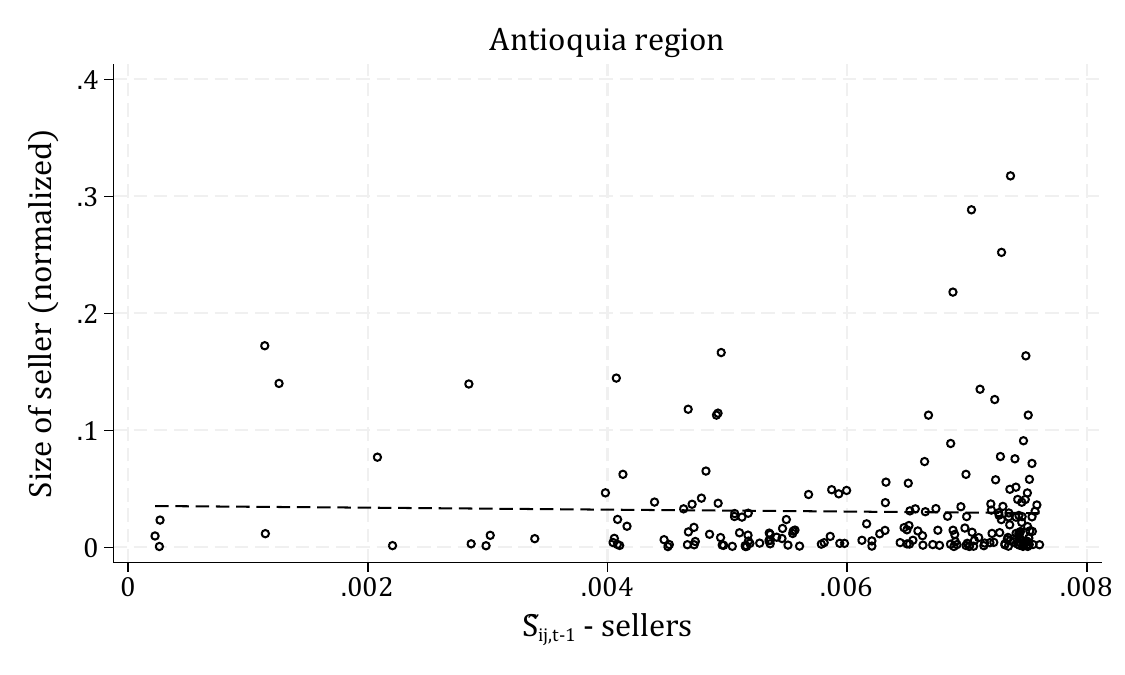}
\hspace*{-0.3in}
\includegraphics[height=1.9in]{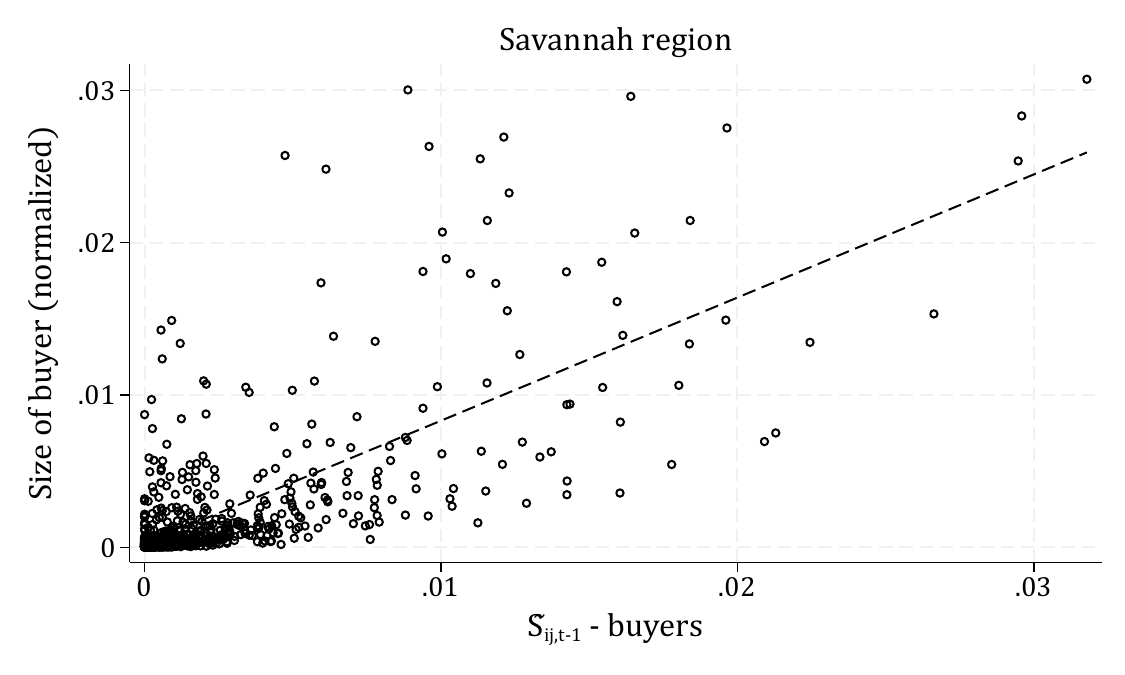}
\includegraphics[height=1.9in]{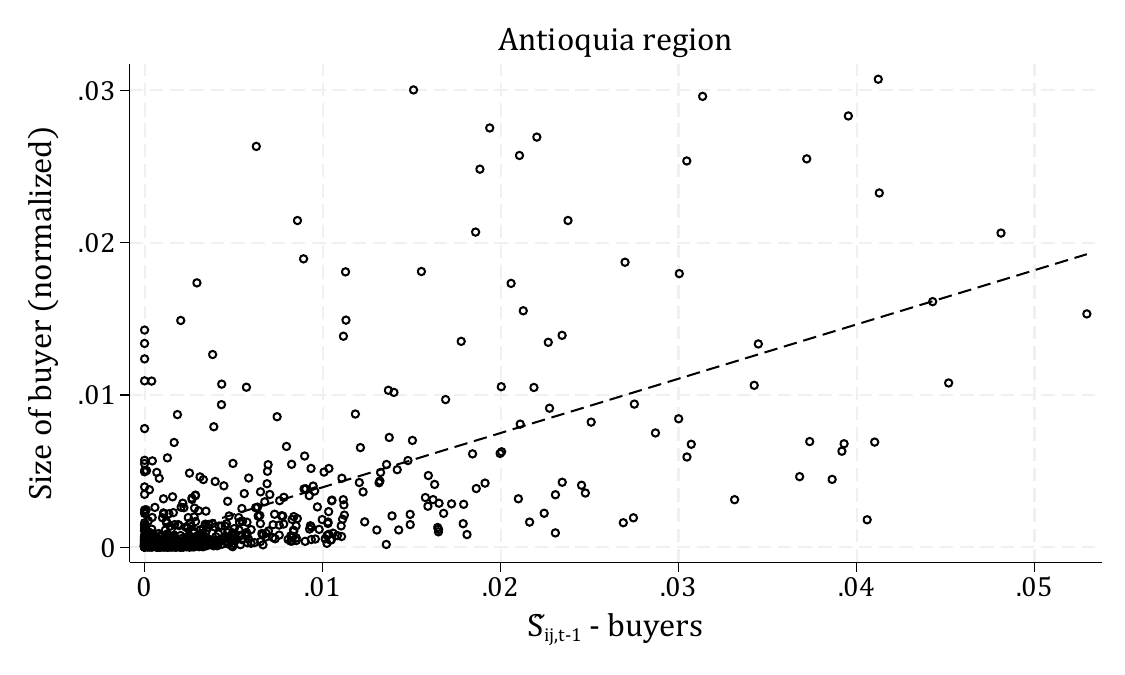}
\vspace*{0.1in}
\caption{Correlation of $\widetilde{S}_{ij}$ with size of buyers and sellers }
\label{fig:corr_size_S}
\vspace*{0.1in}
 \scalebox{0.90}{
\begin{minipage}{1\textwidth}
\advance\leftskip 0cm
	{{\footnotesize{The figures display the correlation between $\tilde{S}$—constructed as the firm-level average—and buyer and seller size, where size is normalized by rescaling to the [0,1] range based on its minimum and maximum values. Observations below the 5th and above the 95th percentiles of the degree distribution are excluded.    }\par } }
\end{minipage}
}

\end{figure}

\begin{figure}[p]
\centering
\vspace*{-0.3in}
   \includegraphics[height=3.3in]{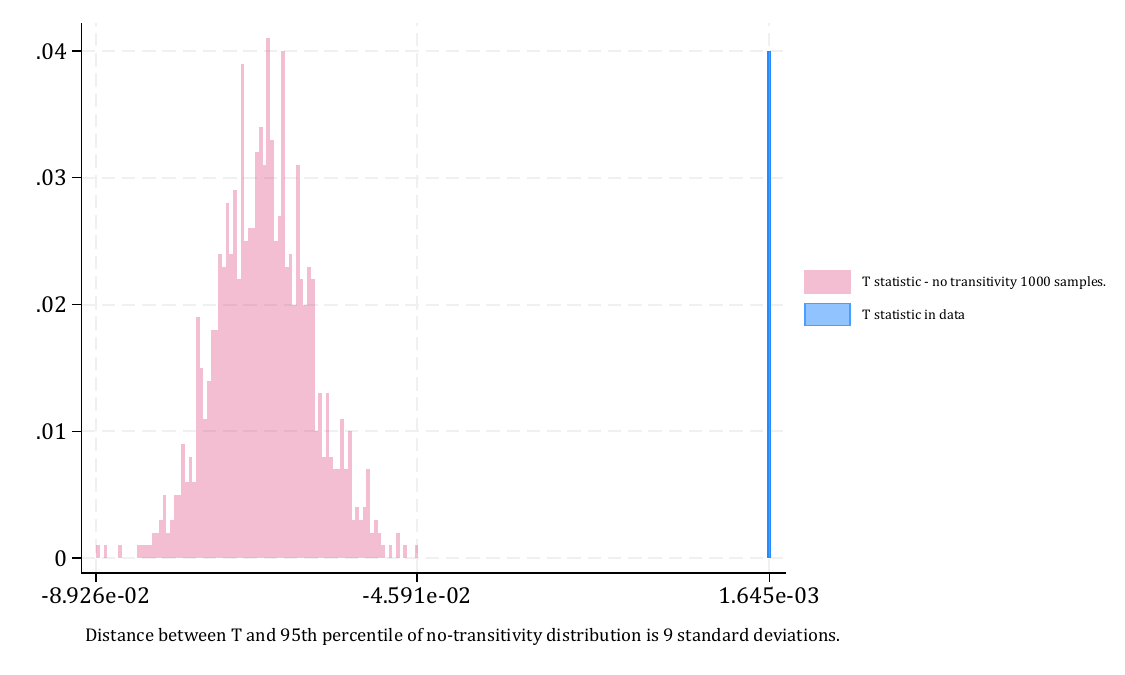}
\vspace*{-0.1in}
\caption{Distribution of alternative $\widetilde{T}$ statistic from our test: Savannah} 
\textbf{(continuous proximity measure)}\vspace*{0.1in}
\label{fig:bog_trans_cont_rob}
\scalebox{0.90}{
\begin{minipage}{0.95\textwidth}
\advance\leftskip 0cm
	{{\footnotesize{The figure shows the distribution of the $\widetilde{T}$ statistic for 1,000 samples without accounting for transitivity in magenta, and the value of the $\widetilde{T}$ statistic obtained from the factual data in blue. The number of importers is 794, and the number of exporters in the Savannah region is 435. The ${\widetilde S}_{ij}$ index is constructed using the rank-normalized proximity measure $z^n_{ki}$.  }\par } }
\end{minipage}
}
\end{figure} 

\begin{figure}[p]
\centering
\vspace*{0.2in}
     \includegraphics[height=3.3in]{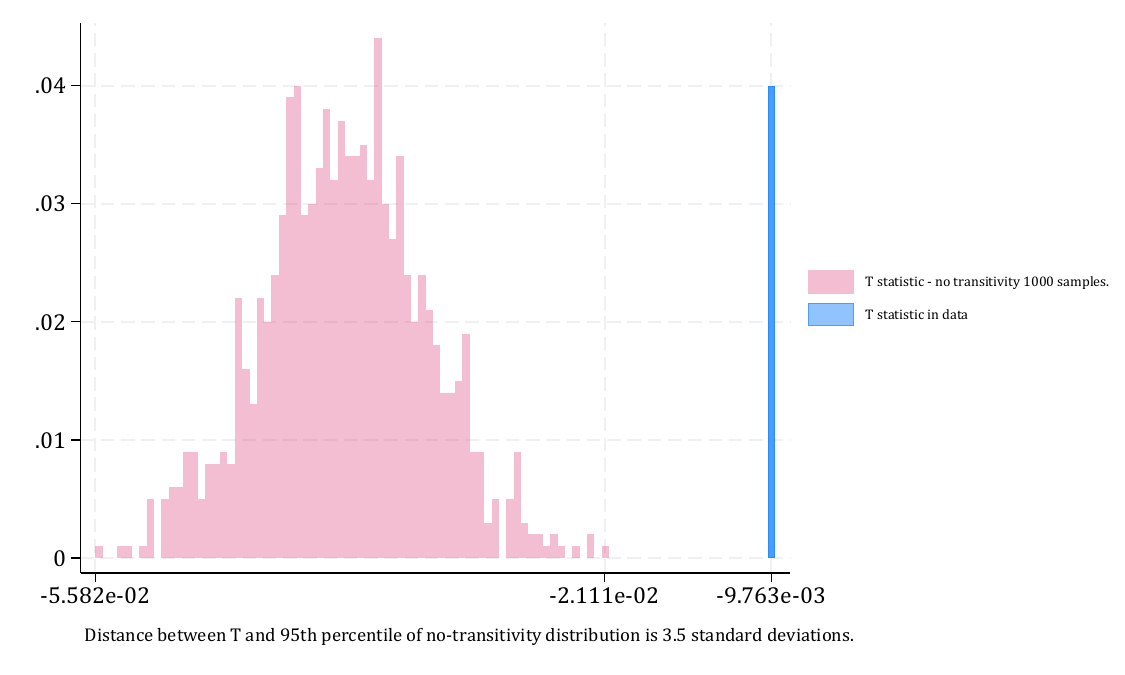}
\vspace*{-0.1in}
\caption{Distribution of alternative $\widetilde{T}$ statistic from our test: Savannah}
\textbf{(discrete proximity measure)}\vspace*{0.1in}
\label{fig:bog_trans_50_rob}
     \scalebox{0.90}{
\begin{minipage}{0.95\textwidth}
\advance\leftskip 0cm
	{{\footnotesize{The figure shows the distribution of the $\widetilde{T}$ statistic for 1,000 samples without accounting for transitivity in magenta, and the value of the $\widetilde{T}$ statistic obtained from the factual data in blue. The number of importers is 794, and the number of exporters in the Savannah region is 435. The ${\widetilde S}_{ij}$ index is constructed using the discrete proximity measure $z^q_{ki}$.  }
    \par } }
\end{minipage}
}
\end{figure} 

\begin{figure}[p]
\centering
\vspace*{-0.3in}
    \includegraphics[height=3.1in]{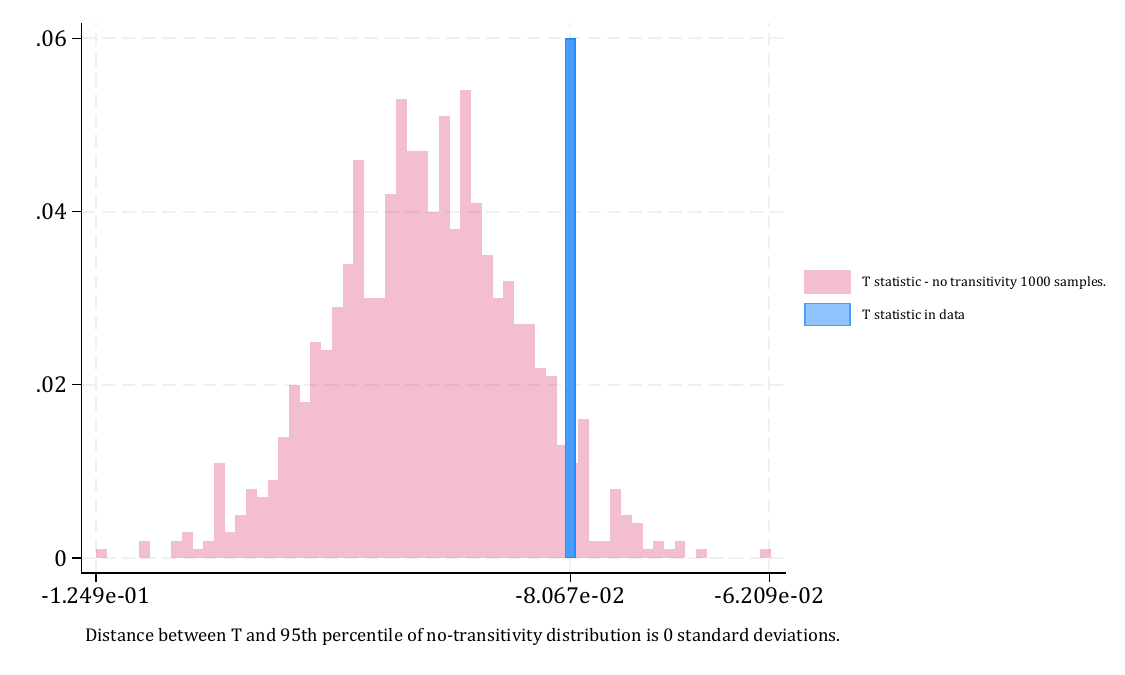}
\caption{Distribution of alternative $\widetilde{T}$ statistic from our test: Antioquia}
\textbf{(continuous proximity measure)}\vspace*{0.1in}
\label{fig:med_trans_cont_rob}
     \scalebox{0.90}{
\begin{minipage}{0.95\textwidth}
\advance\leftskip 0cm
	{{\footnotesize{The figure shows the distribution of the $\widetilde{T}$ statistic for 1,000 samples without accounting for transitivity in magenta, and the value of the $\widetilde{T}$ statistic obtained from the factual data in blue. The number of importers is 794, and the number of exporters in the Antioquia region is 226. The ${\widetilde S}_{ij}$ index is constructed using the rank-normalized proximity measure $z^n_{ki}$.  }
    \par } }
\end{minipage}
}
\end{figure} 

\begin{figure}[p]
\centering
\vspace*{0.2in}
     \includegraphics[height=3.3in]{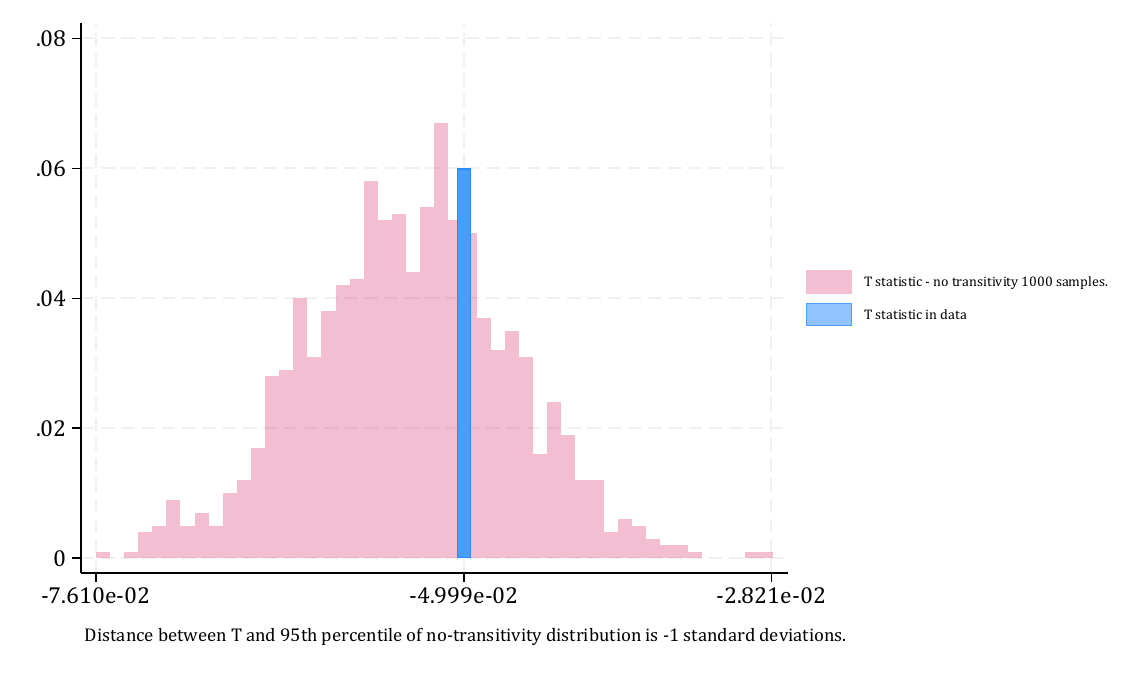}
\caption{Distribution of alternative $\widetilde{T}$ statistic from our test: Antioquia}
\textbf{(discrete proximity measure)}\vspace*{0.1in}\label{fig:med_trans_50_rob}
     \scalebox{0.90}{
\begin{minipage}{0.95\textwidth}
\advance\leftskip 0cm
	{{\footnotesize{The figure shows the distribution of the $\widetilde{T}$ statistic for 1,000 samples without accounting for transitivity in magenta, and the value of the $\widetilde{T}$ statistic obtained from the factual data in blue. The number of importers is 794, and the number of exporters in the Antioquia region is 226. The ${\widetilde S}_{ij}$ index is constructed using the discrete proximity measure $z^q_{ki}$. }\par } }
\end{minipage}
}
\end{figure}

\end{document}